\documentclass[english,12pt,a4paper,oneside]{book}
\usepackage[utf8]{inputenc}
\usepackage[T1]{fontenc}
\usepackage[english]{babel}

\usepackage[
backend=bibtex,
citestyle=numeric-comp,
sorting=none,
maxbibnames=99
]{biblatex}
\AtBeginBibliography{\vspace*{0pt}}

\usepackage{amsmath}
\usepackage{amsfonts}
\usepackage{fancyhdr}
\usepackage{amssymb}
\usepackage{xcolor} 
\definecolor{Prune}{RGB}{99,0,60} 
\definecolor{B1}{RGB}{49,62,72} 
\definecolor{C1}{RGB}{124,135,143}
\definecolor{D1}{RGB}{213,218,223}
\definecolor{A2}{RGB}{198,11,70}
\definecolor{B2}{RGB}{237,20,91}
\definecolor{C2}{RGB}{238,52,35}
\definecolor{D2}{RGB}{243,115,32}
\definecolor{A3}{RGB}{124,42,144}
\definecolor{B3}{RGB}{125,106,175}
\definecolor{C3}{RGB}{198,103,29}
\definecolor{D3}{RGB}{254,188,24}
\definecolor{A4}{RGB}{0,78,125}
\definecolor{B4}{RGB}{14,135,201}
\definecolor{C4}{RGB}{0,148,181}
\definecolor{D4}{RGB}{70,195,210}
\definecolor{A5}{RGB}{0,128,122}
\definecolor{B5}{RGB}{64,183,105}
\definecolor{C5}{RGB}{140,198,62}
\definecolor{D5}{RGB}{213,223,61}
\usepackage{mdframed}
\usepackage{multirow} 
\usepackage{multicol} 
\usepackage{tikz}
\usepackage{graphicx}
\usepackage[absolute]{textpos} 
\usepackage{colortbl}
\usepackage{array}
\usepackage{geometry}
\usepackage{titlesec} 
\usepackage{hyperref}
\hypersetup{ 
    colorlinks=true,
    linkcolor=black,
    urlcolor=purple,
    citecolor=black
}

\begin{document}
\begin{titlepage}



\color{white}

\begin{picture}(0,0)
\put(-152,-743){\rotatebox{90}{\Large \textsc{THESE DE HDR}}} \\
\put(-120,-743){\rotatebox{90}{NNT : }}
\end{picture}

\flushleft
\vspace{10mm} 
\color{Prune}
\fontfamily{cmss}\fontseries{m}\fontsize{22}{26}\selectfont
\centering{\LARGE {Statistical physics of complex  systems}:\\
\Large glasses, spin glasses, continuous constraint satisfaction problems, high-dimensional inference and neural networks}
\normalsize

\color{black}


\fontfamily{fvs}\fontseries{m}\fontsize{8}{12}\selectfont

\vspace{1.5cm}

\normalsize
{Th\`ese de HDR} \\
\textbf{}
 
\vspace{6mm}


\vspace{6mm}

\vspace{15mm}

\bigskip
\normalsize {\color{Prune} \textbf{Pierfrancesco URBANI}} \\
\normalsize{Université Paris-Saclay, CNRS, CEA, \\ Institut de Physique Théorique, \\ 91191, Gif-Sur-Yvette, France}

\vspace{\fill} 

\bigskip
%

\end{titlepage}



\thispagestyle{empty}
\newgeometry{top=1.5cm, bottom=1.25cm, left=2cm, right=2cm}
\fontfamily{rm}\selectfont

\lhead{}
\rhead{}
\rfoot{}
\cfoot{}
\lfoot{}

\noindent 
\titleformat{\chapter}[hang]{\bfseries\Huge\bf\color{Prune}
}{\thechapter\ -}{.1ex}
{\vspace{0.ex}
}
[\vspace{1ex}]
\titlespacing{\chapter}{0pc}{0ex}{0.5pc}

\titleformat{\section}[hang]{\bfseries\Large\bf}{\thesection\ }{0.5pt}
{\vspace{0.1ex}
}
[\vspace{0.1ex}]
\titlespacing{\section}{1.5pc}{4ex plus .1ex minus .2ex}{.8pc}

\titleformat{\subsection}[hang]{\bfseries\large\bf}{\thesubsection\ }{1pt}
{\vspace{0.1ex}
}
[\vspace{0.1ex}] 
\titlespacing{\subsection}{0pc}{4ex plus .1ex minus .2ex}{1pc}

\titleformat{\subsubsection}[hang]{\bfseries\normalsize\bf}{\thesubsubsection\ }{1pt}
{\vspace{0.1ex}
}
[\vspace{0.1ex}] 
\titlespacing{\subsubsection}{3pc}{2ex plus .1ex minus .2ex}{.1pc}

\newgeometry{top=4cm, bottom=4cm, left=2.5cm, right=2.5cm}

\fontfamily{palatino}\fontseries{ppl}\fontsize{11}{11}\selectfont

\newpage
\begingroup
\let\clearpage\relax
\begin{figure}
\includegraphics[width=\columnwidth]{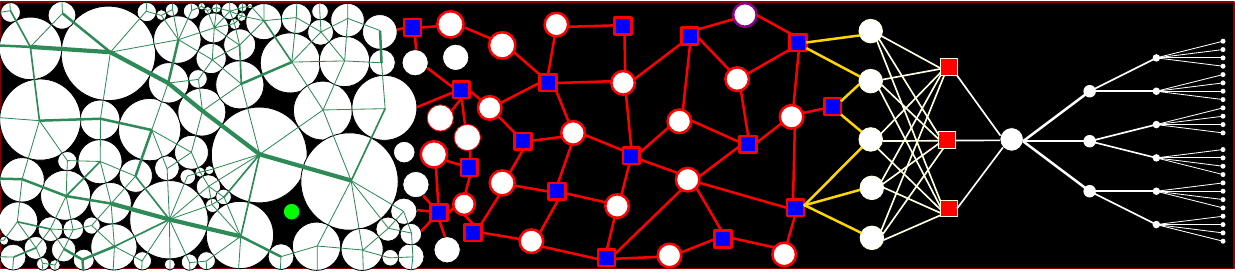}
\end{figure}
\tableofcontents
\endgroup
 
\newgeometry{top=4cm, bottom=4cm, left=2.5cm, right=2.5cm}

\linespread{1.3}

\fontfamily{palatino}\fontseries{ppl}\fontsize{11}{11}\selectfont

\chapter*{Preface}

\section*{\underline{The scope}}
The present thesis has been submitted to obtain the Habilitation \'a Diriger de R\'echerche (HDR).
Its scope is to review my work focusing on my research activity since 2014.
In these years I have been working on several topics. They range from the construction of a theory of structural glasses in the limit of infinite spatial dimensions to high-dimensional disordered optimal control problems.
The thesis will focus on a subset of these topics that covers three, quite diverse, research fields. 
I will try to highlight my contributions to them and how I think that the results that I have obtained may have changed the perspective on a certain number of open problems.
The exposition will not describe the technical details behind the results that I will present. These details are mainly left to the original papers.
Rather, I will try to focus on the main ideas, physical mechanisms, and theoretical methods (when they are of particular interest) that have been introduced and/or used and studied.  I will devote a finite fraction of the exposition to the questions that I believe are fundamental and still open, and I will describe my point of view on them and on how, I think, the progresses I contributed to can be helpful towards a final solution.

The style of the presentation of the thesis will be rather simple. It may be possible that the content may seem superficial and I apologize in advance with the expert reader for that: however I believe that explaining concisely the elementary - but essentially important - aspects of the problems I have been working on may be useful to give a broader picture. Moreover I believe that the scope of this manuscript is not to give a review of all the results on a given topic or to make a detailed description of a selection of the important ones which would be of limited interest, but to review the evolution of my research activity, the spread of my research interests and contributions, my own way of thinking to complex problems and, equally importantly, my activity in collaboration with young researchers. On this particular point I would like to mention that this manuscript covers some topics that have been developed also with a number of PhD and Master students who I would like to acknowledge for their time, collaboration and interest in these research directions.

\section*{\underline{The content}}
\noindent The content of the thesis focuses on three topics: 
\begin{itemize}
\item the problem of the spin glass transition in a field in finite dimensional spin glasses;
\item  the theory of rigidity transitions in zero temperature amorphous solids and how we can model them via continuous constraint satisfaction problems;
\item recent results on the dynamics of optimization and learning algorithms in high-dimensional inference and supervised learning problems.
\end{itemize}
\vspace{0.2cm}

In the first Chapter I will focus on the problem of the spin glass transition in a magnetic field.
This is a longstanding problem in statistical physics of disordered systems which has become central due to the fact that structural glasses, in infinite spatial dimensions, can undergo, in particularly relevant circumstances,  a so called Gardner transition which is in the same universality class as the spin glass transition in a field.
I will review a recently introduced model, the KHGPS model, and show that it suggests a new scenario for the appearance of the spin glass transition in a field. 

\vspace{0.2cm}

In Chapter 2 I will review my activity on the characterization of rigidity/jamming transitions in amorphous systems. 
This line of research started with the works on glassy phases of hard spheres in the limit of infinite spatial dimensions and I kept developing it along the years 
to address the question of how much the results obtained in this setting extend to other problems such as non-spherical particles or very different models like confluent tissues.
While a theory in the infinite dimensional limit is possible in principle in all these cases,
I have been more interested in developing simple models, often abstract and mean-field in nature to describe these problems and their underlying physics. 
The class of models I have been focusing on is the one of continuous
constraint satisfaction and optimization problems which may be relevant in the field of non-convex optimization and may be, in some cases, connected to models of supervised learning in artificial neural networks.
This perspective on these problems is particularly useful because it allows to get a fast theoretical treatment and to perform numerical simulations in some mean-field settings to address also algorithmic questions which have become more and more important in my research activity. 
{The material covered in this chapter overlaps with some review lectures I have given at the IPhT in Saclay in 2017 and at the ICTP-SAIFR School on Disorder Elastic Systems in 2022.}
\vspace{0.2cm}

In Chapter 3, I will give an overview of my activity on high-dimensional statistical inference and learning problems. 
This chapter contains a discussion about two main results. The first is the sub-optimality of gradient based algorithms with respect to message passing ones for solving prototypical high-dimensional statistical inference problems. The second set of results concerns the development of a dynamical mean-field theory analysis of the Stochastic Gradient Descent (SGD) algorithm used to train deep neural networks. I will highlight the connections between the dynamics of SGD and the physics  of active and driven systems and I will discuss how SGD compares with other optimization algorithms in some prototypical settings.
{The material reviewed in this chapter has some overlap with my lectures at the Les Houches School of Physics in 2020.}
\vspace{0.2cm}

%

\section*{\underline{What is original}}
Despite the thesis reviews previously published material, there are a few sections that contains original ideas, results and considerations. Among them I would like to mention the following points that seem  particularly important to me.

\vspace{0.2cm}
In Sec. \ref{math_jamming} I will explore a connection between the scaling solution of the jamming transition and jamming critical systems and the theory of self-intermediate asymptotic solutions of the second kind which is found in the analysis non-linear partial differential equations. This theory has a strong connection with the renormalization group and I will propose some evocative analogies which would suggest further investigations. 

\vspace{0.2cm}
Sec. \ref{off_equilibrium_dynamics} reviews in a critical way the importance of off-equilibrium/gradient descent dynamics. 
I will argue that while off-equilibrium gradient descent dynamics is certainly important in physics, I think that focusing too much on that hides a huge elephant in the room: 
when dealing with optimization problems, gradient descent is never employed as it is written. For example, to get packings of spheres at jamming one rarely employs pure gradient descent
but practical optimization methods rely on a wide range of algorithms whose nature makes them difficult to track analytically.
I will argue that even a detailed analytical solution of gradient descent dynamics, while desirable, does not provide an answer to universality in high dimension which is typically robust and algorithm independent. 
Landscape computations are again insightful but not resolutive because it is difficult to argue what is the right measure over the local minima that are relevant for optimization algorithms. 
I will instead argue that one needs to follow a different approach based on marginal stability and stochastic stability ideas out-of-equilibrium.

\section*{\underline{What is missing}}
This manuscript does not contain technical details about the results that are presented unless they are essential (as in Sec.~\ref{math_jamming} for example).
Therefore technical steps and derivations are left to the original papers. I will often limit myself to the main steps of the derivations only when this is important to present the results.

I want to emphasize that I will not review in detail the construction of the theory of simple glasses in the limit of infinite spatial dimensions. 
The activity on this subject has taken a finite fraction of my work in the last ten years and an account for the results obtained would have been a legitimate part of this report. However a detailed review of this theory has been written in a book \cite{parisi2020theory} published recently together with G. Parisi and F. Zamponi and I feel that the interested reader should take this book as the main reference on this subject. Therefore this thesis contains mostly the results that are not reviewed extensively in that book or that are part of the parallel research activities that I developed during these years and which I feel are as much important as the one on infinite dimensional structural glasses.

Among the research lines I have been focusing recently, I will not cover the one on high-dimensional optimal control problems that attracted my attention in recent years \cite{urbani2021disordered}. 
These problems underly some research directions at the crossroad of the theory of large deviations, instantons in mean-field spin glasses, mean-field Games and the control of many body active matter systems \cite{sinha2023optimal}. 
I started to study the large dimensional limit of disordered optimal control problems in \cite{urbani2021disordered}.  I will not review this topic here but my research activity is still ongoing on this subject.

One of the main directions I have been developing since 2017 is the study of high-dimensional inference and learning problems and this will be reviewed in the third chapter. 
Among the questions I have been exploring on this subject and I cannot discuss in this thesis, one concerns the generalization of Approximate Message Passing algorithms to systems showing replica symmetry breaking (RSB). 
In order to include the effects of RSB, one can develop an Approximate Survey Propagation algorithm as done in \cite{antenucci2019approximate}. This is an example of 
a set of algorithms recently proposed to optimize mean-field spin glasses \cite{subag2021following, montanari2021optimization,alaoui2020algorithmic}.  While I believe that, on mathematical grounds, there is little to add to \cite{subag2021following, montanari2021optimization, alaoui2020algorithmic}, 
I think that the main advantage of the approach developed in \cite{antenucci2019approximate} is that it can be used and extended beyond the problems where it is supposed to be exact. For example one can employ it as a practical approximate algorithm for problems in finite dimension
or optimization problems on random tree-like graphs in the region where glassiness and RSB effects are important and for which there is no exact treatment for the moment.

\fontfamily{palatino}\fontseries{ppl}\fontsize{11}{11}\selectfont

\let\a=\alpha \let\b=\beta \let\g=\gamma \let\d=\delta
\let\e=\varepsilon \let\z=\zeta \let\h=\eta \let\k=\kappa
\let\l=\lambda \let\m=\mu \let\n=\nu \let\x=\xi \let\p=\pi
\let\s=\sigma \let\t=\tau \let\f=\varphi \let\ph=\varphi\let\c=\chi
\let\ps=\psi \let\y=\upsilon \let\si=\varsigma \let\G=\Gamma
\let\D=\Delta \let\Th=\Theta\let\L=\Lambda \let\X=\Xi \let\P=\Pi
\let\Si=\Sigma \let\F=\Phi \let\Ps=\Psi \let\Y=\Upsilon
\let\ee=\epsilon \let\r=\rho \let\th=\theta \let\io=\infty
\let\om=\omega \def\ie{{i.e. }}\def\eg{{e.g. }}
\renewcommand{\vec}[1]{\mathbf{#1}}
\def\vr{{\vec r}}
\def\vk{{\vec k}}

\def\PP{{\cal P}}\def\EE{{\cal E}}\def\MM{{\cal M}} \def\VV{{\cal V}}

\def\CC{{\cal C}}\def\FF{{\cal F}} \def\HH{{\cal H}}\def\WW{{\cal W}}
\def\TT{{\cal T}}\def\NN{{\cal N}} \def\BB{{\cal B}}\def\II{{\cal I}}
\def\RR{{\cal R}}\def\LL{{\cal L}} \def\JJ{{\cal J}} \def\OO{{\cal O}}
\def\DD{{\cal D}}\def\AA{{\cal A}}
\def\GG{{\cal G}} \def\SS{{\cal S}}
\def\KK{{\cal K}}\def\UU{{\cal U}} \def\QQ{{\cal Q}} \def\XX{{\cal X}}
\def\YY{{\cal Y}}\def\ZZ{{\cal Z}}

\def\Re{{\rm Re}\,}\def\Im{{\rm Im}\,}

\def\hh{{\bf h}} \def\HHH{{\bf H}} \def\AAA{{\bf A}} \def\qq{{\bf q}}
\def\BBB{{\bf B}} \def\XXX{{\bf X}} \def\PPP{{\bf P}} \def\pp{{\bf p}}
\def\vv{{\bf v}} \def\xx{{\bf x}} \def\yy{{\bf y}} \def\zz{{\bf z}}
\def\aaa{{\bf a}}\def\bbb{{\bf b}}\def\hhh{{\bf h}}\def\III{{\bf I}}
\def\uu{{\bf u}}
\def\kmin{\kappa_{\rm min}}
\def\kmax{\kappa_{\rm max}}
\def\qm{q_m}
\def\qM{q_{M}}

\def\de{\mathrm{d}}
\def\ul{\underline} \def\olu{{\overline{u}}} \def\erf{\text{erf}}
\def\ol{\overline}

\def\wh{\widehat}
\def\wt{\widetilde}
\newcommand{\sumN}{\sum_{i=1}^N}
\newcommand{\sumn}{\sum_{a=1}^n}
\newcommand{\M}{\mathcal{M}}
\newcommand{\qt}{\tilde{q}}
\newcommand{\dpart}[2]{\frac{\partial #1}{\partial #2}}
\newcommand{\thav}[1]{\left< #1 \right>}
\newcommand{\thavo}[1]{\left< #1 \right>_0}
\newcommand{\thavn}[1]{\left< #1 \right>_n}
\newcommand{\kappat}{\tilde{\kappa}}
\newcommand{\qh}{\hat{q}}
\newcommand{\dbar}{{\,\mathchar'26\mkern-12mu d}}

\newcommand{\beq}{\begin{equation}} 
\newcommand{\eeq}{\end{equation}}
\newcommand{\ba}{\begin{eqnarray}}
\newcommand{\ea}{\end{eqnarray}}

\chapter{The KHGPS model, soft modes in glasses and the spin glass transition in a field}

\section{Introduction}
The simplest examples of spin glasses are non-magnetic materials (Manganese for example) surrounding magnetic impurities (Copper for example) placed at random in space.
The magnetic impurities interact through the Ruderman-Kittel-Kasuya-Yosida (RKKY) potential which describes a long-ranged dipolar interaction \cite{ruderman1954indirect, kasuya1956theory, yosida1957magnetic}. Since the position of the impurities is random in space, the RKKY potential couples their magnetic moments either ferromagnetically or antiferromagnetically. 
The reader interested in the experimental aspects of spin glasses can find a review in the monographs  \cite{mydosh1993spin, mydosh2015spin, vincent2022spin}.

When cooled down to low temperatures, spin glasses undergo a phase transition to a spin glass phase
where the magnetic moments of the impurities, the relevant degrees of freedom, are, on average, frozen into random directions.
While there is little doubt about the emergence of a spin glass transition under certain conditions (zero external magnetic field),
the nature and structure of the spin glass phase is still controversial.
In the past fifty years an intense research activity combining theoretical approaches, experiments and dedicated
numerical simulations has been dedicated to try to understand this problem and lift the degeneracy between competing theoretical scenarios.

A parallel but rather complementary and independent line of research was started long time ago to understand
the mechanism behind the glass transition and the nature of structural glasses at low temperature. 
These physical systems appear to be rather different from spin glasses and, as such, are modeled in very different ways.
Spin glasses are in general described by spins, possibly of various nature, from Ising to continuous spins, whose interaction potential
is described in terms of a set of random couplings extracted from a probability distribution. Therefore, spin glass
models contain {\it quenched disorder} in the form on random interactions between the degrees of freedom.
On the contrary, structural glasses are made of molecules or particles and therefore have to be modeled as particle
systems whose interaction potential is free of any form of disorder or randomness. 
Despite this crucial difference, these two research lines have deeply influenced each other and got in contact many times
in the past forty years. Indeed, both systems, at low temperature, are frozen into
amorphous structures and we would like to understand their statistical properties.

It is fair to say that nowadays a mean-field theory of both spin \cite{MPV87} and structural glasses \cite{parisi2020theory} has emerged. 
At the mean-field level, both kinds of systems are understood in terms of the low temperature structure of pure states of the Gibbs-Boltzmann (GB) measure. This is described by the replica
symmetry breaking theory which is a body of concepts to precisely characterize complex landscapes of pure states. 
The difference between structural and spin glasses is mainly found in the pattern of replica symmetry breaking that characterizes the spin glass and the structural glass transitions.
This difference has an important impact in the critical properties of the transitions themselves and in the nature of the physical responses of these systems in the low temperature phase.

One of the main outcome of the mean-field theory of structural glasses is that in a certain number of cases and under  a set of external drives, 
these systems can undergo a so called {\it Gardner transition}.
At the mean-field level, this transition is in the same universality class of the spin glass transition in an external magnetic field
and this provides a natural point of contact between the physics of spin and structural glasses.
The Gardner transition has emerged as a new theoretical ingredient to understand many universal but rather elusive aspects of glasses
at low temperature.
Therefore, understanding better its nature as well as the one of the spin glass transition and the spin glass phase
beyond mean-field has become an important problem which may be relevant for the whole field of amorphous materials broadly speaking.

The purpose of this chapter is to review the mean-field theory of spin and structural glasses explaining
the connection between the two through the Gardner transition. We will then discuss how the predictions of these mean-field theories
work to describe finite dimensional systems analyzing the successes and, most importantly, the failures.
In an attempt to overcome some of the latter, we will discuss a new spin glass model, the so called KHGPS model, originally introduced to 
understand the density of states of non-phononic vibrations in structural glasses.
We will argue that this model and its physical behavior provide a new perspective on the fate of the finite dimensional spin glass transition in a magnetic field.

\section{Theory of spin and structural glasses}
In this section we will review the mean-field theory of both spin and structural glasses.
This is meant to be a summary of the main findings and the interested reader
can look at more extended books on these subjects, see \cite{MPV87, parisi2020theory}.

The simplest spin glass Hamiltonian describes the so called Edwards-Anderson (EA) model \cite{edwards1975theory}.
Consider $N=L^d$ Ising spins $\sigma_i=\pm 1$ arranged on a $d$-dimensional lattice ${\mathbb Z}^{d}$
of size $L^d$. The Hamiltonian of the EA model is given by
\beq
H_J[\underline \sigma] = -\frac{1}{\sqrt{z(d)}}\sum_{i\sim j} J_{ij}\sigma_i\sigma_j  - B \sum_{i}\sigma_i\:.
\label{HEA}
\eeq
The first sum runs only on spins that are nearest neighbors on the lattice and the couplings $J_{ij}$ are random
and extracted independently from a fixed probability distribution (the same for each edge on the lattice).
For convenience we assume that $J_{ij}$ are normal random variables (Gaussian with zero mean and unit variance).
The prefactor of the nearest neighbor coupling term in Eq.~\eqref{HEA} depends on $z(d)$ that is the connectivity of the underlying lattice.
For a square lattice in $d$ dimensions, $z(d)=2d$, but it is clear that the Hamiltonian in Eq.~\eqref{HEA} can be generalized to arbitrary interaction topologies. 
This rescaling of the coupling strength between spins is such that the model has a well defined limit when $d\to \infty$.
The EA Hamiltonian contains also a ${\mathbb Z}_2$-symmetry breaking term which is given by the interactions between the spins and an external magnetic field $B$. 

We are interested in studying the properties of the Gibbs-Boltzmann (GB) measure defined as
\beq
\begin{split}
P_J(\underline \sigma; \beta, L) &= \frac 1 {Z_J(\beta, L)}\exp\left[-\beta H_J[\underline \sigma]\right]\\
Z_J(\beta, L) &= \sum_{\underline \sigma} \exp\left[-\beta H_J[\underline \sigma]\right]\:.
\end{split}
\label{GB_EA}
\eeq
This is a random probability measure in the sense that it depends on the random variables $J_{ij}$. It is also a function of $\beta$ which denotes the inverse of the temperature $T$, $\beta=1/T$ (we set the Boltzmann constant to 1).

The spin glass problem consists in trying to understand and describe what is the structure of the GB measure defined in Eq.~\eqref{GB_EA} at low temperature.
When the spatial dimension is infinite, a solution of this problem has been established both via theoretical physics methods \cite{MPV87} and on mathematically rigorous grounds \cite{talagrand2003spin, panchenko2013sherrington}. This solution is based on the replica symmetry breaking picture which describes the organization and the properties of the pure states of the GB measure. Given its importance we will review the essential aspects of this solution in the next section.
However, as soon as the dimension is finite, the structure of the GB measure is still unclear and in the following we will discuss the research activity on this problem. 

\subsection{{Mean-field theory of spin glasses}}\label{Sec_MFT_spin_glasses}
\begin{figure}
\centering
\includegraphics[width=0.5\columnwidth]{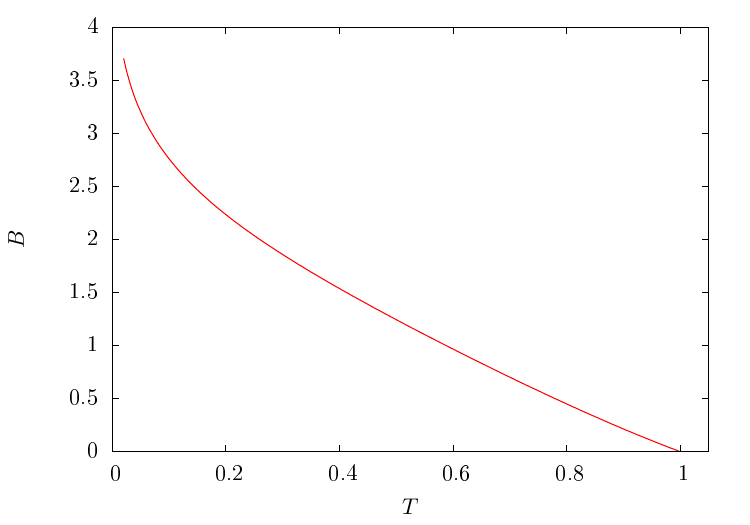}
\linespread{0.8}
\caption{\footnotesize{The phase diagram of the SK model as a function of the temperature and external magnetic field. Above the red line the model is in the RS phase while below
it replica symmetry is broken and a complex free energy landscape of pure states is found. The transition line diverges at infinite values of the external field in the zero temperature limit.}}
\label{FIG_SK}
\end{figure}

After the introduction of the Edwards-Anderson spin glass Hamiltonian, Sherrington and Kirkpatrick considered
a mean-field version of the model in which the spins interact on a complete graph, see  \cite{sherrington1975solvable, kirkpatrick1978infinite}. 
The corresponding Hamiltonian is
\beq
H_{J,B}^{SK}[\underline \sigma] = -\frac{1}{\sqrt N}\sum_{i<j} J_{ij}\sigma_i\sigma_j  - B \sum_{i}\sigma_i
\label{HSK}
\eeq
and we have denoted by $N$ the total number of spins.
We are interested in studying this model in the thermodynamic limit where $N\to \infty$.
Note that the scaling of the interaction term in Eq.~\eqref{HSK} is coherent with the one of Eq.~\eqref{HEA}.
Sherrington and Kirkpatrick proposed a solution of their model based on the replica method. 
This formalism allows one to study the average free energy defined as
\beq
\begin{split}
f(\b, B) &= -\frac 1\b\lim_{N\to \infty} \overline{\ln Z_J(\b, B)}\\
Z_J(\b,B) &= \sum_{\underline \s}e^{-\b H_{J,B}^{SK}[\underline \sigma]} 
\end{split}
\eeq
where the overline denotes the average over the random couplings.
In the $N\to \infty$ limit, the average free energy is controlled by a saddle point over a non-trivial order parameter
which describes the different phases of the model.
Sherrington and Kirkpatrick solved the corresponding saddle point equations
assuming a particular form for the solution called the replica symmetric (RS) ansatz.
However this form of the solution turned out to be physically inconsistent: for sufficiently low temperatures
this implied a negative value of the entropy\footnote{The entropy must be positive semidefinite for systems with discrete degrees of freedom.}.
It was then understood by de Almeida and Thouless \cite{de1978stability} that this inconsistency was essentially due to the form of the SK mean-field solution 
which breaks down at a critical temperature. Below a critical line in the temperature-external field phase diagram, see Fig.~\ref{FIG_SK},
the replica symmetric structure of the corresponding variational problem is not correct and a new solution must be found.
This was obtained by Parisi who, in a series of works \cite{parisi1980sequence,parisi1979infinite,parisi1980order}, introduced the RSB scheme
which was later found to have a deep physical significance \cite{parisi1983order,mezard1984replica} and proven to be correct \cite{talagrand2003spin, panchenko2013sherrington}.

The essence of the RSB solution is that below the critical line identified in \cite{de1978stability} the structure of the pure states of the GB measure becomes complex. 
A way to characterize it is by introducing a non-trivial order parameter: given two configurations $\underline \sigma_1$ and $\underline \sigma_2$ 
both extracted from the same GB measure with the same disorder, the overlap $q$ between them is a measure of their similarity and it is defined by $q=\underline \sigma_1\cdot \underline \sigma_2/N$.
The overlap is a random variable and it makes sense to study its probability distribution $P_J(q)$ given by
\beq
P_J(q)=\sum_{\underline \sigma_1, \underline \sigma_2}\frac{e^{-\beta (H_{J,B}^{SK}[\underline \sigma_1]+H_{J,B}^{SK}[\underline \sigma_2])}}{Z_J(\beta,B)^2} \delta \left(q-N^{-1}\underline \sigma_1\cdot \underline \sigma_2\right)\:.
\label{sample_Pq}
\eeq
We note that $P_J(q)$ is disorder-dependent and therefore it reasonable to study its behavior when averaging 
over the random couplings $J$ \footnote{Sample-to-sample fluctuations of $P_J(q)$ turn out to be very interesting too and are controlled by a set of non-trivial relations called Ghirlanda-Guerra identities \cite{MPV87,ghirlanda1998general}.}:
\beq
P(q)=\overline{P_J(q)}\:.
\eeq
Above the critical line in the temperature-field plane, a detailed computation \cite{parisi1983order} gives
\beq
P(q) = \delta (q-q_{EA})
\label{RS_Pq}
\eeq
with $q_{\rm EA}\geq 0$ the so called Edwards-Anderson order parameter.
This quantity depends on both the temperature and the external magnetic field. 
Indeed, when $B=0$ for example, it is easy to show that if the ${\mathbb Z}_2$ symmetry is not spontaneously broken,
then $q_{EA}=0$. Conversely, as soon as $B>0$ one finds $q_{EA}>0$.
The physical meaning of Eq.~\eqref{RS_Pq} is that in the replica symmetric phase the GB measure is composed by one pure
state (a single valley in the free energy landscape) and, given the mean-field nature of the model, this pure state is entirely
described by the overlap which concentrates in the large $N$ limit.

\begin{figure}
\centering
\includegraphics[width=0.5\columnwidth]{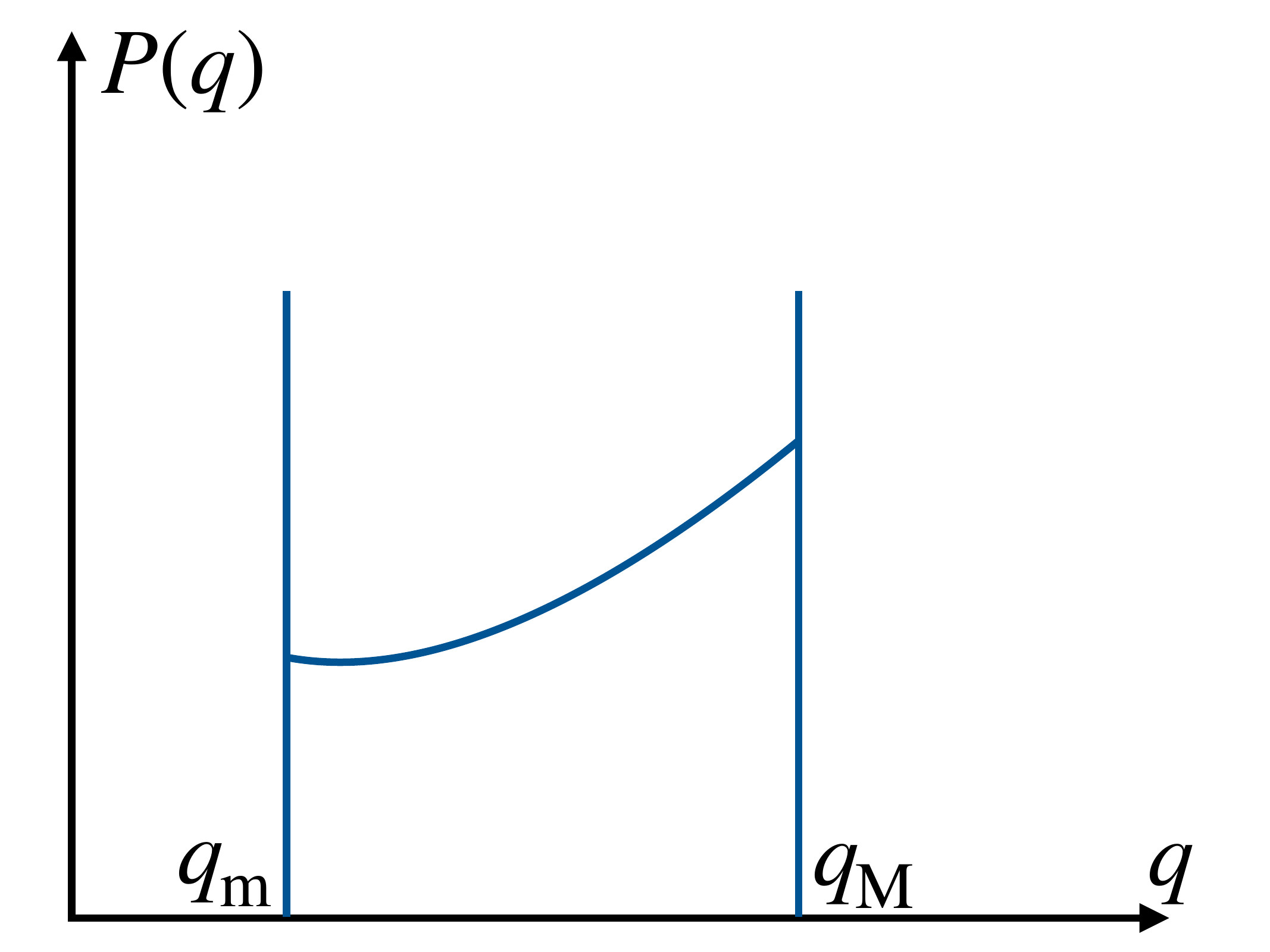}
\linespread{0.8}
\caption{\footnotesize{A sketch of the for of $P(q)$ in the RSB phase for a model displaying full replica symmetry breaking. The value of $q_m$ is sensible to $B$. In particular for $B=0$ one has $q_m=0$ and  $P(q)$ contains an additional symmetric branch for $q\in [-q_M,0]$.}}
\label{sketchP_q}
\end{figure}

This simplicity in the structure of pure states changes completely at low temperatures, below the spin glass transition, in the RSB region.
Here there exists two values, $q_{\rm m}$ and $q_{\rm M}$, $0\leq q_{\rm m}<q_{\rm M}$, such that $P(q)>0$ for $q\in [q_{\rm m},q_{\rm M}]$
(a sketch of the shape of $P(q)$ can be found in Fig.\ref{sketchP_q}).

The physical meaning of these findings is that in this case the GB measure can be decomposed into a non-trivial structure of pure states.
The geometry of the corresponding space has been shown to posses an ultrametric structure \cite{mezard1984replica, panchenko2013parisi} controlled by $P(q)$.
We note that when the external field is absent, $B=0$, the form of $P(q)$ is symmetric around the origin.
This is just due to ${\mathbb Z}_2$ symmetry and it implies that for each pure state, one can find another one by flipping all the spins (or, better, their average local magnetization \cite{thouless1977solution}).
This symmetry in the structure of $P(q)$ is lifted as soon as $B>0$.

At this point two important remarks are mandatory.
Firstly, the RSB phase transition is found also for $B>0$. The Hamiltonian in Eq.~\eqref{HSK} has no symmetry left as soon as the magnetic field is positive and therefore the phase transition is driven by an unconventional pattern of symmetry breaking (RSB) which encodes for the statistical structure of pure states in the low temperature phase.
Secondly, it is found that in the SK model the transition line in the $T-B$ plane diverges for $T\to 0$.
Therefore, at zero temperature, the SK model is in a RSB phase regardless the value of the external field.
Both observations turn out to be relevant for the rest of this chapter.

\subsection{Finite dimensional spin glasses and the problem of the transition in a field}
The emergence of RSB in finite dimensional systems and the nature of the low temperature phase in spin glasses
 is still a debated matter in the field.
A natural way to clarify this issue is to understand whether a phase transition can arise when an external
magnetic field is present. We now clarify better this point.

As we discussed in the previous section, a non-trivial prediction of the RSB picture is that a complex structure of pure states appears also
when ${\mathbb Z}_2$ symmetry is explicitly broken by $B>0$.
Renormalization Group (RG) studies have established that a perturbative fixed point of the RG flow  below the upper critical
dimension (which is supposed to be\footnote{In recent studies \cite{angelini2022unexpected} it has been suggested that the upper critical dimension should be 8 if the transition is dominated by a zero temperature critical point.} 6) is found at criticality within an $\epsilon$-expansion \cite{wilson1974renormalization} at zero external field.
However, Bray and Roberts \cite{bray1980renormalisation} have shown that the RG flow 
has no perturbative fixed point close to the Gaussian one as soon as $B>0$.
In this case, below the upper critical dimension the Gaussian fixed point becomes unstable and no other perturbative fixed points are found.
This analysis has been confirmed by more extended studies, see \cite{temesvari2002generic, temesvari2002replica, pimentel2002spin}.
In recent years, such RG computations have been revisited and in particular, in \cite{charbonneau2017nontrivial},
a two loop RG computation has been performed. The result of this analysis shows that the RG flow has a 
new fixed point below the upper critical dimension but this is found in a non-perturbative region in 
the space of relevant coupling constants. Therefore this analysis is of limited validity.
The absence of a perturbative fixed point below the upper critical dimension
was soon taken as an indication that the RSB transition does not take place in three dimensions.

In the eighties, an alternative picture to the RSB one, the {\it droplet theory}, was proposed \cite{fisher1986ordered, fisher1988equilibrium, 
moore2011disappearance}. 
According to it, the pure state structure of the low temperature, finite dimensional
GB measure of the EA Hamiltonian is composed by two pure states at $B=0$ one of which is lifted if $B>0$.
This implies that as soon as the external magnetic field is finite, there is no spin glass transition upon decreasing temperature. 
Therefore, a way to discriminate between the droplet and the RSB pictures is to investigate what happens 
to the spin glass transition in a field beyond mean-field theory.

In the last forty years an extensive amount of studies has been done to clarify the tension between these two scenarios.
In particular an entire research program has been set up to investigate the problem using numerical
simulations. The interested reader can look at a series of review papers published recently \cite{martin2022numerical}.
From the theoretical viewpoint, in recent years, Parisi and Temesvari \cite{parisi2012replica} have proposed
that if a phase transition in a field is found in finite dimensions, this may be controlled in the RG-flow sense,
by a zero temperature critical point. A very similar picture happens in the Random Field Ising Model (RFIM) where 
a ferromagnetic phase transition appears below a critical line in the $T-\Delta B$ plane, being $\Delta B$ the variance
of the random fields. In this case one can show that the $\Delta B=0$ critical point is unstable upon the RG flow
as soon as $\Delta B>0$. Under renormalization, the flow of the RG is attracted by the zero temperature critical point. This is the reason why, for the RFIM, a number of studies
have been focusing on the behavior of the critical theory at the $T=0$ critical point, see \cite{parisi1979random, tissier2011supersymmetry, parisi2014critical, kaviraj2022parisi, rychkov2023four}.

Parisi and Temesvari suggested that a similar picture may happen for the spin glass transition in a field.
Therefore they proposed to look directly at the zero temperature critical point in the (T-B) plane and check its robustness away from mean-field theory. 
In practice this would suggest to consider a mean-field spin glass model with a zero temperature critical point and study what happens when finite dimensional fluctuations are switched on.
The main obstacle to this program is that the principal mean-field spin glass model, the SK model, lacks a phase transition at zero temperature 
being the model always in the RSB phase regardless the value of $B$, see Fig.\ref{FIG_SK}.

To overcome this difficulty, mean-field models defined on finitely connected lattices have been studied.
The simplest one is the Viana-Bray model \cite{viana1985phase}: this is an Ising spin glass defined on random
locally tree-like graphs. At zero temperature, this model has a RSB transition at a finite value of the external magnetic field.
The critical theory of the model has been studied in great details along the corresponding finite temperature critical line \cite{parisi2014diluted}.
More recently, a RG program has been set up to study finite dimensional fluctuations starting from this model. 
The main idea of \cite{Parisi:2011oZ, efetov1990effective, sacksteder2007sums} is to use this model as the starting point of a loop expansion.
A detailed investigation of this procedure has been started in \cite{angelini2022unexpected} where a careful study of the upper critical dimension has been performed.
Despite all these research efforts, establishing whether an RSB transition can take place in finite dimensions
is still an open issue. Even more open is the nature of the possible RSB phase and how to describe its properties \cite{perrupato2022ising}.

In Sec.\ref{Sec_KHGPS} we  discuss a bona fide fully connected model in which local heterogeneities are introduced
and which shows a spin glass transition in a field at zero temperature. 
We show that the transition can be in two different universality
classes and that one of them suggests a more plausible mechanism for the appearance of the RSB transition
in finite dimensional systems.
Before discussing this model it is useful to explain why this problem is important for the physics of amorphous solids
and how this recent research line has been a drive to define the model that we discuss in Sec.\ref{Sec_KHGPS}.

\subsection{Mean-field theory of structural glasses and the Gardner transition}\label{sec_Gardner}
One of the reasons why the spin glass transition in a field has seen a renewed interest in recent years 
is that within mean-field theory and under some circumstances, structural glasses
can undergo a phase transition called the Gardner transition, which is in the same (mean-field) universality
class as the spin glass transition in a field. 
The emergence of the Gardner transition has been crucial to describe a lot of interesting findings in finite dimensional glasses, and signatures
of the transition have been found in both simulations and experiments, mainly on colloidal glasses \cite{berthier2016growing,seguin2016experimental, hammond2020experimental}.
The purpose of this section is to review the main results of this research line and to link them to the activity
around the spin glass transition problem.

The construction of a mean-field theory of structural glasses  started long time ago when Kirkpatrick, Thirumalai and Wolynes
understood that a certain class of mean-field spin glasses (spherical $p$-spin glasses and Potts spin glasses) have 
the same phenomenological phase diagram that is found in experiments on structural glasses 
\cite{kirkpatrick1987dynamics, kirkpatrick1987stable, kirkpatrick1987connections, kirkpatrick1988mean, kirkpatrick1987p}.
This observation generated a stream of works and set the basis of the so called Random First Order Transition (RFOT) Theory
of structural glasses \cite{kirkpatrick1989scaling, xia2000fragilities, lubchenko2007theory,berthier2011theoretical}.
However it is fair to say that such a theory was heavily based on an analogy rather than a mapping between structural and spin glass models
and it took a long time and a series of breakthroughs to show that such analogy can be put on a more formal and concrete ground.
In the last ten years this research activity culminated in the formulation of a mean-field theory of structural glasses based on the limit
of infinite spatial dimensions.
The full construction can be found in \cite{parisi2020theory} and here we review just the main ideas behind it
stressing the main outcomes. The interested reader can find a similar, more concise presentation in a series of recent reviews on the subject
\cite{charbonneau2017glass, berthier2019gardner, urbani2022low}.

The main difference between structural glasses and spin glasses is that the latter contain quenched disorder.
Paradoxically, this seems a major problem with respect to the former but actually it turns out to be a simplification. Models
with quenched disorder can be treated via the replica method \cite{MPV87} and the spin glass phases in which spins are frozen
in amorphous structures can be straightforwardly understood since quenched disorder provides the driving mechanism
to the emergence of such locally stable (metastable) structures.
Conversely, the way in which structural glasses get stuck in amorphous configurations is more complicated
and disorder is dynamically {\it self-generated}. How to develop a theory of amorphous phases in models that do not have  quenched
disorder was a major challenge in the development of the theory. 
Conceptually a solution to this problem was proposed in a few ways, with different
but rather equivalent constructions \cite{monasson1995structural, franz1995recipes, mezard1999thermodynamics}. 
Here we review the construction based on the so called Franz-Parisi potential \cite{franz1995recipes}.

To a first approximation, glasses are amorphous solids in which the degrees of freedom, or particles, are caged by their neighbors.
Inside their cages particles move and vibrate and this motion can be understood as a thermal motion around a set of equilibrium
average positions which form an amorphous structure of lattice points statistically equivalent to a configuration of the system itself
\footnote{Here we mean that the static structure factor and radial correlation function of the equilibrium lattice points and of the glass itself are essentially the same.}. 
From this point of view these equilibrium positions can be seen as the quenched, self-generated, disorder which pin the dynamics in the glass phase. 
The main difference with respect to spin glasses is that this disorder is also self-sustained by the particles themselves.
In other words when the temperature is sufficiently low, 
the configurations of phase space explored by particle vibrations are not sufficient to destroy (at least on short timescales)
the amorphous lattice around which particles move. Conversely it is the vibration of the particles that cage neighboring particles
and generate frustration in the system.

This simple idea can be translated into a well defined mathematical setting. 
Let us consider a system of $N$ particles in $d$ spatial dimensions. Let us denote by $\underline x_i$ a $d$-dimensional vector coding for the position of particle $i$ and 
by $X=\{\underline x_1,\ldots,\underline x_N\}$ the collective coordinates of the system in phase space. 
Let us assume that the Hamiltonian of the system is written in terms of a local interaction potential $v(r)$ so that
\beq
H[X] = \sum_{i<j}v(|\underline x_i- \underline x_j|)\:.
\eeq
We introduce the mean square displacement $\D(X,Y)$ as a measure of the distance  between two configurations in phase space
\beq
\D(X,Y) = \frac 1N \sum_{i=1}^N |\underline x_i -\underline y_i|^2\:.
\eeq
We are interested in studying the free energy of a glass and, according to the discussion above, this is meant to be approximately the free energy 
of particle vibrations around equilibrium configurations statistically equivalent to the typical positions of the particles themselves.
Therefore we introduce the following free energy
\beq
V_Y(\Delta_r,\beta_s) =-\frac 1N\log \int \de X e^{-\beta_s H[X]}\delta (\Delta_r - \Delta(X,Y))\:.
\label{constrained_FE}
\eeq
This function is a random quantity since it depends on the configuration $Y$ which plays the role of quenched disorder
and pins the phase space where the configurations $X$ are sampled. 
The configuration $Y$ represents also the amorphous lattice around which particles vibrate.
The free energy in Eq.~\eqref{constrained_FE} assumes that the configuration $X$ is sampled at inverse temperature $\beta_s$ which is therefore a control parameter in the problem.

Since we are interested in the average properties of glassy structures
it makes sense to average $V_Y$ over the configuration $Y$ so that we arrive at
\beq
\begin{split}
V_{FP}(\Delta_r,\beta_r, \beta_m) &=-\frac 1N \int \de Y \frac{e^{-\beta_m H[Y]}}{Z[\beta_m]}\log \int \de X e^{-\beta_s H[X]}\delta (\Delta_r - \Delta(X,Y)) \\
Z[\beta_m]&=\int \de Y e^{-\beta_m H[Y]}\:.
\end{split}
\eeq
It is important to note that the configuration $Y$ is sampled from a GB measure at inverse temperature $\beta_m$ which is different, in general, from $\beta_s$.
The function $V_{FP}$ is called the Franz-Parisi (FP) potential \cite{franz1995recipes}
and can be computed when the number of spatial dimensions goes to infinity.
Here we just review the outcome of this computation and the interested reader can look at \cite{parisi2020theory,rainone2015following} for more details.

We start by considering the case in which $\beta_s=\beta_m=\beta$. 
As a function of $\D_r$, the FP potential  is a convex function with a global minimum at $\Delta_r\to \infty$ for $\beta<\beta_d<\infty$.
This corresponds to the high temperature liquid phase where the configuration $Y$ is not able to pin the configuration $X$
which is sufficiently hot to diffuse in phase space.
At $\beta=\beta_d$ the FP potential develops a saddle at a finite value of $\Delta_r=\Delta_r^*$ and this corresponds to the so called dynamical/Mode Coupling transition \cite{gotze2009complex}.
For $\beta>\beta_d$ this saddle becomes a local minimum. 
The physical significance of this local minimum is clear: for sufficiently low temperatures
the free energy landscape develops metastable states which are nothing but glassy states.
It can be shown that the number of such states is exponential in $d$ and 
therefore the probability to find the system in one of them is exponentially small. 
However if $Y$ is a typical configuration sampled from one of these glasses
it can pin the configurations $X$ so that they stay close to $Y$ at a finite value of the mean square
displacement $\Delta_r$ that corresponds to the one for which the FP potential has a local minimum.

It must be noted that a non-convex free energy can be meaningful only in an infinite dimensional setting where metastable states become truly stable,
see Chapter 1 of \cite{parisi2020theory} for a detailed discussion on this point.
In any finite dimensions, such local minimum disappears due to the Maxwell construction\footnote{This reflects the fact that the free energy is formally a Legendre transform and therefore it must be convex.} and the system has a global minimum for $\Delta_r\to\infty$.
It turns out that the local minimum at $\Delta_r^*$ becomes the global minimum of the FP potential at an inverse critical temperature $\beta_K$.
This is the so called Kauzmann or ideal glass transition \cite{kauzmann1948nature}. 
At this point the number of metastable glassy states contributing to the GB measure become
sub-exponential. Correspondingly the probability to find the system in one of them becomes finite
and therefore the system is thermodynamically stable\footnote{Unless there is a crystalline state with lower free energy.}. 

Despite the physics of ideal glasses is an important and interesting subject we do not investigate it here (see the discussion below).
Instead we focus on the regime in which typical glassy states sampled from the GB measure are exponentially many and metastable.
We are interested in understanding the properties of these states when they are followed in temperature (but other perturbations
such as compression or shear strain \cite{rainone2015following}, can be studied as well).
The FP formalism offers a way to investigate this problem: the system $Y$ essentially selects one of the glassy states at intermediate inverse temperatures
$\beta_m\in[\beta_d,\beta_K]$ and the inverse temperature $\beta_s\geq \beta_m$ is then increased to probe how these glasses behave upon cooling.
It can be also shown that this way of proceeding as a clear physical significance: indeed this {\it mismatched} FP potential can be shown to 
describe the adiabatic cooling dynamics for example, see \cite{agoritsas2019out}.

In order to study the glassy state selected by $Y$ one can repeat the same argument as followed in spin glasses.
Imagine to consider to configurations $X_1$ and $X_2$ both coupled to the same configuration $Y$ and extracted according to the constrained
measure in Eq.~\eqref{constrained_FE} where $\Delta_r^*$ is the value of $\Delta_r$ in the local minimum of the FP potential. 
It makes sense to study the probability distribution of their mean square displacement
\beq
\begin{split}
P_Y(\hat \D,\beta_s) &= \int \de X_1\, \int \de X_2 \frac{e^{-\beta_s (H[X_1]+H[X_2])}}{Z_Y[\beta_s]}\delta(\hat \D - \D(X_1,X_2))\prod_{i=1,2} \delta(\D_r^*-\D(X_i,Y))\\
Z_Y[\beta_s]&= \int \de X e^{-\beta_sH[X]} \delta(\D_r^*-\D(X,Y))
\end{split}
\eeq
The distribution of the mean square displacement $P_Y(\hat \D)$ plays the same role as the distribution of the overlap
between two configurations in spin glasses, see Eq.~\eqref{sample_Pq}.
It makes sense to consider its average over the distribution of $Y$
\beq
P(\hat \D,\beta_s) = \int \de Y \frac{e^{-\beta_m H[Y]}}{Z[\beta_m]}P_Y[\hat \Delta]\:.
\eeq
A detailed computation \cite{parisi2020theory} shows that for sufficiently small $\beta_s$ close to $\beta_m$ one has
\beq
P(\hat \D) = \delta(\hat \Delta - \Delta_{EA}(\beta_m, \beta_s))
\eeq
and in particular $\Delta_{EA}(\beta_m, \beta_m) = \Delta_r^*$.
However under some circumstances \cite{parisi2020theory}, there exists a transition point $\beta_G$
such that for $\beta_s>\beta_G$ the distribution of $P(\hat \D)$ is not a Dirac delta but it has a continuous shape
as it happens in the RSB phase of spin glasses. 
The critical point is called Gardner transition and it corresponds to an out-of-equilibrium version of a transition to RSB found in certain spin glasses
in \cite{gardner1985spin} and \cite{gross1985mean}, see also \cite{rizzo2013replica, montanari2003nature, montanari2004instability}.
The transition signals the point where the internal free energy landscape of a glass metabasin
starts to have a structure which is controlled by the RSB picture.
At the mean-field level this phase transition has the same properties of the spin glass transition in a field where the distribution of the overlap changes from a Dirac
delta to a broad function \cite{urbani2015gardner}.

\subsubsection{The consequences of the Gardner transition}
The Gardner transition in structural glasses was clearly identified in the context of infinite dimensional hard spheres at high pressure \cite{kurchan2013exact}
and then it was found in many other mean-field computations of simple glass formers under several perturbations, see \cite{parisi2020theory} for a review.
The consequences of the Gardner transition on the properties of infinite dimensional low temperature glasses can be summarized as follows:
\begin{itemize}
\item At the Gardner transition elasticity breaks down and the response of the system becomes intermittent and driven by avalanche-like events. 
This can be understood by looking at the non-linear elastic moduli of glasses: these are susceptibilities that describe the response of solids to 
shear deformations and in glasses they can fluctuate from one glassy state to another. 
In \cite{biroli2016breakdown} it has been shown that while such quantities have a typical value that is smooth and finite across the Gardner point, their sample
to sample fluctuations may diverge at the transition. This means that, crossing the transition, different glasses can respond in a rather different way and 
therefore in some glassy states the response to a strain can be very large. This is an alternative way to interpret avalanche-like/spatially heterogeneous responses.
Moreover the average response can be history dependent \cite{jin2017exploring}.
Correspondingly the system becomes highly correlated and diverging length scales can be defined \cite{berthier2016growing}.
Low temperature amorphous solids are typically found to have intermittent behavior under deformations and appear to be critical in the sense
that the can have large responses to small perturbations \cite{nicolas2018deformation,bonn2017yield}. The Gardner transition offers a way to understand this phenomenology in terms
of the corresponding low temperature structure of glass metabasins.
\item One of the most spectacular predictions of Gardner physics is the values of the critical exponents of the jamming transition of hard spheres and jamming critical systems.
This is a topic that will be reviewed in Chapter \ref{Ch_jamming}. In a nutshell, when hard sphere glasses jam into random packings, they are described by universal features controlled by a set of critical exponents. In \cite{charbonneau2014fractal, charbonneau2014exact} these exponents have been computed from the infinite dimensional solution of hard sphere glasses at jamming. They are also found to control the physics of jamming-critical systems far from jamming \cite{franz2019critical} and appear to be in very good agreement with numerical simulations.
\end{itemize}

\subsection{Finite dimensional glasses: in and out-of-equilibrium}
The mean-field theory of structural glasses predicts that their phase diagram is described by a series of phase transitions.
Here we summarize what is know about their fate in finite dimensions.

\begin{itemize}
\item Dynamical/Mode-Coupling transition: this becomes a crossover in any finite dimensions. Indeed metastable glassy states
have always a finite lifetime in finite dimensions if $\beta<\beta_K$. This is due to activated jump processes between them. 
How glasses behave around the Mode-Coupling crossover has been recently investigated in \cite{charbonneau2022dimensional} for the case of hard spheres by changing the spatial dimension and a detailed comparison with mean-field theory has been done to check how this limit is approached when the dimension is increased. 
A dynamical theory of the dynamical crossover has been also developed recently, see \cite{rizzo2015qualitative,rizzo2016dynamical, rizzo2020solvable}. 
Finally, describing this activated regime is an important problem in the field, see the recent works on the subject \cite{lopatin1999instantons, rizzo2021path,ros2021dynamical,ros2020distribution}.
\item Kauzmann/ideal glass transition: its existence is still a matter of debate. One of the main issues in this case is that in finite dimensions, the relaxation time
is supposed to diverge at this point in an exponential way (the Vogel-Fulcher law \cite{berthier2011theoretical}). Therefore it is essentially impossible to probe the ideal glass phase
and my personal point of view is that while an assessment of this problem is important, the physics of glasses will be always the physics of the out-of-equilibrium glassy structures that can be probed experimentally and numerically.
\item Gardner transition: we already pointed out that this transition is in the same mean-field universality class as the spin glass transition in a field. 
It is important to note the Gardner phase for glasses prepared at $\beta<\beta_K$ is metastable as much as glass phase itself.
Therefore, strictly speaking,  it does not make sense to think about it as a true transition.
However, for all practical purposes, the lifetime of glassy states is so large that to a first approximation it can be taken to be infinite with respect to 
the microscopic timescales. Therefore it may be reasonable to probe if the RSB physics arises when out-of-equilibrium glasses are cooled down across the Gardner point. 
This issue has been recently
addressed in a series of works concluding that some indications of Gardner physics are present for 2$d$ and $3d$ colloidal glasses  at high pressures \cite{berthier2016growing,seguin2016experimental, hammond2020experimental} while soft molecular glasses
do not show mean-field signatures of the transition \cite{albert2021searching, scalliet2017absence}. 
\end{itemize}
The reasons why Gardner physics is more visible in colloidal systems
will be discussed at the end of this chapter. This is probably due to the incipient jamming transition which is always present for these systems. 

Soft (molecular) glasses are instead found to be in disagreement with some quantitative predictions of the infinite dimensional theory.
For them, jamming is irrelevant because either they do not have a proper jamming point or are far from jamming.
The purpose of the next part of this chapter is to review one specific aspect of finite dimensional low temperature glasses, the behavior of the density of states
of soft glasses, whose quantitative  behavior is not captured by the theory in infinite dimensions.
To solve this problem we will introduce a new model, the KHGPS model, outline its solution and discuss a new physical mechanism
for the spin glass transition in an external magnetic field.

\section{Soft linear modes in glasses and the KHGPS model}\label{Sec_KHGPS}
In the previous sections we have briefly reviewed the theory of spin and structural glasses
and highlighted that they come very close in specific regions of the phase diagrams of these systems
since at the mean-field level structural glasses
undergo a Gardner transition which is in the same universality class as the spin glass
transition in an external magnetic field. 
However, while Gardner physics seems
visible in the case of colloidal glasses and, in particular, in hard spheres, it remains more elusive in the case of soft glasses.

One of the main point of quantitative disagreement between mean-field computations and three dimensional numerical
simulations of model systems of soft glasses is the behavior of the vibrational density of states at low frequencies.
Naive mean-field theory \cite{franz2015universal} predicts in fact that the Gardner phase gives rise to a non-phononic vibrational density
of states with arbitrarily soft vibrational modes. 
Numerical simulations of three dimensional soft glasses \cite{lerner2021low} find soft non-phononic vibrational states but their density
at low frequencies follows a robust (and seemingly universal) behavior which is not quantitatively captured by mean-field theory.
In other words, while Gardner physics captures the emergence of an abundance of soft-vibrational modes, it is not able
to capture its quantitative features.

To cure this problem new ideas have emerged \cite{bouchbinder2021low, rainone2021mean, folena2022marginal, urbani2022field}.
As we will review in the following, in a nutshell one can refine the mean-field theory of structural glasses by introducing local heterogeneities.
These heterogeneities naturally emerge in finite dimensions due to lack of concentration of statistical properties across space,
 and can be put by hand in mean-field models.
It can be shown that they are sufficient to produce a mean-field theory of the density of states in structural glasses which is in agreement
with numerical simulations.
However while these ideas clearly reconcile mean-field theory with finite dimensional numerical simulations,
we will see that an interesting offspring of them is to provide a fresh view on the problem of the spin glass transition in a field.

\newpage

\subsection{Soft modes in glasses: low energy linear excitations}
A way to probe the physics of amorphous systems is to look at their low energy excitations.
The simplest ones are harmonic excitations which are encoded in the vibrational density of states.
Consider a glass well equilibrated at a temperature $T_m$ and call a typical configuration
of this glass $X_m$.
Starting from $X_m$, one can find an inherent structure of the potential energy landscape by minimizing
the Hamiltonian through gradient descent\footnote{In practical numerical simulations one can use other, more efficient, minimization algorithms
like conjugate gradient descent.}
\beq
\begin{cases}
\underline{\dot x}_i(t) &=-\frac{\partial H[X]}{\partial \underline x_i}\\
X(0)&=X_m\:.
\end{cases}
\eeq
At the end of this dynamics we get a configuration of the degrees of freedom, let us call it $X^*$, that is in a local minimum of the Hamiltonian.
The Hessian of the Hamiltonian at this point, also called dynamical matrix,
is given by
\beq
\HH_{i\alpha,j\beta}=\frac{\partial^2 H[X^*]}{\partial x_i^{(\a)}\partial x_j^{(\b)}} \ \ \ \ \ i,j=1,\ldots, N,\ \ \ \a,\b=1,\ldots, d
\eeq
and it approximates the landscape around the local minimum of $H$ by a quadratic form
\beq
\begin{split}
H[X]&\simeq \frac 12 \sum_{i\a,j\b}\HH_{i\alpha,j\beta} \delta x_i^{(\alpha)}\delta x_j^{(\beta)}\\
\underline \delta x_i &= \underline x_i-\underline x^*_i\:.
\end{split}
\eeq
The spectrum of the Hessian contains a description of the normal modes of vibrations of the system.
In general this is a random object since it depends on the local minimum selected
by gradient descent which is in general an amorphous configuration. The Hessian also depends on the initial configuration $X_m$ and on the control
parameters that define macroscopically the statistics of the initial glass states that are considered (for example their temperature).
Therefore it is reasonable to consider the spectrum of the Hessian averaged over different initial conditions $X_m$ extracted
from configurations of glasses prepared with the same protocol and control parameters.

Let us call the eigenvalues of the Hessian $\l_i$ and the corresponding eigenfrequencies $\omega_i =\sqrt{\l_i}$.
We are interested in studying
\beq
D(\omega)= \frac{1}{d(N-1)}\left\langle\sum_{i=1}^{d(N-1)}\delta(\omega-\omega_i)\right\rangle
\eeq
and the brackets denote the average over the different glasses and local minima.
Note also that the factor $d(N-1)$ is essentially the total number of eigenmodes\footnote{Note that in a generic numerical simulation, periodic boundary conditions are imposed to confine the system so that $d$ eigenmodes are always zero and are trivial space translational modes which can be removed from the analysis.}.
The form of $D(\omega)$ for $\omega\to 0$ encodes the softest vibrational modes in the system.
Since on long wavelengths glasses are homogeneous materials, the low frequency tail of $D(\omega)$
is always populated by phonons which are approximate plane wave eigenvectors.
Their density is  given by the Debye law, $D(\omega)\sim \omega^{d-1}$, and the corresponding eigenvectors are delocalized.

In recent years however extensive numerical simulations have shown that, together with phononic modes,
the low frequency tail of the dynamical matrix is populated by soft quasi-localized modes, see
\cite{lerner2021low} for a recent review. In particular it has been shown that the density of such soft modes follows a rather robust
law, namely 
\beq
D_{\rm loc}(\omega)\sim A_4 \omega^4.
\eeq
The prefactor $A_4$ is in general very dependent on the details of the system preparation, but the quartic tail is found to be rather robust.

Given that the density of these vibrational modes in three dimensional systems is much smaller than the phononic
one controlled by Debye theory, several numerical strategies have been put in place to detect such modes. 
Here we list some of them:
\begin{itemize}
\item {\it Analysis of the Inverse Participation Ratio of the eigenmodes}: A way to carefully distinguish between
phononic modes that are mostly plane waves and soft localized quartic modes is to look at the localization
properties of the corresponding eigenvectors. In particular, one can measure the participation ratio of the eigenvectors 
as a function of their frequency. A precise numerical study in a large variety of model systems has been done
in \cite{mizuno2017continuum, shimada2018spatial, shimada2018anomalous}.
\item {\it Finite size analysis}. The density of phonons in a finite size system is bounded from below
by a minimal frequency that scales as $\omega_{\rm ph}^{\rm min}\sim 1/L$ being $L$ the linear size of the system. Therefore if the quartic modes
have a localization length that is smaller than $L$ for sufficiently small $L$, one can use a small finite size to push
up in frequency the phononic part of the density of states so that for $\omega\ll \omega_{\rm ph}^{\rm min}$ one is left with the quartic part of the $D(\omega)$.
This strategy has been used in \cite{lerner2016statistics,kapteijns2018universal, paoluzzi2020probing}.
\item {\it Random Pinning}. A similar way to push up the phononic part of the vibrational spectrum is to freeze at random
the position of some particles. This breaks translational invariance explicitly and therefore one achieves essentially the same effect of taking a finite system size.
This strategy has been used in \cite{angelani2018probing, shiraishi2023non}.
\end{itemize}
The result of all these numerical studies, see also \cite{das2020robustness, bonfanti2020universal, ji2020thermal, guerra2022universal, kumar2021density, mizuno2020anharmonic, shimada2020vibrational}, is that model glasses, regardless their preparation, always display soft quasi-localized modes
whose density follows a quartic tail. The prefactor $A_4$ instead depends on the glass preparation and details: in general, the more annealed the glass
the smaller $A_4$ \cite{wang2019low}. 
The only exception to this picture is found in harmonic soft spheres right above their jamming point. In this case, jamming physics imposes that $D(\omega)\sim {\rm const}$ for $\omega\to 0$ due to isostaticity \cite{liu2010jamming}.

A mean-field theory of soft harmonic excitations in glasses based on the Gardner phase was discussed in \cite{franz2015universal}
where a simple and abstract model of harmonic soft spheres, was analyzed. This model will be reviewed in Chapter \ref{Ch_jamming}.
This analysis concluded that in the Gardner phase, far from the jamming point, the vibrational density of states contains non-phononic soft modes but they follow a different law, $D(\omega)\sim \omega^2$
while at jamming $D(0)>0$.
It must be noted that the mean-field model analyzed in \cite{franz2015universal}
does not contain phonons by construction and therefore the $D(\omega)$ should be understood as the non-phononic part
of the density of states.
Furthermore, the eigenvectors corresponding to the low frequency tail of $D(\omega)$ are delocalized which
is a remnant of the mean-field (fully connected) nature of the model considered in that study.
Therefore while the Gardner phase provides a mechanism for the generation of non-phononic soft excitations, their density and properties
differ from what is found in $3d$ models of structural glasses.

\subsection{The KHGPS model}
A theoretical framework to explain the soft non-phononic part or the spectrum was started long time ago.
This was triggered by a parallel research line in glass physics aiming at understanding the low temperature (quantum) properties of molecular glasses.
Here we briefly review this activity and link it to the problem of soft harmonic modes in glasses.

Early in the seventies, Zeller and P\"ohl \cite{zeller1971thermal}  discovered that structural glasses display an anomalously large specific heat as compared to their crystalline counterpart.
The Debye theory would predict that for a $3d$ system, phononic excitations give rise to a specific heat which should follow a cubic behavior at low temperature, $C_V\sim T^3$.
However, glasses display a much larger specific heat which is found to follow a linear in temperature behavior: $C_V\sim T$. 
It is clear that adding the non-phononic part of the spectrum of harmonic excitations to the Debye analysis would not increase $C_V$ significantly since this would contribute only through a term of order $T^5$ in the specific heat.
Therefore Anderson Halperin and Varma \cite{anderson1972anomalous,leggett2013tunneling} suggested that in addition to soft linear excitations, glasses should contain soft {\it non-linear} excitations
which should be the crucial ingredients to understand the low-T anomalous specific heat.

In the simplest setting, these non-linear excitations consist in tunneling two-level systems (TLS). These can be thought as groups of degrees of freedom
that rearrange collectively and can be found in two locally stable configurations separated by an energy barrier.
If one assumes certain statistical properties for these TLS, it is easy to provide a phenomenological mechanism
for the generation of the linear specific heat.
During the eighties and nineties a phenomenological theory for the statistics of TLS has emerged under the name of Soft Potential Model, see \cite{esquinazi2013tunneling} for an exposition of this approach.
Within this approach, it is assumed that these groups of degrees of freedom are described by a collective coordinate which is subjected to a random confining quartic potential.
The theory essentially fits the statistics of the random quartic potentials with the data on the specific heat and other thermodynamic quantities coming from experiments. 

Crucially, in order for this approach to be manageable, the TLS excitations are assumed to be  weakly interacting. 
However it is natural to think that some interactions must be included. 
After all, when a rearrangement is triggered in of these mesoscopic regions (imagine that a group of degrees of freedom jumps from one locally stable configuration to another), it locally deforms the system and  creates a local stress which must be relaxed via elasticity.
In the early 2000, a systematic study of interacting TLS has been started via simple models \cite{gurevich2003anharmonicity, gurevich2005pressure, parshin2007vibrational}.
Their Hamiltonian can be written by introducing a set of $N$ real degrees of freedom $x_i \in {\mathbb R}$ parametrizing the collective coordinates of TLS excitations. They interact on a lattice in $d$ dimensions so that the Hamiltonian can be written as
\beq
\begin{split}
H[\underline x] &= -\sum_{i<j} \frac{J_{ij}}{|\underline n_i-\underline n_j|^d} x_ix_j +\sum_{i=1}^N v_i(x_i) \\
v_i(x_i)&=\frac{\k_i}{2}x_i^2 +\frac A{4!} x_i^4 
\end{split}
\label{GPS_model}
\eeq
where $\underline n_i$ is the vector coding for the position of the variable $x_i$ on the lattice, $J_{ij}$ are random variables with positive and negative sign with equal probability, and $\k_i>0$ are random stiffnesses extracted independently from a prescribed probability distribution $p(\k)$. The constant $A$ tunes the hardness of the quartic, anharmonic, part of the local confining potential.

The class of models in Eq.~\eqref{GPS_model} are not solvable from an analytic point of view since they are soft spin glasses in finite dimensions,
but, quite remarkably, numerical simulations and a phenomenological analysis concluded \cite{gurevich2003anharmonicity, gurevich2005pressure, parshin2007vibrational} that local minima of $H$ display
a density of states whose low frequency tail follows a quartic law populated by localized eigenvectors. Therefore TLS models not only provide a way to rationalize the behavior of the specific heat
but can also model the anomalous density of states.

In a parallel stream of works, K\"uhn and Horstmann \cite{kuhn1997random} considered a different point of view: while in these class of models one assumes that mesoscopic regions are subjected to random quartic potentials whose statistics has to be fixed a priori, one can imagine that these properties can emerge from the interactions between the mesoscopic regions themselves \cite{kuhn2003universality, kuhn2007finitely}.
To make this idea more concrete, they considered a simple model in which anharmonic degrees of freedom interact via fully connected random couplings in the same fashion as in the SK model. 
The corresponding Hamiltonian is 
\beq
\begin{split}
H[\underline x] &= -\frac J{\sqrt{N}}\sum_{i<j} {J_{ij}} x_ix_j +\sum_{i=1}^N v(x_i) \\
v(x_i)&=\frac{1}{2}x_i^2 +\frac A{4!} x_i^4 
\label{KH_model}
\end{split}
\eeq
where the couplings $J_{ij}$ are Gaussian with finite mean $J_0/\sqrt{N}$ and unit variance. The control parameter $J$ tunes the strength of interactions.

It can be shown \cite{kuhn1997random} through the cavity method \cite{MPV87} that the solution of this model can be written in terms of an ensemble
of effective TLS whose statistical properties emerge from the interactions between the microscopic degrees of freedom.
However, the fully connected nature of the model allows also for exact solubility within the replica method and the model in Eq.~\eqref{KH_model} can be shown to exhibit
a replica symmetry breaking phase at $T=0$ and for sufficiently strong $J$ where $D(\omega)\sim \omega^2$ at low frequencies.
The nature of the RSB phase and the density of states is the same as the one found in the model analyzed in \cite{franz2015universal}.

From our perspective, these two streams of works show that it may be reasonable to consider soft spin glasses to model the non-phononic spectrum of amorphous solids at low temperature.
After all, if the Gardner picture for low temperature amorphous solids is correct, we should expect that glasses can be modeled by spin glasses.

However a crucial difference between the model in Eq.~\eqref{GPS_model} and the one in Eq.~\eqref{KH_model} is that the density of states of local minima of the corresponding Hamiltonians differ quite strongly.
One may not be very surprised by this fact: the Hamiltonian in Eq.~\eqref{GPS_model}, has short range interactions and therefore one may suspect that the universal $\omega^4$ law in the density of states is the outcome of finite connectivity/finite dimensionality of the model. 
This perspective was further developed in a work \cite{lupo2017critical} where finitely connected mean-field spin glasses, of the same kind of the Viana-Bray model discussed above,
with soft degrees of freedom (XY spins) have been analyzed showing that the $\omega^4$ law can be captured by the finite connectivity of the model.
However, while this is very interesting, finitely connected spin glasses are hard to analyze, especially when RSB effects are at play and therefore it is hard to make progress in such theoretical setting.

To summarize, these works show that the physics of soft localized excitations following a quartic law at small frequencies is
rather universal and can be found generically in models which are either in finite dimensions or for which the topology of interaction has a finite connectivity. 
However all these models cannot be solved analytically and therefore the robustness of these findings can be mostly checked via numerical simulations.

The picture changed in \cite{bouchbinder2021low, rainone2021mean} where an alternative way of thinking was proposed. 
The main idea behind it is suggested by an additional crucial difference between the models in Eqs.~\eqref{GPS_model} and \eqref{KH_model}.
Indeed, in Eq.~\eqref{KH_model} the local potential term $v_i(x_i)$ is not random. However this term may model some local heterogeneities
in the system which are definitely present in finite dimensions.
Therefore if we want to include the effects of local heterogeneities, it makes sense to keep the local quartic confining potential to be random.
Of course this implies that the statistical properties of effective TLS are not modeled only through the interactions between degrees of freedom, as proposed by K\"uhn and Hortsmann.
However we will see that interactions crucially renormalize the statistical properties of the effective TLS and, in the end, are the essential ingredients
that give rise to the quartic tail in the spectrum of linear soft excitations. Therefore one may consider the model in Eq.~\eqref{GPS_model} and tale a mean-field limit
where the interactions are fully connected as in  Eq.~\eqref{KH_model}.

An additional feature of the Hamiltonians in Eq.~\eqref{GPS_model} and \eqref{KH_model} is that they are invariant under ${\mathbb Z}_2$ symmetry
which has no meaning in real glasses \cite{urbani2015gardner}. Therefore it makes sense to add a ${\mathbb Z}_2$ symmetry breaking magnetic field.
We arrive at the following Hamiltonian
\beq
\begin{split}
H[\underline x] &= -\frac J{\sqrt{N}}\sum_{i<j} {J_{ij}} x_ix_j +\sum_{i=1}^N v_i(x_i) \\
v(x_i)&=\frac{\k_i}{2}x_i^2 +\frac 1{4!} x_i^4 - B x_i \:.
\label{KHGPS_model}
\end{split}
\eeq
In \cite{bouchbinder2021low, rainone2021mean} it has been proposed to call the Hamiltonian in Eq.~\eqref{KHGPS_model} the KHGPS model
to credit to K\"uhn, Hortsmann, Gurevich, Parshin and Sch\"ober the idea to use these class of soft spin glasses to address the problem of low
energy linear excitations in glasses. 
However it is important to note that none of these authors studied the model in Eq.~\eqref{KHGPS_model} and the purpose of the rest of this section
is to highlight how this model has a set of crucial ingredients that are necessary to give the right physics (namely producing a quartic tail in the low frequency density of states) without losing exact solubility.
We would like to anticipate that among the crucial ingredients, the main one is represented by the external magnetic field whose introduction was triggered by some experimental findings, see \cite{albert2021searching}. In this work, models of the kind of Eqs.~\eqref{GPS_model}-\eqref{KHGPS_model} were used to fit data on the behavior of glassy glycerol at low temperature: the main outcome of these studies was that the local magnetic field is present and it is typically larger than the spin-spin interaction scale.

\subsection{Solution of the KHGPS model}
In this section we would like to describe briefly how the model in Eq.~\eqref{KHGPS_model} can be solved through the replica method.
The material collected in this section follows closely the publications \cite{bouchbinder2021low, folena2022marginal}.
We will consider the model in Eq.~\eqref{KHGPS_model} and 
set 
\beq
p(\k)=
\begin{cases}
\left(\kmax-\kmin\right)^{-1} & \k\in \left[\kmin,\kmax\right]\\
0 &  \k\notin \left[\kmin,\kmax\right]
\end{cases}
\eeq
The models studied in \cite{bouchbinder2021low} and \cite{rainone2021mean} differ slightly
because of the choice of the value of $\kmin$, while $\kmax=1$ always.
In particular in \cite{rainone2021mean} the model with $\kmin=0$ was considered but we will show
that this model does not have a spin glass transition in a field and actually at zero temperature it is always 
in a RSB phase as much as the SK model. Therefore for the rest of this chapter we will consider always $\kmin>0$ and set 
$\kmin=0.1$ whenever we will need to specify this value to produce a quantitative result (for example the phase diagram of the model). We always set $A=1$
without loosing generality and consider $J_0=0$ so that $J_{ij}$ are just Gaussian random numbers with zero mean 
and unit variance\footnote{The choice of removing the ferromagnetic interaction term is due to the fact that this has no meaning to model structural glasses.}.

The control parameters of the model, once $p(\k)$ is fixed, are the temperature $T$, the strength of the interactions between the degrees of freedom $J$, and the external magnetic field  $B$.
We are interested in computing the free energy of the model given by
\beq
\begin{split}
{\rm f} &= -\frac{1}{\beta N} \overline{\ln Z}\\
Z&=\int \de \underline x \exp[-\beta H]
\end{split}
\label{def_f_KHGPS}
\eeq
Note that in the zero temperature limit, $\beta\to \infty$, the free energy becomes the energy of the 
ground state of the Hamiltonian.
In Eq.~\eqref{def_f_KHGPS} we have denoted by an overline the average over all sources of disorder, namely the random couplings $J_{ij}$
and the stiffnesses $\k_i$ as well.
Performing the average over the logarithm of the partition function $Z$ is a hard task and therefore we resort to the replica method \cite{MPV87}
\beq
\overline \ln Z=\lim_{n\to 0} \partial_n \overline{Z^n}\:.
\eeq
When $n$ is an integer we can compute the last term of the previous equation via a replicated integral
\beq
\overline {Z^n} = \int \left[\prod_{a=1}^n\de \underline x^{(a)}\right] \exp\left[-\beta \sum_{a=1}^n H[\underline x^{(a)}]\right]\:.
\eeq
The idea of the replica method is to perform the computation for integer values of $n$ and then take the analytic continuation of
the explicit formulas in $n$ down to $n\to 0$.

An explicit evaluation, see the appendix of \cite{bouchbinder2021low}, allows to compute the average over the replicated partition
function via a saddle point. Indeed one has
\beq
Z^n\propto\int\de Q \exp\left[N\AA[Q]\right]
\eeq
and $Q$ is the overlap order parameter which, at the saddle point, is given by
\beq
Q_{ab} = \frac 1N \langle\underline x^{(a)} \cdot \underline x^{(b)} \rangle
\eeq
being the brackets the average over the replicated GB measure.
The action $\AA[Q]$ takes the form
\beq
\AA[Q] = -\frac 14 \beta^2J^2\sum_{ab}Q^2_{ab}+\ln \ZZ
\eeq
and $\ZZ$ is the partition function of a so called impurity problem, given by
\beq
\ZZ = \int \de \k p(\k) \int \prod_{a=1}^n \de x_a \exp\left[\frac{\beta^2 J^2}{2}\sum_{ab}Q_{ab}x_ax_b-\beta \sum_{a=1}^n v_\k(x_a)\right] 
\eeq
where 
\beq
v_\k(x) = \frac \k 2 x^2 + \frac{x^4}{4!} - Bx\:.
\eeq

At this point one needs to compute the solution of the saddle point equations for $Q$ and take the analytic continuation in $n$ to compute the $n\to 0$ limit.
The action $\AA$ is invariant under permutations of replicas (replica symmetry).
This suggest that if this symmetry is unbroken, the simplest saddle point solution must respect it and therefore
\beq
Q_{ab}=\begin{cases}
q_d & a=b\\
q & a\neq b
\end{cases}
\eeq
and this is the so-called replica symmetric (RS) ansatz of the very same kind of the one used by Sherrington and Kirkpatrick to analyze the SK model. 
The value of $q_d$ and $q$ are then fixed variationally in the $n\to 0$ limit.
It is useful to write the corresponding equations in the following way.
We first introduce the Gaussian measure
\beq
P(q,h) = \frac{1}{\sqrt{2\pi J^2 q}}\exp\left[-\frac{(h-B)^2}{2J^2q}\right]\:.
\eeq
We also define
\beq
\begin{split}
f(q,h) &= \frac 1\beta \ln \int \de x\, \exp[-\beta v_{\rm eff}(x|\k,B)]\\
v_{\rm eff}(x|\k,h) &= \frac 1 2\left(\k-\beta J^2 (q_d-q)\right) x^2 + \frac{x^4}{4!} -h x\:.
\end{split}
\eeq
Then we have that the RS equations can be written as
\beq
\begin{split}
q &=\int \de \k p(\k) \int \de h P(q,h) \left(\frac{\partial f(q,h)}{\partial h}\right)^2\\
\beta(q_d-q) &=\int \de \k p(\k) \int \de h P(q,h) \frac{\partial^2 f(q,h)}{\partial h^2}\:.
\end{split}
\label{RS_SP_KHGPS}
\eeq
The physical meaning of these equations is rather clear. The function $f(q,h)$ is (minus) the free energy
of an effective degree of freedom $x$ subjected to the effective (random) potential $v_{\rm eff}(x|\k, h)$.
Then, $q$ and $q_d$ can be written as 
\beq
\begin{split}
q_d &= \overline{\langle x \rangle^2}\\
q &= \overline{\langle x^2 \rangle}
\end{split}
\label{RS_eq_q}
\eeq
and the brackets denote the average over a GB measure defined by $v_{\rm eff}$ while the overline denotes the average over the effective local quenched disorder, $\k$ and $h$, affecting $v_{\rm eff}$.
The solution of the Eqs.~\eqref{RS_eq_q} is not known explicitly but can be easily found via a numerical integration.

The consistency of this solution, namely the fact that the replica symmetric form is the right saddle point of the action $\AA$, is controlled by the so called replicon eigenvalue
(which is the singular value of the Hessian matrix which controls the behavior of $\AA$ around this saddle point \cite{de1978stability})
\beq
\l_R = 1 - J^2 \int \de \k p(\k) \int \de h P(q,h)  \left[\frac{\partial^2 f(q,h)}{\partial h^2}\right]^2\geq 0\:.
\label{replicon_KHGPS}
\eeq
When the bound in Eq.~\eqref{replicon_KHGPS} is violated, the RS solution is said to be unstable and the model enters in a RSB phase.
Therefore Eqs.~\eqref{RS_SP_KHGPS} and \eqref{replicon_KHGPS} give access to the part of the phase diagram where the replica symmetric solution
is correct. We will now discuss in more details this phase diagram.

\subsection{The phase diagram of the KHGPS model}
\begin{figure}
\centering
\includegraphics[width=0.45\columnwidth]{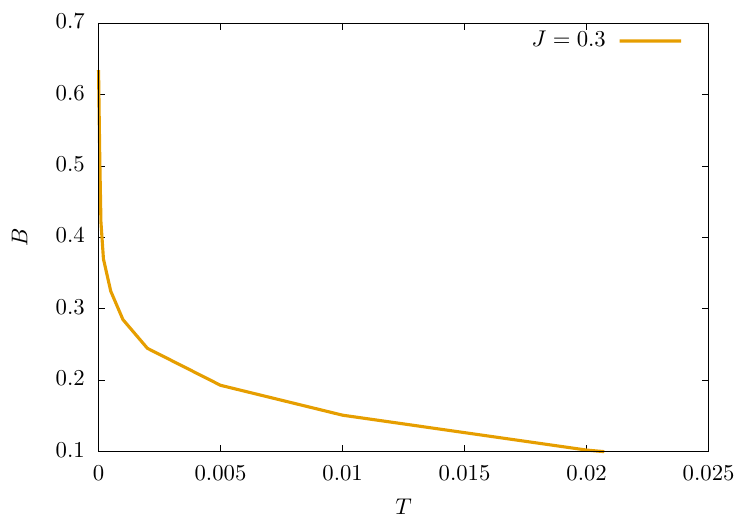}
\includegraphics[width=0.45\columnwidth]{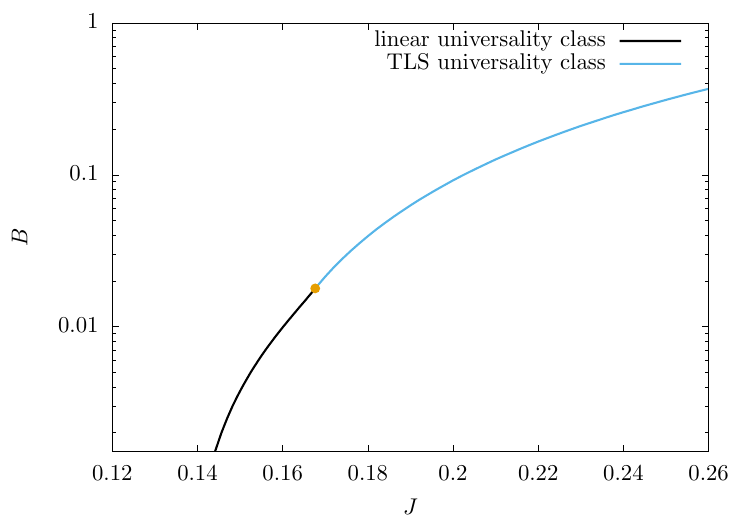}
\linespread{0.8}
\caption{\footnotesize{{\it Left Panel}: Phase diagram in the temperature-external field plane for two different values of the interaction strength $J$. Below both lines the model is
RSB and above them it is RS. The model presents a zero temperature spin glass transition as a function of the external field. Close to the critical line, one can compute the solution of the RSB equation in perturbation theory following the same strategy as in \cite{parisi1980magnetic} and it has been shown that the solution is of fullRSB type \cite{folena2022marginal}. {\it Right Panel}: The phase diagram at zero temperature as a function of the external field and interaction strength. The zero temperature transition to RSB is controlled by two different universality classes depending on the value of the external field. These phase diagram have been presented in \cite{folena2022marginal} (left) and \cite{bouchbinder2021low} (Right).}}
\label{Fig_KHGPS_PD}
\end{figure}

One can integrate numerically Eqs.~\eqref{RS_SP_KHGPS} to get the values of $q_d$ and $q$ and check with Eq.~\eqref{replicon_KHGPS}
whether the RS ansatz is consistent.
The resulting phase diagram as a function of the control parameters is reported in Fig.\ref{Fig_KHGPS_PD}.
  
At fixed interaction strength $J$, if the external magnetic field is sufficiently small, the model has a spin glass transition
at sufficiently low temperatures. In \cite{folena2022marginal} it has been shown that the nature of the RSB transition is the same as in the SK model 
in the sense that for temperatures right below the transition one, the model displays a fullRSB phase\footnote{This means that the $P(q)$ right beyond the transition is a non trivial function with support in a unique interval $[\qm,\qM]$. Whether this is true up to zero temperature can be checked only via a numerical integration of the fullRSB equations.}. 

The picture changes when we look at the phase diagram in the zero temperature limit, as shown in Fig.\ref{Fig_KHGPS_PD}.
The model is in a RS phase for either sufficiently strong external field $B$ or for sufficiently weak interaction strength $J$.
To discuss this phase diagram more carefully it is useful to look at the saddle point equations directly for $T\to 0$.
Defining the linear magnetic susceptibility as
\beq
\chi=\lim_{\beta\to \infty}\beta (q_d-q)
\eeq
we can look at the zero temperature limit of the effective potential, namely
\beq
v^{(0)}_{\rm eff}(x|\k,h) = \frac 12\left(\k- J^2\chi\right) x^2 + \frac{x^4}{4!} -h x\:.
\eeq
This implies that 
\beq
\lim_{\beta\to \infty}\frac{\partial f(q,h)}{\partial h} = x^*(\k,h)
\eeq
being $x^*(\k,h)$ the ground state of $v^{(0)}_{\rm eff}(x|\k,h)$. 
In the RS phase at zero temperature, the shape of $v^{(0)}_{\rm eff}(x|\k,h)$ is always convex and one can show that this implies that the RS solution describes properties of the unique ground state of the Hamiltonian \cite{folena2022marginal}.
However, differently to what happens at any finite temperature where the transition to RSB is controlled by the saturation of the bound in \eqref{replicon_KHGPS}, at $T=0$ the transition to the RSB can happen in two different ways.
In the first case, one has that, as it happens for $T>0$, the transition coincides with the saturation of the bound in Eq.~\eqref{replicon_KHGPS}.
Correspondingly, the shape of $v^{(0)}_{\rm eff}(x|\k,h)$ is convex up to the phase transition point. This mechanism happens for small values of the external field $B$ and it coincides with the one found in other models at zero temperature, see  \cite{kuhn1997random, franz2015universal}. As we will see, the transition to RSB is driven by an abundance of soft linear excitations and therefore when this happens, we dub the corresponding critical point to be in the  {\it linear universality class}.

The picture changes qualitatively for larger values of $B$. In this case the shape of $v^{(0)}_{\rm eff}(x|\k,h)$ becomes non-convex at the transition point.
This happens when $\k-J^2\chi<0$ for some $\k\in[\kmin,\kmax]$. In particular, for large enough external magnetic field, increasing $J$ one has that $\chi$
grows up to the point where $\kmin-J^2\chi=0$. Beyond this point, the shape of $v^{(0)}_{\rm eff}(x|\k,h)$
is a double well for all values of $\k$ such that $\k-J^2\chi<0$. Correspondingly, we have that 
\beq
\lim_{\beta\to \infty}\frac{\partial^2 f(q,h)}{\partial h^2} =\frac{x^*(\k,h)}{\partial h}=2x^*(\k,0^+)\delta(h) + \left(\left.\frac{\partial^2v^{(0)}_{\rm eff}(x|\k,h)}{\partial h^2}\right|_{x=x^*(\k,h)}\right)^{-1}\:.
\label{divergence_delta}
\eeq
and therefore the integral in Eq.~\eqref{replicon_KHGPS} is divergent for the values of $\k$ such that $\k-J^2\chi<0$.
This implies that the RS solution is inconsistent and the model is RSB.
It is essential to stress that in this case the bound in Eq~\eqref{replicon_KHGPS} is not saturated up to the RSB point where $\l_R>0$. The condition determining the phase transition is rather the point where $\kmin-J^2\chi=0$.
Since at this point double wells effective potentials appear, we dub this mechanism the {\it TLS universality class}.
The zero temperature phase diagram is summarized in the right panel of Fig.~\ref{Fig_KHGPS_PD}. 

At this point it is mandatory to comment on how one can reconcile that the finite temperature transition point
is always driven by the saturation of the bound in Eq.~\eqref{replicon_KHGPS} while at zero temperature
this is not always the case.
A natural mechanism \cite{folena2022marginal} for this to happen is that the form of $\l_R$ as computed from Eq.~\eqref{replicon_KHGPS} at small temperatures is 
\beq
1-\l_R\sim A(B)/T+{\rm regular\ terms}\:.
\eeq
It turns out that $A(B)=0$ at the point where $\kmin-J^2\chi=0$ exactly at $T=0$ and therefore in the zero temperature limit one can have a phase transition to a RSB with $\l_R>0$. This is confirmed by a numerical analysis, see \cite{folena2022marginal}.

Therefore the analysis of the RS equations gives that there are two mechanisms for breaking the replica symmetry at zero temperature.
In one case the shape of the local effective potential  $v^{(0)}_{\rm eff}(x|\k,h)$ induces the saturation of the bound in Eq.~\eqref{replicon_KHGPS}.
In the second case instead $v^{(0)}_{\rm eff}(x|\k,h)$ develops a double well shape and this instantaneously and discontinuously violates the bound Eq.~\eqref{replicon_KHGPS}.
It is important to observe that the model where $k_{\rm min}=0$ is always RSB at zero temperature regardless the strength of the interactions, as far as $J>0$, and the value of the external field.
Indeed for this model one always has that double wells are present and therefore the replicon condition is always violated. 

Finally, it must be noted that for both universality classes, the RSB transition is a {\it topology trivialization} \cite{ros2023high, fyodorov2013high, fyodorov2014topology} transition in the sense that in the RS phase
the landscape of the Hamiltonian is convex with a single global minimum while in the RSB phase one has exponentially many (in $N$) minima.
We will come back to this at the end of this chapter.

\subsection{Two mechanisms for the spin glass transition in a field}
In the previous section we highlighted that, strictly at zero temperature, the spin glass transition in a field can happen in two different ways.
In this section we would like to clarify what the two mechanisms correspond to in terms of the physics of the model.
To do this, we first focus on the spectrum of the Hessian of the global minimum of the Hamiltonian in the RS phase.
Given the unique ground state in the RS phase, we can look at the corresponding Hessian matrix given by
\beq
\HH_{ij} = -J J_{ij} + \delta_{ij}\left(\k_i + \frac{\left(x_i^*\right)^2}{2}\right)
\eeq
and we have denoted with $x_i^*$ the coordinates of the system in the ground state.
The matrix $\HH$ is a random matrix given by the sum of a GOE matrix $JJ_{ij}$ and a random diagonal
matrix whose elements depend on the coordinates of the ground state.
The average density of state of this matrix can be studied and we refer to \cite{bouchbinder2021low}
for a detailed analysis.
Here we report the main conclusions.

In the RS phase, the Hessian in the ground state is gapped in the sense that the lower edge of the support of the corresponding density of eigenvalues $\rho(\l)$,
say $\l_{\rm min}$ is strictly positive. At the zero temperature spin glass transition in a field instead the spectrum becomes gapless $\l_{\rm min}=0$.
However the behavior of $\rho(\l)$ close to $\l=0$ changes depending on the mechanism of RSB.

In the linear universality class, one finds that $\rho(\l) \sim \l^{1/2}$ for $\l\to 0$. 
In frequency domain, this corresponds to $D(\omega)\sim \omega^2$ and therefore at the transition 
one gets a gapless spectrum as in the KH case \cite{kuhn1997random} and in the case of \cite{franz2015universal}.

However in the TLS universality class, one finds that $\rho(\l)\sim \l^{3/2}$ which, in frequency domain, means a low-frequency tail
of the form $D(\omega)\sim \omega^{4}$. We notice that this is precisely the quartic tail that we are aiming at modeling!
One can also show that while in the linear universality class, the low frequency eigenvectors are delocalized,
in the TLS case they are localized and this reflect a random matrix mechanism that was studied in detail in \cite{lee2016extremal}.
 
In order to understand what these different behaviors in the density of state mean, it is mandatory at this point to
recall what the saturation of the bound in Eq.~\eqref{replicon_KHGPS} practically implies.
Let us introduce the spin glass susceptibility defined as
\beq
\chi_{\rm SG} = \frac{1}{N} \sum_{ij}\left(\left.\frac{\partial \langle x_i \rangle}{\partial B_j}\right|_{B_i=B}\right)^2
\label{def_chi_SG}
\eeq
where we implicitly promoted the external field to be site dependent. It can be shown by Landau type arguments \cite{MPV87}
that the saturation of the bound in Eq.~\eqref{replicon_KHGPS} corresponds to a divergence of $\chi_{SG}$.
Eq.~\eqref{def_chi_SG} can be rewritten at $T=0$ as
\beq
\chi_{\rm SG}=\int \de \l \frac{\rho(\l)}{\l^2}\:.
\label{chi_rho}
\eeq
and we have assumed that the integral runs over the support of $\rho(\l)$.
It is clear that as soon as the spectrum of the Hessian is gapped at $T=0$, which happens in the RS phase, $\chi_{\rm SG}<\infty$.
At the zero temperature spin glass transition, the spectrum of $\rho(\l)$ becomes gapless but
this leads to a diverging spin glass susceptibility only when the transition is in the linear universality class. 
Therefore in this case the phase transition is driven by the emergence of an abundance of soft {\it linear} modes which make
the ground state of the Hamiltonian marginally stable to linear perturbations.
Instead in the case of the TLS universality class, the spin glass susceptibility stays finite up to the transition point since the integral in Eq.~\eqref{chi_rho}
is not divergent.
Therefore while in the TLS universality class the ground state develops soft modes, 
their density is not sufficiently large to trigger a diverging spin glass susceptibility.
Therefore we still need to understand what drives the transition in the TLS case.

In order to clarify this point we consider the solution of the model in the RSB phase.
The RS ansatz is inconsistent and one needs to develop a RSB formalism.
We just outline its form in a nutshell, the interested reader can look at \cite{folena2022marginal} for more details.
As the RSB line in the $T-B$ plane is crossed, the model becomes fullRSB and is described by a function $x(q)$ which
represents the cumulative of the average overlap distribution $P(q)$ described in Sec.\ref{Sec_MFT_spin_glasses}.
This function is defined in some interval $q\in[\qm,\qM]$ and the free energy $f(q,h|\k)$ and the probability distribution
of the field $h$, $P(q,h|\k)$ both depend on $\k$ and obey a flow equation. The free energy $f(q,h|\k)$ is determined by the Parisi's equation 
\beq
\begin{split}
\frac{\partial f(q,h|\k)}{\partial q} &= -\frac{J^2}{2}\left[\frac{\partial^2 f(q,h|\k)}{\partial h^2} + x(q) \left(\frac{\partial f(q,h|\k)}{\partial h}\right)^2\right]\\
f(\qM,h|\k) = &= \frac 1\beta \ln \int \de x\, \exp[-\beta v_{\rm eff}(x|\k,B)]\\
v_{\rm eff}(x|\k,h) &= \frac 1 2\left(\k-\beta J^2 (q_d-\qM)\right) x^2 + \frac{x^4}{4!} -h x
\label{eq_fk}
\end{split}
\eeq
which must be solved backward from $q=\qM$ to $q=\qm$. Instead the equation for $P(q,h|\k)$ is a Fokker-Planck
equation \cite{sommers1984distribution} given by
\beq
\begin{split}
\frac{\partial P(q,h|\k)}{\partial q} &= \frac{J^2}{2}\left[\frac{\partial^2 P(q,h|\k)}{\partial h^2} -2x(q)\frac{\partial}{\partial h} \left(P(q,h|\k)\frac{\partial f(q,h|\k)}{\partial h}\right)\right]\\
P(q,h|\k) &= \frac{1}{\sqrt{2\pi J^2 \qm}}\exp\left[-\frac{(h-B)^2}{2J^2\qm}\right]\:.
\end{split}
\eeq
Both equations depend on $x(q)$ which must be fixed according to the self consistent equation
\beq
q = \int \de \k p(\k) \int \de h P(q,h|\k) \left(\frac{\partial f(q,h|\k)}{\partial h}\right)^2\:.
\label{eq_q}
\eeq
A solution of Eqs.~\eqref{eq_fk}-\eqref{eq_q} is in principle possible via an iterative scheme but for the moment it has not been done yet.
The consistency of the RSB solution is again controlled by a set of conditions. To stay as simple as possible,
one of them is a generalization of Eq.~\eqref{replicon_KHGPS} to the RSB case
\beq
\l_R(\qM) = 1 - J^2 \int \de \k p(\k) \int \de h P(\qM,h)  \left[\frac{\partial^2 f(\qM,h)}{\partial h^2}\right]^2\geq 0\:.
\label{Rep_RSB}
\eeq

As for the RS case, we can look at the zero temperature scaling limit of the  RSB equations.
Indeed, $\chi$ is still defined as
\beq
\chi=\lim_{\beta\to \infty}\beta\left(q_d-\qM\right)
\eeq
and we will assume that the solution of the RSB equations gives
a value of $\chi$ which is smooth as a function of the control parameters in the problem. 
It is simple to check that any RSB construction that is not continuous\footnote{By continuous or fullRSB we mean that the $P(q)$ is not a sum of Dirac deltas in the corresponding range of $q$ and this implies that $x(q)$ has no jumps in the same range.} for $q\to \qM$ violates the bound in Eq.~\eqref{Rep_RSB}.
Indeed the form of $P(q,h|\k)$ for any RSB which is not continuous, can be seen as a set of re-weighted Gaussian convolutions \cite{charbonneau2014exact} which gives a $P(\qM,h=0)>0$
and this stays true in the zero temperature limit. 
However when we cross the TLS universality class transition line at zero temperature, the effective potential in Eq.~\eqref{eq_fk} develops a double well structure
for $\k\in[\kmin, J^2\chi]$ and correspondingly a divergent structure for the derivatives of $f(\qM,h)$ as in Eq.~\eqref{divergence_delta} can be found leading to a violation 
of the bound in Eq.~\eqref{Rep_RSB}, precisely as in the RS case.
The solution to this problem can come only with a continuous, or fullRSB solution, for $q\to \qM$. 
In this case one can show that a direct consequence of the RSB equations is
\beq
1= J^2\int \de \k p(\k) \int \de h P(\qM,h|\k) \left(\frac{\partial^2 f(\qM,h|\k)}{\partial h^2}\right)^2
\label{repl_qM}
\eeq
A careful analysis, see \cite{folena2022marginal} shows that this can happen only if the density of the field $h$ is pseudogapped and follows a linear behavior
close to $h\sim 0$
\beq
P(\qM,h|\k)\sim \g_\k |h| \ \ \ \ \ h\sim 0
\eeq
for some come coefficient $\g_k$ that is $\k$ dependent.
Therefore this means that crossing the TLS universality class zero temperature transition line
the model develops double well potentials (TLS) but the RSB
analysis implies that the density of almost degenerate double wells vanishes so that the bound in Eq.~\eqref{Rep_RSB} is marginally satisfied.
The physics is very similar to the one of the SK model where at zero temperature the density of spins
having a local field close to zero vanishes \cite{MPV87}. The main differences with respect to the SK model are two: 
first, the KHGPS model is a soft spin model and therefore one can have a finite linear susceptibility $\chi$ (while this vanishes
in the SK model \cite{MPV87} due to the hardness of Ising spins) and (ii) the density of double wells depends on the linear susceptibility itself.
In other words, in the KHGPS model, there is only a fraction of spins, the one for which $\k_i\in[\kmin,J^2\chi]$ which can develop
almost degenerate double wells while in the SK model all spins are of this kind\footnote{In the language of the KHGPS model, the SK model is an extreme case in which all spins are symmetric, infinitely rigid double wells, centered at $x_i=\pm1$.}. 
Effective potentials with two degenerate wells encodes non-linear excitations. Indeed these spins could flip from one
to the other well with almost zero cost. However such "flip" could induce a global rearrangement in the system (an avalanche)
and therefore the density of spins having such symmetric effective potential must be suppressed.
"Spin flips" from one well to the other are non-linear excitations. However these excitations cannot be triggered by a deformation at strictly zero temperature.
In order to explore them one needs to relax the system to the ground state for each perturbation and therefore they are triggered only in the zero temperature limit.

Therefore this analysis shows that there are two ways to break replica symmetry at zero temperature:
\begin{itemize}
\item in the linear universality class case, the ground state of the Hamiltonian develops an abundance of soft linear modes
which lead to a divergence of the spin glass susceptibility.
\item in the TLS universality class instead, the transition is driven by the appearance of a finite density of non-linear excitations
whose density is controlled by marginal stability.
\end{itemize}

It is clear that the only mechanism that can survive in finite dimension is the second one.
Indeed, soft delocalized non-phononic excitations are hardly found in finite dimension\footnote{With the exception of jamming physics.} since in this case
the edges of the spectrum are typically made of localized excitations whose density is controlled by the quartic law.
Instead the analysis above shows that the spin glass transition is associated to the the birth of
soft non-linear excitations whose density is pseudogapped to make the system marginally stable.

The emergence of the TLS universality class driven by a critical point with a finite spin glass susceptibility has interesting consequences
if it is incorporated in the Parisi and Temesvari picture for the fate of the spin glass transition in finite dimension.
According to this picture the transition, if it is there, is controlled by the zero temperature critical point.
We have shown in this chapter that this point may be characterized by a finite spin glass susceptibility.
Therefore it may well be that numerical simulations or experiments \cite{albert2021searching} trying to detect the transition
may fail in doing so if they focus on the spin glass susceptibility.

\section{Towards finite dimensions: field theory for the zero temperature spin glass transition in a field}
The KHGPS model is a mean-field fully connected model where one can understand the physics of the spin glass transition
in a field in great detail. With respect to other more complicated models, the main advantage of this model is that
we have access the solution both in the RS and RSB phases\footnote{Note that this is not the case for models defined on locally tree like graphs where the RSB solution is essentially unknown apart from the special case where it is 1RSB \cite{mezard2002analytic}.}.
In this section we would like to follow the route proposed by Parisi and Temesvari and start to develop a field theory for the transition in a field at zero temperature from the KHGPS model.
The content of this section is a summary of a detailed work \cite{urbani2022field}.

A way to construct a field theory to investigate the properties of phase transitions is to start from a fully connected model and generalize the degrees of freedom to vectorial spins \cite{moore20121}.
To be more precise, we consider a model defined on a $d$ dimensional lattice.
On each lattice point we put an $M$ dimensional real vector $\underline x_i=\{x_{i,1}, \ldots, x_{i,M}\}\in {\mathbb R}^M$
and we define the following Hamiltonian
\beq
\begin{split}
H[\underline x] &= \frac{J}{\sqrt{2dM}}\sum_{ij}c_{ij}\sum_{\a \b} J^{\a\b}_{ij}x_{i,\a}x_{j,\b} +\sum_i \sum_\a v(x_{i,\a}|\k_{i\a})\\
v(x_{i,\a}|\k_{i\a}) &= \frac{\k_{i,\a}}{2}x_{i,\a}^2 + \frac 1{4!}\left(x_{i,\a}\right)^4 - Bx_{i,\a}\:.
\end{split}
\eeq
The matrix $c_{ij}$ encodes the connectivity of the underlying lattice. Its elements are zero unless two sites are connected. In this case we fix $c_{ij}=1$.
The random couplings $J_{ij}^{\a \b}=J_{ij}^{\b \a}$ are Gaussian random numbers with zero mean and unit variance.
Furthermore the local stiffnesses $\k_{i,\a}$ are all independent and identically distributed according to $p(\k)$. 
One can show that regardless the connectivity of the lattice, the limit $M\to \infty$ is a mean-field limit \cite{urbani2022field}. 
In other words, in this limit the free energy of the model can be obtained and the corresponding saddle point equations are the
same as the ones of the KHGPS model. We note that the model is not $\OO(M)$ invariant as it happens in standard
phase transitions where the generalization to $\OO(M)$ symmetry is a way to compute critical exponents
at fixed dimension \cite{bray1974self, parisi1992field}.

The main idea to construct a field theory for the zero temperature spin glass transition in a field is to consider an expansion around the $M\to \infty$ limit.
The details on this procedure can be found in \cite{urbani2022field}, here we report just the main steps and results.

Starting again from the replica method, one can consider the average of the replicated partition function.
This is given in terms of an integral over a space dependent overlap matrix which represent the fluctuating field of the theory. 
For $M\to \infty$, fluctuations are suppresed and the field concentrate on its average, space-independent, value.
Therefore one can consider an expansion around the mean-field limit. This gives a field theory for the fluctuations of the overlap field
around its saddle point value. Denoting this quantity by $\delta q_i^{ab}$ and the index $i$ runs on the sites of the lattice of the model.

The probability distribution of $\delta q_i^{ab}$ is controlled by a field theory defined by an action which can be written at the lowest order in the $1/M$ expansion
as composed by a kinetic term and an interaction term.
The form of the kinetic term is given by
\beq
\SS_{\rm kin}  = \frac 12 \sum_{\underline p} \delta q^{ab}(\underline p) \MM_{ab;cd}(\underline p)\delta q^{cd}(-\underline p)
\eeq
and we have used the Fourier representation of the field $\delta q_i^{ab}$. The mass matrix $\MM$ can be diagonalized.
At the zero temperature critical point in the TLS universality class, all the eigenvalues are positive indicating that there is no critical behavior in this sector of the action.
This is a consequence of the fact that in this universality class the spin glass susceptibility stays finite at the phase transition point.

The interaction sector of the theory for $\delta q$ is controlled by a series of cubic terms. 
They can be computed in detail but in \cite{urbani2022field} a shortcut has been taken.
Instead of looking to all of them, one can project the coupling terms of the fluctuations of the field $\delta q$ which are the eigenspace of $\MM$ which is controlled by the replicon eigenvalue.
This is not what one should do  in principle given that this sector of fluctuations is perfectly regular at the transition.
However it can be shown anyway that the phase transition is manifest in the behavior of a set of coupling constants whose bare behavior becomes singular at the critical point.
The form of the interaction terms is given by
\beq
S_3 = \frac 1{\sqrt M}\sum_{i=1}^N \left[g_1  {\rm Tr}[\phi_i^3] + g_2\sum_{ab}\left(\phi_i^{ab}\right)^3\right]
\eeq
and we have indicated with $\phi_i$ the projection of $\delta q_i$ on the replicon eigenspace of $\MM$.
The coupling constants $g_1$ and $g_2$ can be computed at the bare level as a function of the control parameters. 
In other words they are found by expanding the original action around the saddle point solution and should be considered as the parameters defining the microscopic theory.
Their behavior under renormalization is the main subject of the theory of critical phenomena. 
However an interesting outcome of \cite{urbani2022field} is that, already at the bare level, they have an interesting behavior.
Approaching the zero temperature spin glass transition in a field one can show that they both diverge with a power law with the distance from the critical point.
At this point it is unclear how to proceed given that one has a theory where criticality is manifested in the interaction sector. 
Naively this may suggest some first order behavior but further investigations are needed to clarify this point.
However I believe that the most important missing ingredient is that one needs to include in the analysis (or at least try to isolate from it)
the size of non-linear excitations. As I will discuss in the conclusions of this chapter, this is an important point which must be properly included in the analysis.

\section{Perspectives}
In this chapter we have reviewed the problem of the spin glass transition in a field
and its connection with the physics of structural glasses out-of-equilibrium.
We have shown that the KHGPS model offers a way to understand the mechanism for the zero temperature spin glass transition in a field in terms of 
the appearance of a finite density of pseudogapped non-linear excitations.
We have also discussed how one can construct a field theory for the transition in finite dimensions starting from this model.
Despite these progresses a final conclusion on the fate of the transition in finite dimensions is still to be written. 
In the rest of this section I would like to outline what are the main research directions that could be helpful in this regard.
Finally I will also discuss some important questions that are interesting within the realm of the mean-field theory itself.

As we already discussed in Sec.\ref{sec_Gardner}, the Gardner transition is a crucial ingredient
to describe the criticality of the jamming transition in hard sphere glasses. In the infinite dimensional limit, the fullRSB phase that
appears at the Gardner point develops a scaling regime close to jamming which is responsible for the criticality
of the transition. Quantitatively, the fullRSB solution captures the critical exponents that characterize
jamming and these exponents are compatible with what is measured in numerical simulations.
Therefore it is natural to conjecture that the scaling theory emerging from the fullRSB analysis is correct also in three dimensional
packing of hard spheres. As we will see in the next chapter these conclusions seem to hold also in other systems
that are far from jamming but display the same kind of criticality.
Therefore for jamming critical system, one has that fullRSB physics seems to be very visible in three dimensional numerical simulations.
Correspondingly signatures of the Gardner transition, namely a diverging spin glass susceptibility, have been found.

A natural question is therefore the reason why colloidal glasses such as hard spheres are more prone to display Gardner/fullRSB physics 
with respect to molecular glasses.
A major difference between hard or soft spheres close to jamming and soft molecular glass models at zero temperature is that
in the former, the systems are barely mechanically stable. Indeed the physics of the jamming transition is deeply affected
by the fact that close to this point, sphere packings are are almost isostatic. Isostaticity is the condition such that the number of touching spheres
 equals the number of degrees of freedom in the system. This condition seems to be exactly verified at
jamming and one can extract a set of diverging correlation lengths that control the suppression of fluctuations away
from isostaticity \cite{hexner2018two, hexner2019can}.
As we will discuss in the next chapter, isostaticity implies that non-linear excitations are bond breaking and can be system spanning.
To trigger a global rearrangement of the system of hard spheres at jamming, it is tyically sufficient to open a contact between two spheres.
These excitations are of the same nature as the spin flip excitations in the SK model \cite{wyart2012marginal} and in the KHGPS model.
As in the KHGPS model, they are the main drive of the statistical properties of the response of hard sphere packing to external perturbations.

However in the SK and KHGPS model, the fact that a single spin flip is sufficient to generate an avalanche is only due to the fully connected
nature of the interactions and therefore this property is lost as soon as the model is on a finitely connected lattice.
It is possible that marginal stability in finite dimensional spin glass models is restored as soon as one looks at excitations which have a fractal
dimension that is positive. In other words, it may well be that finite dimensional spin glasses are stable for single or few simultaneous
spin flip excitations and become marginally stable when one flips a number of spins that scales with a sublinear power of the system size.
Therefore probing marginal stability of the RSB phase at zero temperature becomes more problematic in spin glasses
while hard spheres at jamming have the isostatic sum rule that makes excitations well localized on contacts between spheres.
Therefore an important direction to understand better the spin glass transition in a field is to construct models where one can have access
to the size of non-linear excitations in a controlled way. In my opinion this is the most important missing ingredient that is left in the theory.
Having this under control probably will lead to a better theory of the spin glass phase.

While the conclusions above naturally lead to an interesting direction to understand the role 
of finite dimensional effects, I believe that there are also interesting directions within mean-field theory itself out-of-equilibrium. 
In the context of the SK model, it has been shown that equilibrium avalanches, namely abrupt changes in the ground state
when the magnetic field is slightly increased, have the same statistics as out-of-equilibrium avalanches \cite{le2012equilibrium} in which local minima above the ground state  are destabilized in the same way.
This is mainly due to the fact that in the SK model, linear excitations are frozen by the hard nature of the spins.
In the context of the KHGPS model instead, one may have equilibrium avalanches obeying different statistics than the out-of-equilibrium ones.
The main reason is that while equilibrium avalanches are induced by the chaotic nature of the energy landscape upon changing the local 
external field, in the case of out-of-equilibrium avalanches, this is not true anymore. Indeed it is reasonable to expect that given the fact
that the degrees of freedom are continuous variables, local minima get destabilized when they become saddles at strictly zero temperature.
The avalanches induced by this kind of events may be rather different from the ones happening at equilibrium and a careful
investigation, both numerical and theoretical is mandatory.

A related question concerns the properties of the energy landscape in the KHGPS model.
Crossing the zero temperature RSB transition line in the phase diagram, one passes from a convex landscape to a non-convex
one with exponentially minima in the system size.
It would be very interesting to understand the behavior of the corresponding complexity, defined as the number of minima of the Hamiltonian at a given energy, and 
particularly its behavior close to the transition point \cite{gershenzon2023site}. It may be possible that the behavior
of the landscape changes depending on whether the corresponding transition is 
in the linear or TLS universality class. 

Finally, we recall that the KHGPS model has its roots in a series of models that were introduced
to understand the low temperature behavior of the specific heat in glasses.
The study of the specific heat at low temperature requires the introduction of quantum fluctuations in the model
and it would be very interesting to understand whether a detailed analysis is doable within the KHGPS model.
Spin glass models have been investigated at low temperature adding quantum fluctuations, see \cite{cugliandolo2023quantum} for a recent review.
However it has been claimed that the saturation of the bound in \eqref{replicon_KHGPS} naturally leads to a specific heat which scales as $C_V\sim T^3$.
However, we have seen that at the critical point, in the TLS universality class, the bound in  \eqref{replicon_KHGPS} is not saturated and therefore the mechanism
leading to a cubic specific heat is not found. It would be interesting therefore to check whether the KHGPS model gives rise to a linear in $T$ specific heat at low temperature.

\fontfamily{palatino}\fontseries{ppl}\fontsize{11}{11}\selectfont

\chapter{{Rigidity transitions, continuous constraint satisfaction and optimization problems}}\label{Ch_jamming}

\section{Introduction}
In the last nine years I have dedicated a substantial fraction of my research activity to
study questions around the jamming transition and its relation with constraint satisfaction
and optimization problems. 
After the discovery in \cite{charbonneau2014fractal} that the critical properties of the jamming transition of hard spheres
are described by replica symmetry breaking physics, at least within the infinite dimensional treatment of glassy hard sphere states,
several questions were raised and I dedicated some efforts to understand them.
The purpose of this chapter is to review my activity on this research subject and the progresses that I contributed to make.

The study of the jamming transition has its roots in the physics of granular materials \cite{jaeger1996granular}.
As much as molecular glasses, granular materials are typically found in amorphous structures and we are interested in understanding 
their statistical properties.
Granular matter differs from molecular glassy materials because in the former case thermal (and quantum) fluctuations are essentially irrelevant
and the main driving forces that are responsible for their physical properties are of entropic origin.

A key critical point which is essential to understand and organize the phase diagram of these systems is the jamming transition \cite{liu1998jamming}.
Considering for example frictionless hard spheres, the jamming transition is found at the density at which hard sphere packings cannot be compressed further.
This point also marks the onset of rigidity of soft sphere packings at zero temperature: if spheres can overlap, below the jamming density they can be moved 
without paying an energy cost while above jamming they are found in local minima of the potential energy landscape and form an athermal amorphous solid 
which responds with a finite shear modulus upon deformations.

One of the reasons why jamming has attracted a lot of attention, is that it is a critical point characterized by diverging lengthscales and timescales,
both controlled by a set of critical exponents \cite{liu2010jamming, charbonneau2017glass}. 
A fully consistent theory for the jamming transition of three dimensional systems is still lacking.

The present chapter is dedicated to the review the physics of jamming and rigidity transitions, focusing mostly on the systems that I have been working on.
As in the previous chapter, rather than developing a microscopically grounded first principle mean-field theory 
of such systems, I will mainly discuss a mapping between the jamming transition of several finite dimensional models
and the phase diagram of random infinite dimensional constraint satisfaction and optimization problems.
This has two advantages: on the one hand it will be useful to present a solution of the corresponding infinite dimensional models in a unifying way.
On the other hand the mapping to abstract models is useful and paradigmatic and it can be helpful beyond the soft matter context where
these models originated from.
This is actually one of the main reasons I have been interested in this research subject: indeed the particular class  of models
that share the physics of jamming points are {\it Continuous} Constraint Satisfaction Problems (CCSP).
These problems have been less studied in computer science and probability with respect to their discrete counterparts (think about $K$-satisfiability for example \cite{mezard2009information, moore2011nature, monasson1996entropy, monasson1997statistical, monasson1999determining, mezard2002analytic, semerjian2003relaxation, krzakala2007gibbs, montanari2007solving, montanari2008clusters, semerjian2008freezing, ding2015proof}) due to the fact that the latter are essential  to define computational
complexity classes. However it turns out that CCSP are very important given
that they appear naturally is simple models of artificial neural networks.

This chapter is organized as follows.
In Sec.\ref{sec_finite_dimensions} we review the problem of the jamming transition in finite dimension and in several different settings; we will 
describe the critical properties of the transition and how they change when changing the nature of the problem.
In Sec.\ref{sec_perceptrons} we will map these finite dimensional systems to a set of infinite dimensional
random constraint satisfaction and optimization problems. This section will present the mains steps for a parallel solution of all these models 
which will allow a detailed comparison between them, both technical and conceptual.
In Sec.\ref{Sec_iso_criticality} we will instead discuss the critical behavior arising is such systems and review the mathematics behind it
and in Sec.\ref{off_equilibrium_dynamics} we will discuss how this critical behavior emerges also in off-equilibrium dynamics
and we will review the main open questions in this context.

\section{The jamming/rigidity transition: models}\label{sec_finite_dimensions}
The jamming transition is an ubiquitous phenomenon: when a system of objects, interacting with a hard core excluded volume potential,
is sufficiently compressed, it jams at a critical density. The jamming transition has been investigated extensively in the last
30 years especially because it provides an original point of view on the physics of amorphous materials \cite{liu1998jamming}.
In this section I will review some interesting aspects of jamming in various settings. 
It is important to say that this review is not meant to be exhaustive. I will limit myself to present the topics that have attracted my attention the most
and that I contributed to study.

\subsection{Packing Spheres}\label{sec_spheres_general}
The simplest model of a granular material is a packing of spheres. 
Consider a set of $N$ $d$-dimensional spheres in a volume $V$, whose centers are denoted by $d$-dimensional vectors $\underline x_i$.
The spheres are frictionless and interact with a finite range interaction potential which is controlled by their radii $R_i$.
If the radii are all equal the system is said to be monodisperse, conversely we have a polydisperse system.
In this section we consider a finite range interaction potential defined out of the gaps between two spheres, $i$ and $j$, as
\beq
h_{ij} = |\underline x_i - \underline x_j| - R_i -R_j\:.
\label{gap_spheres}
\eeq
It is evident that $h_{ij}>0$ if the spheres do not overlap and $h_{ij}<0$ otherwise. Two spheres for which the corresponding gap is equal to zero are said to be in contact.
We are interested in studying the statistics of assemblies of spheres that do not overlap.
Finding non-overlapping configurations of spheres is a constraint satisfaction problem.
The natural parameter controlling the hardness of this problem is the density of the spheres or, equivalently, the packing fraction $\phi$ defined as
the fraction between the total volume of the spheres and the total volume available to the system.
One can expect that when $\phi$ is sufficiently small, non-overlapping configurations of spheres do exist and can be found. 
Increasing $\phi$, such configurations progressively rarefy in phase space up to the point where
they do not exist anymore.
The densest packings of non-overlapping monodisperse spheres are known in $d=$ 2, 3, 8 and 24, see \cite{conway2013sphere, viazovska2017sphere, cohn2017sphere}. However the problem is unsolved in any other dimension and is in general thought to be hard especially when the system is polydisperse or when particles have non-spherical shapes.

A possible approach to solve this problem is to develop search algorithms that try to find non-overlapping configurations at increasingly higher densities.
Despite the fact that the space of possible search algorithms is huge, here we discuss two of the main ones:
\begin{itemize}
\item {\it Gradient Descent}. In order to remove overlaps between spheres one can define a cost function that penalizes them.
The simplest example is the Hamiltonian of Harmonic Soft Spheres (HSS) defined as as 
\beq
H=\frac 12 \sum_{ij}h_{ij}^2 \theta(-h_{ij})\:.
\label{harmonic_spheres}
\eeq
The minimization of $H$ can be done by using the Gradient Descent algorithm.
One can initialize the position of the spheres at random in space and then update their positions according to the dynamics \cite{o2003jamming}\footnote{In numerical simulations one cannot run true continuous time Gradient Descent. Rather one can discretize Eq.~\eqref{GD_HSS} via an Euler scheme so that $\underline x_i(t+\de t) = \underline x_i(t)-\de t \frac{\partial H}{\partial \underline x_i}$ and run the corresponding dynamics for sufficiently small $\de t$.}
\beq
\underline{\dot x}_i(t) = - \frac{\partial H}{\partial \underline x_i}\:.
\label{GD_HSS}
\eeq
The endpoint of this dynamics can be either a configuration with no overlaps or a local minimum of the corresponding Hamiltonian \eqref{harmonic_spheres}.
In the large $N$ limit, there exists a critical packing fraction $\phi_J$ such that, for $\phi<\phi_J$ with probability one the system reaches a zero energy configuration,
while for $\phi>\phi_J$ the dynamics ends in a local minimum of $H$ at positive energy. The transition at $\phi_J$ is the jamming point of the Gradient Descent algorithm. 
\item {\it Hard Sphere compression.} Since we would like to find non-overlapping particles, it makes sense to construct search algorithms
in which overlaps between spheres are never generated. A way to do this is provided by the so called Lubachevsky-Stillinger compression algorithm \cite{lubachevsky1990geometric, kansal2002computer}.
Let us define a parameter $\gamma(t)$ such that $\gamma(t=0)=0$ and $\gamma(t\to \infty)=1$. The form of  $\gamma(t)$ defines precisely the compression protocol.
At $t=0$, the spheres are thrown at random in space. Their initial radii is fixed to be $\gamma(0)R_i=0$ and with probability 1 they do not overlap so that $h_{ij}(t=0)>0$ for all couples of spheres.  
A dynamics is then assigned to the spheres by giving them independently an initial velocity extracted from a Gaussian distribution. 
The spheres interact as perfectly repulsive frictionless hard spheres which means that their collisions are resolved completely elastically. In this way, the system explores
the phase space available to configurations which have no overlaps. Increasing progressively $\g(t)$, one can have two situations. 
Either the pressure of the system stays finite all along the compression protocol up to $\gamma=1$ in which case an overlap-free configuration is produced at the target packing fraction,
or the pressure diverges at some critical value of $\gamma_J$. In this last case, the target packing fraction is beyond the jamming point whose packing fraction is given by the one at $\gamma=\gamma_J$.
\end{itemize}
Both algorithm are clearly characterized by a critical point $\phi_J$ which is the jamming transition. For $\phi<\phi_J$ with probability one in the thermodynamic limit,
the algorithms do find configurations with no overlaps while for $\phi>\phi_J$ either the configurations are not accessible in the case of the compression of hard spheres, or the algorithm leads
to a local minimum of a cost function as in the case of GD on the Hamiltonian of HSS.

It is known that the jamming packing fraction $\phi_J$ is not universal and numerical simulations have shown for example that $\phi_J$ is a function of the compression protocol $\gamma$. In particular, the larger the compression rate, the smaller the jamming packing fraction \cite{parisi2010mean, chaudhuri2010jamming}.
Furthermore, while the gradient descent descent algorithms is a completely out-of-equilibrium dynamics, the compression algorithm can be mapped to some equilibrium GB measure if the compression is performed in an adiabatic way (namely with a compression rate smaller than the relaxation time of the system to the corresponding stationary state). In this case one finds a phenomenology that is very similar to the one of colloidal glasses and this phenomenology can be fully understood in the limit of infinite spatial dimensions \cite{charbonneau2017glass, parisi2020theory}.

Despite the difference between the gradient descent algorithm and the compression ones, remarkably it is found that the configurations of spheres at jamming have rather universal
properties. 
These properties can be divided in two main classes: (i) bulk properties involving the behavior of thermodynamic quantities such as the energy, the pressure and the shear moduli and (ii) microstructural properties such as the properties of the contact network and the statistics of gaps between spheres. 
A remarkable scaling theory has been developed for bulk properties in \cite{goodrich2016scaling} and we will not review it here.
Instead we will focus now on the microstructural properties.
\vspace{0.2cm}

{\bf Isostaticity }-- At jamming, a set of permanent contacts is established. The corresponding contact network has statistical properties that can be studied in detail.
In particular, the total number of contacts equals the total number of degrees of freedom and this property is called {\it isostaticity}. 
This means that, if we call $z$ the average degree of the contact network, then, in the thermodynamic limit, $z=2d$.
It is also found that fluctuations away from isostaticity are suppressed. For example one can consider GD for $\phi\geq \phi_J$ and run several times the algorithm 
for different random initial conditions. At the end of the dynamics, a finite number $C$ of pairs of spheres will overlap and it is clear that $C$ grows linearly with the system size so that we can define $c$ as $C= c Nd$. The fraction of overlaps $c$ is a random number and therefore one can study its statistics. It is found that the average value $\overline c$ converges to 1 in the unjamming limit where $\phi\to\phi_J^+$. This is nothing but a consequence of isostaticity. However one can also show that the variance of $c$ shrinks to zero in the unjamming limit too \cite{hexner2019can}. This leads to the definition of diverging lenghtscales that control the suppression of the fluctuations of overlaps and that imply that the system is hyperuniform in the contact number, see also \cite{hexner2018two} where a different procedure has been employed with the same conclusions \cite{Diamant2019}.
\vspace{0.2cm}

{\bf Force and Gap distributions }-- Given a configuration at jamming one can look at the distribution of gaps $g(h)$. It is found that for $h\to 0$, $g(h)\sim h^{-\gamma}$ and $\gamma\simeq 0.41$, see \cite{lerner2013low, charbonneau2015jamming} for the most precise recent numerical simulations on this. Furthermore the contact network can be associated to a set of contact forces. This can be understood in two ways. If we imagine that the spheres are hard, at jamming the system cannot be compressed further and therefore any external force applied to the boundary of the system will generate a force network spreading all over the system to sustain the external pressure. Another way to understand and measure the contact forces is by looking again to HSS slightly above jamming. We have already mentioned that in this case the system develops overlaps
namely pairs of spheres such that the corresponding gap is negative. Let us call these gaps $h_o$ and the index $o$ runs over the pairs of spheres for which the corresponding gap is negative. The average absolute value of $h_o$ is given by $\overline h=\frac{1}{C} \sum_o |h_o|$ and if we take the limit $\phi\to \phi_J^+$ we know that $\overline h\to 0$. Therefore we can define the forces as $f_o=|h_o|/\overline h$. The forces have a finite limit in the unjamming limit and we can study their statistics. If bucklers spheres, namely spheres that are poorely connected\footnote{More precisely bucklers are spheres that are in contact with $d+1$ spheres.} are properly removed from the statistical analysis \cite{charbonneau2015jamming}, it is found that the probability distribution of the rest of contact forces follows a universal behavior for $f\to 0$, namely $P_f(f)\sim f^{\theta}$ with $\theta\simeq 0.42$. Note also that force with large values occur but exponentially rarely \cite{o2003jamming, liu1995force}.
\vspace{0.2cm}

{\bf Density of states }-- Considering again HSS slightly above jamming one can investigate the spectral properties of the Hessian matrix in a local minimum of $H$ found by the gradient descent algorithm.
It is found that the density of states $D(\omega)$ follows the following behavior sufficiently close to jamming
\beq
D(\omega)\sim\begin{cases}
\omega^\nu & \omega<\omega^*\\
{\rm const} & \omega^*<\omega<\omega'\\
D_{\rm reg}(\omega) & \omega>\omega'
\end{cases} 
\eeq
where $D_{\rm reg}$ is a smooth function and $\omega'$ is a high frequency. The low frequency part $\omega<\omega^*$ contains mainly phonons and localized quartic excitations \cite{mizuno2017continuum}. 
This behavior is cut off at a scale $\omega^*$ above which the density of states is constant before changing towards a non-universal regular high frequency part.
The frequency $\omega^*$ is a function of the distance from jamming. In particular it is found that $\omega^*\to 0$ when $\phi\to \phi_J^+$ so that at the jamming point 
the density of states is flat and positive at zero frequency. Since the system has a vanishing characteristic frequency, this can be associated to a diverging lengthscale at the jamming point \cite{silbert2005vibrations, wyart2005effects, liu2010jamming, liu2011jamming}.
This legthscale is nothing but a different manifestation of isostaticity and it can be connected to deviation from isostaticity when at finite distance from $\phi_J$.
\vspace{0.2cm}

It can be shown that the physics of packings of frictionless spheres at jamming is essentially described by these universal properties.
In particular, the response of these systems to external perturbation is entirely driven by (i) the properties of the contact network and (ii) the statistics of pairs of spheres which are either weekly in contact
or close to become in contact \cite{liu2010jamming, muller2015marginal}.

\subsection{Jamming far from jamming:  linear spheres}

The physics of packing of spheres beyond the jamming point can be only studied if spheres are not hard and can overlap.
In this case the properties of the systems will depend on the interaction potential and on the algorithm used to minimize it.
The most general form of the cost functions penalizing overlaps between overlapping spheres is
\beq
H = \sum_{i<j} v(h_{ij})\theta(-h_{ij})
\eeq
and the HSS potential defined in the previous section represents the particular case of $v(h)=h^2/2$.
The minimization of $H$ via gradient based algorithms (for example gradient descent or some simulated annealing extensions)
can be easily implemented if $v(h)$ is differentiable and if
\beq
\left.\frac{\de v(h)}{\de h}\right|_{h\to 0^-}=0\:.
\label{continuity_cond}
\eeq
This is the example of HSS and Hertzian soft spheres for which $v(h) = 2 h^{5/2}/5$. However one could be interested in studying the general case
where $v(h)=|h|^a$ and look at how the statistics of local minima of $H$ change when varying the exponent $a$.
The shape of $v(h)$ has a qualitative change when crossing the case $a=1$. Indeed for $a>1$, the condition in Eq.~\eqref{continuity_cond} is satisfied
while this is no more true for $a\leq 1$. While a generic theory for $a<1$ is still lacking, in a recent work \cite{franz2020critical}, the marginal case $a=1$ has been studied in detail.
Since the potential $v(h)$ is linear ramp for $h<0$, the corresponding model has been called {\it linear soft sphere} model.

\begin{figure}
\centering
{
\setlength{\fboxsep}{0pt}
\setlength{\fboxrule}{1.5pt}%
\fbox{\includegraphics[width=\columnwidth]{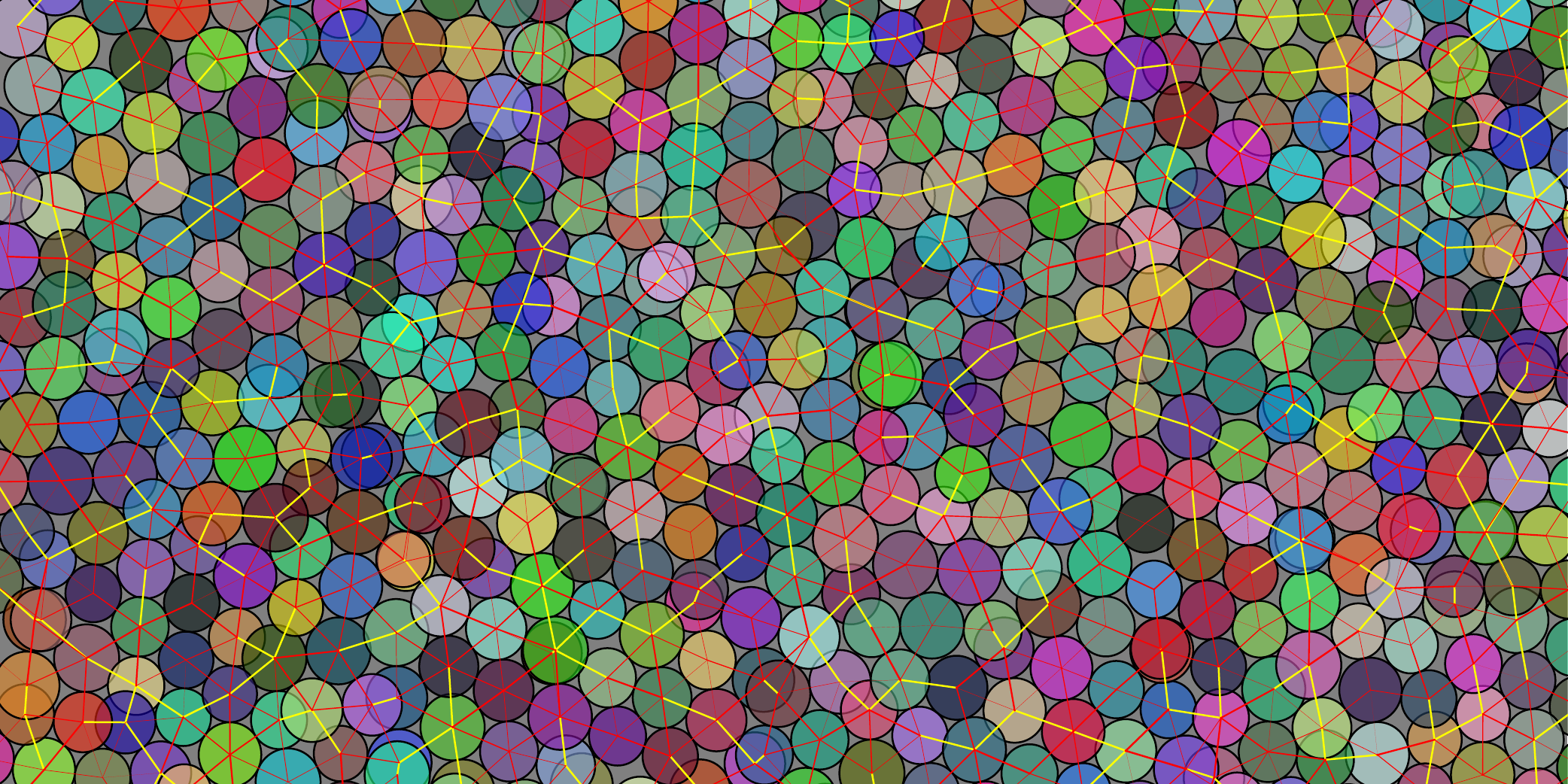}}
}
\linespread{0.8}
\caption{\footnotesize{An example of a configuration of $N=1024$ polydisperse linear soft spheres at $\phi=1$ in a box with periodic boundary conditions. In red we join the centers of spheres in contact. The thickness of the lines is proportional to the force exchanged in the contact. In yellow we join the centers of overlapping spheres and therefore the thickness of the lines is the same for all of them being the force exchanged equal to 1.}}
\label{LS_config}
\end{figure}

Finding local minima of the Hamiltonian of linear soft spheres is definitely a challenge and a precise way to do it can be found in \cite{franz2020critical}. 
Here I will not discuss this non-trivial algorithmic problem but I will mainly focus on the properties of the configurations of the system in local minima of $H$ for $\phi>\phi_J$.

In Fig. \ref{LS_config} I present a typical configuration of polydisperse linear soft spheres above jamming. 
The first interesting remark is that one finds spheres that are in contact.
This is very different from what happens for $a>1$ where as soon as $\phi>\phi_J$ 
there are no contacts and one finds only overlaps.
Another surprising fact comes as soon as one counts the total number of contacts. 
Indeed it is found that this is exactly equal to the number of degrees of freedom in the system.
In other words, even if the system is far from jamming in the dense phase, the corresponding
contact network is isostatic. However the notion of isostaticity here is slightly different from the $a>1$ case at jamming
since now contacts coexists with overlaps.

Having an isostatic system, one can look at the distribution of forces and gaps between spheres in contact.
It turns out that for linear soft spheres, contact forces are defined only in the interval $(0,1)$ \footnote{Instead for Hard Spheres one finds that the forces are in the interval $(0,\infty)$.}. 
While we will not give the precise algorithm to compute them, see \cite{franz2020critical}
for some details, this fact can be understood as follows. 
Starting from two isolated spheres in contact one can easily imagine that in order to make them overlap
one should push one sphere against the other by a force that is strictly larger\footnote{If the potential is defined by $v(h)=A|h|$, then the minimal force is equal to $A$, namely the slope of the potential for $h<0$.} than 1.
Therefore, contacts can sustain forces which are in the interval $(0,1)$ only. Overlapping spheres instead exchange a constant force equal to 1.

Using numerical simulations, one can access the statistics of contact forces $P_f(f)$. 
It is found \cite{franz2020critical} that close to the edges of the support of this distribution, one has
\beq
P_f(f) \sim \begin{cases}
F_-f^{\theta_-} & f\ll 1\\
F_+(1-f)^{\theta_+} & 1-f\ll1
\end{cases}
\label{forces_linear}
\eeq
and the exponents $\theta_-$ and $\theta_+$ are compatible to be the same as the one which is found at jamming of hard spheres, namely $\theta_-\simeq \theta_+\simeq 0.42$. 
The prefactors $F_-$ and $F_+$ appear to depend on many details including the density and the preparation protocol of the packings.
However the power law behavior at the edges seems very robust\footnote{It must be remarked that if the packings are sufficiently close to the jamming density the exponent $\theta_-$ can change. However a careful analysis can show that this shift in the exponent is only apparent and one can recover the scaling in Eq.~\eqref{forces_linear} if localized bucklers excitations are properly removed.}.

At the same time one can focus on the distribution of gaps between spheres that are not in contact.
Calling this density $g(h)$ it is clear that its support will be on both the negative and positive side of the $h$-axis.
In particular it is found that
\beq
g(h)\sim \begin{cases}
H_+h^{-\gamma_+} & 0<h\ll 1\\
H_-|h|^{-\gamma_-} & -1\ll h< 0\\ 
\end{cases}
\eeq
and, again, numerically one observes that $\gamma_-\simeq \gamma_+\simeq 0.41$. This means that the gap distribution is controlled by the same critical exponent
that controls positive gaps of hard sphere packing at jamming. Also in this case the prefactors $H_+$ and $H_-$ are not universal because they are found to depend on the preparation protocol and on the density of the packings.

All in all, these findings point to the fact that dense linear soft spheres far from jamming share many critical properties of packing of hard spheres at the jamming transition.
For this reason they are said to be jamming critical.
Jamming criticality far from jamming came really as a surprise. 
It reveals that there is nothing peculiar about the jamming point itself. 
Criticality is rather associated with emergent isostaticity of local minima and 
the physics of linear soft spheres is a manifestation of {\it self-organized criticality}.
 
\begin{figure}
\centering
{
\setlength{\fboxsep}{1pt}
\setlength{\fboxrule}{1.5pt}%
\includegraphics[width=\columnwidth]{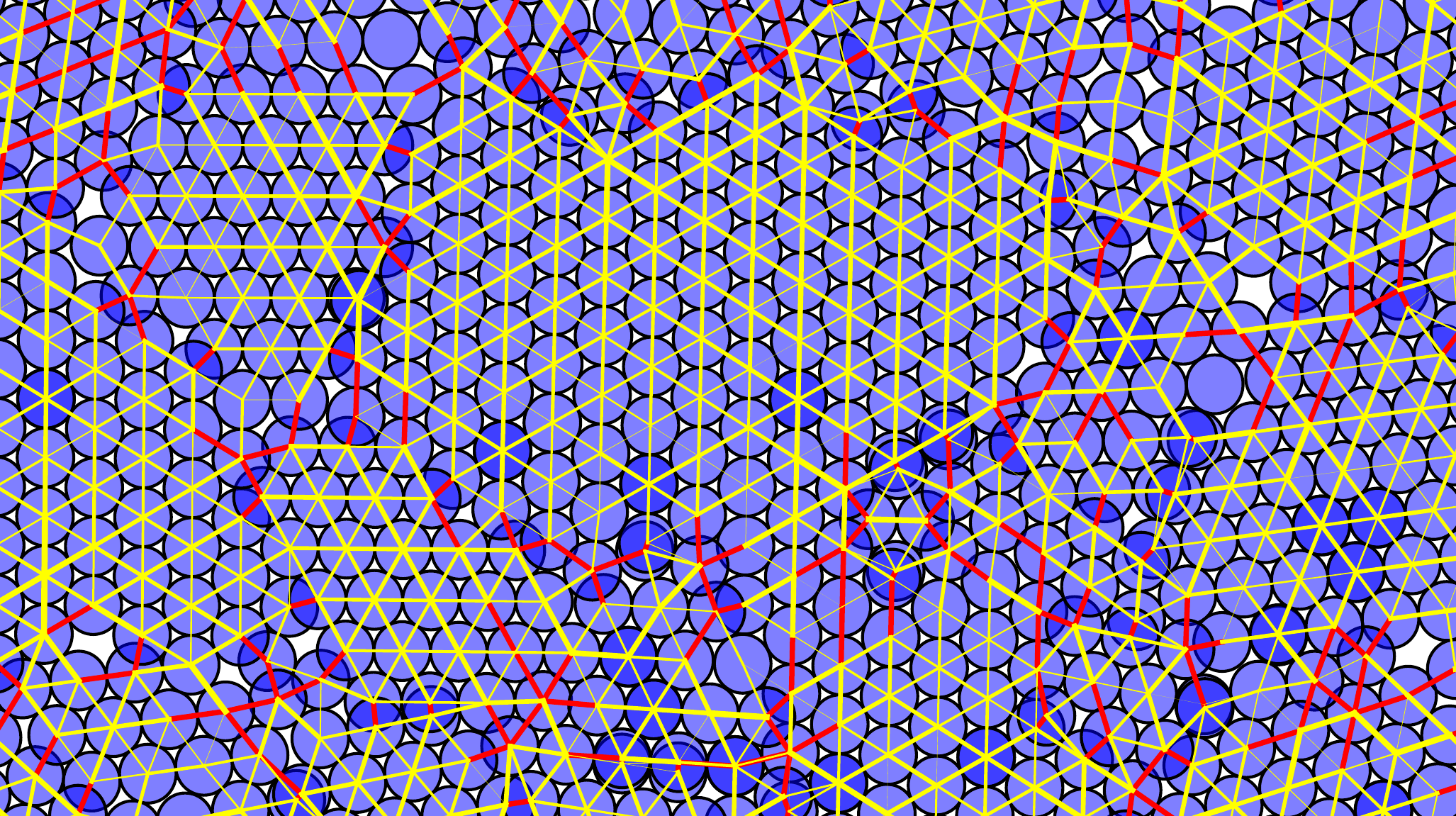}
}
\linespread{0.8}
\caption{\footnotesize{A configuration of monodisperse linear spheres above jamming. The system forms crystalline patches with rather compact but disordered interfacial boundaries.}}
\label{Mono_spheres}
\end{figure}

Finally it is important to note that as soon as the system is monodisperse, local minima of linear spheres deviate from the isostatic universality class.
In Fig.~\ref{Mono_spheres} we plot a typical locally stable configuration of linear spheres above jamming.
We clearly see that crystalline patches form and are arranged in a disordered way with disordered but compact boundaries. The force network is also disordered and the stress propagating through crystalline patches is not uniform. A precise study of the statistics of these configurations has not been done but it is likely that, given that the configurations are closer to hyperstatic crystals, the force and gap disributions are not singular as in the isostatic case.
Moreover, numerical experiments also show that infinitesimal polydispersity brings instantaneously the system to the isostatic class and this is analogous to what has been observed at jamming \cite{charbonneau2019glassy}.
The ground state structure of soft linear spheres above jamming is also interesting. It is easy to show that an hexagonal lattice of spheres is an unstable structure given that it contains only overlaps, a configuration that is stable only if not perturbed (it is a saddle of the energy landscape). Therefore soft linear spheres provide interesting examples of systems with frustrated ground states.
Finally, an open question is to understand if anything happens in the structure of local minima when increasing the packing fraction: in other words it may be that monodisperse linear spheres undergo phase transition in the dense phase. This possibility has not been investigated in detail.

\subsection{Jamming of non-spherical particles}\label{sec_ellipses}
Up to now we have reviewed the essential features of the jamming transition of frictionless spherical particles
and the properties of the dense phase of linear soft spheres.
The purpose of this section is to review briefly what changes in the case of non-spherical particles. 
It is clear that once the spherical geometry is left an infinite number of possible deformations can be done and giving a review on all this research activity goes beyond the scope of this manuscript. Therefore I will limit myself to consider the simplest example of non-spherical particles, namely ellipsoidal
objects. However I would like to emphasize that the same physics that I will describe below has been observed also in other non-spherical particle models \cite{williams2003random, blouwolff2006coordination, wouterse2007effect, wouterse2009contact, marschall2018compression, jiao2010distinctive, delaney2010packing, vanderwerf2018hypostatic}. 

In the case of the ellipsoids, the coordinates of the center of a particle are not sufficient to define its position given that also the orientation of the particle must be given.
Numerical simulations on systems of ellipsoids have been extensively performed in the literature \cite{van2009jamming, torquato2010jammed, donev2004improving, man2005experiments, delaney2005random, donev2007underconstrained, mailman2009jamming, zeravcic2009excitations, schreck2012constraints}.
The main big difference between such systems and spheres is that, at jamming, the latter are isostatic (the number of contacts equals the number of degrees of freedom) while in the former case the system is found to be hypostatic (the number of contacts is smaller than the number of degrees of freedom). In particular, ellipsoids can be characterized by a parameter $\Delta$ that describes the strength of asphericity of the particles. If $\Delta\to 0$, ellipsoids become spheres. It is found that the coordination of the contact network $z$ for frictionless packings of ellipsoids at jamming is found to follow a law given by 
\beq
z - 2d \sim \Delta^{1/2} \ \ \ \ \ \Delta\to 0\:.
\eeq
This would suggest that packings are hyperstatic, but in fact, the total number of degrees of freedom is actually $2(d+d_{ex})$ where the extra ones, $2d_{ex}$,  come from the orientational degrees of freedom. 
Numerical simulations show that $z<2(d+d_{ex})$ and therefore such packings are hypostatic.

Together with the statistical properties of the contact network, numerical simulations have focused on the characterization of the density of states $D(\omega)$ of such packings close to jamming.
This is found to be characterized by three branches. 
The lowest frequency branch contains $\NN_0=N(d_{ex}-\delta z/2)$ vibrational modes, being $\delta z =z-2d$. Their typical frequency is given by $\omega_0\sim (\Delta p)^{1/2}$ being $p$ a small pressure which tunes the distance from jamming.
The intermediate branch is composed instead by a set of $\NN_1=N \delta z/2$ modes with characteristic frequency $\omega_1\sim \Delta$. 
Finally the part of the spectrum corresponding to the highest frequency band is composed by $\NN_2=Nd$ translational modes whose typical frequency $\omega_2\sim \Delta^{1/2}$.
Therefore the structure of the density of states differs rather substantially from the one of spherical particles. 

In \cite{brito2018universality, ikeda2020infinitesimal, ikeda2019mean} a theory for these scaling relations as well as for the statistics of the contact network has been proposed.
In this chapter I will review it only partially. In particular I will consider the mapping between the properties of packing of non-spherical objects and a set of continuous constraint satisfaction problems
in which some internal degrees of freedom are added to mimic the orientational degrees of freedom which characterize  non-spherical particle models.

\subsection{Rigidity transition in confluent tissues}
Granular materials may have very little in common with soft biological tissues at the first sight.
However in recent years a parallel line of research has shown that in fact one can rationalize the physics
of a set of models of confluent biological tissues according to rigidity phase transition points whose nature
is very close to jamming \cite{farhadifar2007influence, nagai2001dynamic, angelini2010cell, angelini2011glass, nnetu2012impact, schoetz2013glassy, haeger2014cell, sadati2013collective,garcia2015physics, sepulveda2013collective, bi2015density, bi2016motility, bi2014energy, merkel2018geometrically, merkel2019minimal}.

In confluent biological tissues cells tesselate space leaving no hole in the system. 
The simplest example of these systems is the epithelial tissue that forms the human skin.
Understanding the physical properties of confluent tissues is an important problem which connects the metabolisms
of cells with their collective mechanical response to perturbations.
Among the general questions of interest in this research field we can mention
\begin{itemize}
\item {\it Morphogenesis}: the formation of organs within organisms and organisms themselves, follows a path in which cell-cell interactions (both metabolic and mechanical)
shape the macroscopic forms of multicellular aggregates. It is clear that the biology of cells interacts with large scale elasticity of the corresponding aggregates and therefore understanding the whole process
of morphogenesis requires to have under control the mechanical properties of cell aggregates \cite{turing1990chemical,heisenberg2013forces,guillot2013mechanics, etournay2015interplay}.
\item {\it Tumor growth and metastasis generation: } it is well known that tumor masses have a core-periphery structure. Cells at the boundaries of tumors 
have a higher mobility with respect to the corresponding ones in the core. This higher mobility, together with the interaction between healthy and cancer cells is a key player in metastasis generation \cite{li2020cooperation, sinha2020spatially}.
\end{itemize}

A growing body of literature has therefore concentrated in modeling confluent biological tissues from a statistical mechanics perspective to investigate the questions above.
Confluent tissues tessellate space and therefore any reasonable model of such systems should involve a way to cover space via approximate polytopes (polygons in $d=2$).
While there are many models that do so, I will mostly focus on a set of models called {\it Voronoi models} \cite{bi2016motility}. 

In this class of models, one has $N$ cells, whose centers are described by $d$-dimensional vectors $\underline x_i\in {\mathbb R}^d$. 
In order to define a tessellation one can use the Voronoi construction to partition the space.  A cell $i$ is defined by the set of points satisfying the following condition
\beq
i\equiv\{x\in {\mathbb R}^d: |x_i-x|< |x_k-x| \  \forall k\neq i\}\:.
\eeq

Given the Voronoi partition, one can define the Hamiltonian of the cell-cell interaction in the following way.
Because of biological/metabolic reasons, one can expect that cells try to attain a target volume $V_0$ and a target surface $S_0$.
Therefore, calling $V_i$ and $S_i$ the volume and surface of the cell $i$, the Voronoi model is defined via the following cost function
\beq
H=\frac 12 \sum_i \left[k_V(V_i-V_0)^2 + k_S (S_i-S_0)^2\right]
\label{def_Voronoi}
\eeq
where $k_V$and $k_S$ are elastic coefficients.
The Hamiltonian in Eq.~\eqref{def_Voronoi} is rather general and its form has not been constructed out of biological data, as far as I know. 
However it has been extensively studied in a series of works, see for example \cite{bi2016motility, merkel2018geometrically, sussman2018no}.

An interesting feature of Eq.~\eqref{def_Voronoi} is that the Hamiltonian is a sum of local terms. However interactions between cells
are contained in the form of the functions $V_i$ and $S_i$: indeed the volume and surface of a given cell crucially depends on the position of the neighbors.
Therefore, the nature of cell-cell interactions is clearly multibody and models of confluent tissues represent therefore a case where
pairwise interactions are not enough to describe the microscopic interactions between degrees of freedom.

The Hamiltonian in Eq.~\eqref{def_Voronoi} depends on an adimensional control parameter called the {\it shape index} which is defined as
\beq
p_0=S_0 V_0^{1/d-1}
\eeq
and tunes the degree of compactness of cells. Intuition suggests that having $p_0$ large enough pushes cells to have elongated forms where the surface is large compared to the volume.
On the other hand, if $p_0$ is small, the shape of cells is pushed towards a more compact and regular form.
Therefore one can think of $p_0$ as a way to inject {\it tension} locally in the system since driving cells towards small values of $p_0$ forces them to take shapes that are not compatible 
with the geometry of the problem.

Given the Hamiltonian \eqref{def_Voronoi}, one can define a dynamics of the system in which elasticity, mediated by Eq.~\eqref{def_Voronoi},
is mixed with both thermal noise and some sort of active drive. Cells can indeed self-propel and therefore it is natural to consider self-propulsion in modeling tissues.
Therefore one arrives to the following class of dynamical models
\beq
\dot{\underline x}_i(t) = -\frac{\partial H}{\partial \underline x_i} +\underline \eta_i(t)+\underline v_i(t)
\label{dyn_tissues}
\eeq
where $\underline \eta_i(t)$ is a standard Langevin noise at temperature $T$ and $\underline v_i(t)$ is a self-propulsion velocity whose dynamics is model dependent
and can be autonomously fixed.

A study of Eq.~\eqref{dyn_tissues} can be done via numerical simulations \cite{bi2016motility} which show that if the self-propulsion and the noise in the problem are sufficiently small, one can induce a phase transition
in the collective behavior of cells from a fluid like regime at large $p_0$ where cells can diffuse in the system, to a solid, glassy, regime at small $p_0$ where cells are caged.
The nature of this rigidity transition has attracted a lot of attention recently. 
One way to try to understand it is to focus on the unusual driving control parameter, the shape index, and set to zero both the noise and self-propulsion terms.
Therefore one can study gradient descent dynamics starting from a random initial configuration of the centers of the cells
\beq
\underline{\dot x}_i (t) = -\frac{\partial H}{\partial \underline x_i}
\eeq
and describe how the properties of the configurations depend on the shape index $p_0$ when the dynamics stops in a local minimum of $H$.

This procedure allows to study how the potential energy landscape defined by $H$ depends on $p_0$. In 3$d$ it is found that for large enough values of $p_0$, gradient descent
is able to find configurations of the centers of the cells at zero energy \cite{merkel2018geometrically}. This implies that all cells attain their target volume and surface.
Moreover, the Hessian of the Hamiltonian, computed in such configurations, has a spectrum with an extensive number of exactly zero modes.
This means that the local minima found by gradient descent are not isolated but are rather connected along a {\it zero energy manifold} of configurations, which takes the shape of a {\it canyon}
in phase space.

The situation changes completely if $p_0$ is decreased below a critical value $p_J$ where, with probability one in the thermodynamic limit,
gradient descent will not be able to find a configuration where cells attain their target shape\footnote{Actually, for $p_0<p_J$ all cells reach a shape which differs from the target one.}.
Therefore in this case, gradient descent dynamics stops in a local minimum of $H$ at finite energy density. The corresponding Hessian does not contain any zero mode\footnote{Apart the trivial ones given by global translational and rotational invariance.} and the system is trapped in a glassy local minimum of the potential energy landscape.

These findings therefore imply that the critical point $p_J$ is a zero temperature rigidity transition, as far as the potential energy landscape is concerned.
More abstractly, it marks the point such that for $p_0>p_J$ one can find a solution to a set of non-linear equations given by
\beq
V_i = V_0 \ \ \ \ \ \ \ S_i=S_0 \ \ \ \forall i=1,\ldots, N
\label{eq_constraints}
\eeq
These equations can be seen as {\it equality constraints} on the degrees of freedom represented by the centers of the cells and therefore, 
from this point of view, the zero temperature rigidity transition is nothing but a satisfiability transition for a constraint satisfaction problem defined by Eq.~\eqref{eq_constraints}.
This resonates with what we have discussed so far for the jamming transition of particle systems: however in the case of spheres, the non-overlapping constraint
on pairs of spheres, for example, turns out to define a set of {\it inequality constraints} while in the present case, {\it the constraints are non-linear equalities} imposed on the degrees of freedom.

The picture described so far holds in all dimensions $d\geq 3$. The $2d$ case instead requires a particular care.
Indeed it has been shown in \cite{sussman2018no} that at zero temperature there is in fact no rigidity transition for the Hamiltonian in Eq.~\eqref{def_Voronoi}.
The reason why this is so can be easily understood via constraint counting. 
Consider again Eq.~\eqref{def_Voronoi} in $d$ dimensions. 
The Hamiltonian defines a set of $2N$ non-linear constraints for $dN$ degrees of freedom. 
On general grounds one can expect that for $d\geq 3$, the system is underconstrained and therefore it is possible to observe a phase where $H=0$.
On the other hand, $d=2$ represents a marginal case and it has been shown in \cite{sussman2018no} that, under certain conditions, no transition is possible.
One way to induce a transition also in $2d$  is to change the Hamiltonian in Eq.~\eqref{def_Voronoi} with a simpler one
\beq
H=\frac 12 \sum_i (S_i-S_0)^2\:. 
\eeq
This reduces by a factor two the total number of constraints and opens the possibility for having a transition at a critical value $p_J$ \cite{sussman2018no}\footnote{Note that in this case the $V_0$ entering in the definition of the shape index is $V_0=V/N$ where $V$ is the total volume accessible to the system. We note also that given that Voronoi models are defined out of tessellations, one always considers $V_0=V/N$.}.

All in all these considerations suggest that, at zero temperature, the rigidity transition of confluent tissues is a sort of satisfiability transition point
for a continuous constraint satisfaction problem as much as the jamming transition of spheres. 
The main difference between the two transitions is in the nature of the constraints imposed to the degrees of freedom.

A clarification of the critical properties of the transition and how they differ with jamming transition of particle systems is still in the infancy of this field \cite{li2021softness}.
As an example, the nature of the low-temperature harmonic excitations is still unclear and a precise study of the density of states as done in standard
models of simple glasses is still not done yet, see \cite{sussman2018anomalous} for some preliminary numerical simulations.

In the following I will suggest a mean-field and abstract model to study the rigidity transition of confluent tissues and show that it has the same qualitative
phase diagram of finite dimensional models described by Eq.~\eqref{def_Voronoi}. This will provide a benchmark to study the critical properties of the transition
itself and to compare them with numerical simulations in finite dimensional models.

\subsection{Mean-field theory in the infinite dimensional limit}
In the last ten years I have been working extensively to try to construct a theory of simple glass models in the exactly soluble limit of infinite spatial dimensions \cite{parisi2020theory}.
This theoretical framework has been applied only partially to the problems discussed in the previous sections and in the present section I  review the outcome of this approach where
this has been developed.

A theory of the jamming transition of hard and harmonic soft sphere glasses in infinite dimensions has been presented in a series of works \cite{charbonneau2014exact, charbonneau2014fractal, biroli2016breakdown, rainone2015following, biroli2018liu, rainone2016following, yoshino2014shear, scalliet2019marginally}.
The outcome of this approach is the following:
\begin{itemize}
\item sufficiently close to the jamming point in the density-temperature plane, such system always undergo to a Gardner transition \cite{biroli2016breakdown, biroli2018liu}. 
This is always the case no matter the detail of the preparation of the glassy states and the protocol used to adiabatically drive them close to jamming.
\item Below the Gardner transition point one finds RSB within a glass. The form of RSB at jamming is typically continuous at least when it comes
to describe the organization of the ultrametric tree of pure states sufficiently close to the leaves.
\item The continuous (full) RSB solution close to the leaves of the ultrametric tree implies that the structure of configurations at jamming are isostatic.
At the same time, the distribution of contact forces and gaps displays the critical features discussed above and are controlled by a set of critical exponents
that can be computed exactly, see \cite{charbonneau2014exact, charbonneau2014fractal}, and appear to be in agreement with numerical simulations \cite{charbonneau2021finite} and experiments \cite{perrin2021nonlocal}.
\end{itemize}
Therefore the theory in infinite dimensions provides a detailed description of glassy states close to jamming and it works remarkably well to describe finite dimensional data.

The same theoretical approach could be developed, in principle, to study linear soft spheres in the infinite dimensional limit. Close to the jamming point at zero temperature, for $\phi<\phi_J$, 
this would lead exactly the same result as the theory for hard spheres, given that the critical behavior approaching jamming from the SAT side
does not depend on the interaction potential between spheres.
However a derivation of the $d\to \infty$ phase diagram at any temperature has not been done but it follows directly from the analysis in \cite{parisi2020theory}.

The situation becomes even harder for the case of non-spherical particles and, even more, for confluent tissues.
Here the theory in infinite spatial dimensions could be in principle developed but technically it is very challenging and has not been done yet.

In the following section I will instead describe a different approach to the rigidity transitions of such systems which is based on the introduction of simplified models
of continuous constraint satisfaction and optimization problems. This approach is less first principle in nature but allows to get many results assuming that
the class of models that we consider is in the same universality class as the corresponding particle-based models in infinite dimensions.

Despite the fact that we loose a first principle derivation of the results, there are anyway several advantages to consider this approach:
\begin{itemize}
\item the mean-field models that we will consider are sufficiently simplified to allow for an almost straightforward theoretical analysis. 
This analysis will not be able to give a global account for the shape and properties of the phase diagrams of the corresponding particle models in infinite dimensions. 
However it will be enough if we focus on the properties (critical and general) of the jamming point or (in the case of linear soft spheres) the configurations right beyond it. 
Moreover, for what concerns critical and universal properties,
this approach will lead to scaling theories that correspond to the same ones arising for particle models. 
This has been checked in the case of hard and soft spheres but it has not been investigated in details in the case of soft linear spheres. However
in this last case we have strong evidences that this is correct, given that also finite dimensional linear soft spheres seem to follow, within the precision of numerical simulations,
the scaling theory derived whithin abstract models.
\item I believe that the main advantage to develop simplified and abstract models, is that there is the chance that they can turn out to be paradigmatic and useful also
beyond the context where they have been developed and this is the main reason why they are attractive for me from an intellectual point of view.
\end{itemize}

Therefore in the following section I will introduce a set of simple and abstract models for the rigidity transitions discussed in this section and I will also present the main steps of a parallel solution of all of them.

\section{Random continuous constraint satisfaction and optimization problems}\label{sec_perceptrons}
In the previous sections we emphasized that the rigidity/jamming transitions of many model systems can be regarded as (algorithmic) satisfiability transitions for CCSP.
The purpose of this section is to introduce a series of CCSP and optimization problems which turn out to be mean-field models
for the transition of the particle-based models described in the previous sections. After defining the main ingredients of the models, we will proceed by presenting their
parallel solution. 
We will not be exhaustive in all the technical details but we will try to underline the crucial points where the solutions of different models differ
and how this depends on the physical ingredients of the models themselves. The details about the formal solution of the models can be found in the original papers.
Furthermore, since they rely on standard tools the interested reader can look at the lecture notes of a course I have given in 2017 at the IPhT on this subject \cite{urbani2018statistical}.

\subsection{Models}\label{sec_ab_models}
The first step to define an abstract model is to fix the corresponding degrees of freedom.
All the models we have been discussing so far involve continuous variables.
They represent: the coordinates of the centers of the spheres; the center and orientation 
of the ellipsoids (for the simplest case of non-spherical particles); the centers of the cells for the Voronoi model of confluent tissues. 
Of course we are interested in sending to infinite the system size or, equivalently, the number of degrees of freedom.
Therefore, from an abstract point of view, we will consider an $N$-dimensional vector $\underline x = \{x_1,\ldots, x_N\}$
as the vector describing a collection of degrees of freedom that we are allowed to change. The components of this vector are real variables
and in order to have a finite phase space we will consider this vector to have a fixed norm. Without loosing generality we will consider
$|\underline x|^2=N$ so that the phase space is an $N-1$-dimensional sphere. The thermodynamic limit consists then in taking the limit $N\to \infty$.

Given the degrees of freedom, we need to define a set of constraints over them.
To do so we introduce a set of $M=\alpha N$ functions $h_\mu(\underline x)$, with $\mu=1,\ldots, M$.
According to the discussion above we will consider two types of constraints: inequality and equality constraints.
The simplest form of inequality constraints will be
\beq
h_\mu(\underline x)\geq 0 \ \ \ \ \ \forall \mu=1,\ldots, M\:.
\label{ineq}
\eeq
This constraints will model the non-overlapping constraints of both spherical and non-spherical objects.
Equality constraints instead emerge naturally when considering confluent tissues and they will take the form
\beq
h_\mu(\underline x)=0 \ \ \ \ \ \forall \mu=1,\ldots, M\:.
\eeq
In view of an analogy with particle based models, we call the variables $h_\mu$, the gap variables.

It will be also useful to define a cost function or Hamiltonian which penalizes violated constraints.
The most general form of such Hamiltonian is given by
\beq
H[\underline x] = \sum_{\mu=1}^M l(h_\mu(\underline x))
\label{generic_H}
\eeq
We will always assume that $l(h)$ is a continuous function but its particular form will depend on the nature of the constraints.
If inequality constraints are considered then Eq.~\eqref{ineq} can be enforced by choosing $l(h)=v(|h|)\theta(-h)$ being $\theta$ the Heaviside step function.
Therefore the function $v(x)$ defines the interaction between degrees of freedom when constraints are violated and must represent a penalty cost. In other words it should be a continuous positive definite function of its argument but its precise shape is not constrained further.

In order to model equality constraints instead, we will need to consider $l(h)$ such that it is a continuous positive definite function with a unique zero at $h=0$.
Again, the form of this function away from $h=0$ is rather arbitrary.

If the cost function $H$ and $l(h)$ is differentiable one can define a gradient descent algorithm to find solutions of the CCSP.
The algorithm takes the form of
\beq
\dot x_i(t) = -\nu(t)x_i(t) -\frac{\partial H}{\partial x_i}
\label{GD_generic_models}
\eeq
and the Lagrange multiplier $\nu(t)$ is self-consistently fixed in order to enforce that at all times $|\underline x(t)|^2=N$. The algorithm finds solution as soon as it relaxes to zero energy.

It is important to note that the particular form of $H$ as defined by $v$ and $l$ enters in two important aspects. 
First it changes the global potential energy landscape of the problem. However, crucially, zero energy phase space configurations  are left invariant if we change $v(h)$. 
The second important point is that the precise form of $H$ crucially changes the properties of gradient descent dynamics.

Up to now we did not specify neither the precise form of the functions defining the constraints, $h_\mu(\underline x)$ neither the ones of $l(h)$ and $v(h)$. In the following we remove this degeneracy and define a
set of specific models which directly connect with the particle models discussed before.

\subsubsection{Perceptron and hard and soft (harmonic and linear) spheres}
It is convenient to begin the discussion with the problem of packing non-overlapping spheres in space.
In this context, we would like to find an arrangement of the centers of the spheres such that there is no overlap between them.
This implies that for each pair of spheres we would like to 
have that their corresponding gap is non-negative
\beq
h_{ij}\geq 0\ \ \ \ \ \forall i,j=1,\ldots N
\label{constraint_spheres}
\eeq
and the gap variable $h_{ij}$ is defined in Eq.~\eqref{gap_spheres}.
The main idea contained in this section is to "replace" the constraint in Eq.~\eqref{constraint_spheres} with a simpler one defined by an abstract $h_\mu(\underline x)$.
The simplest form for $h_\mu(\underline x)$ is a linear function:
\beq
h_\mu(\underline x) = \frac 1{\sqrt{N}}\underline \xi^\mu\cdot \underline x - \sigma
\label{def_h_perc}
\eeq
and therefore we can define the following CCSP
\beq
h_\mu(\underline x)\geq 0 \ \ \ \ \forall \mu=1,\ldots, M=\alpha N\:.
\label{const_percp}
\eeq
The vectors $\underline \xi^\mu=\{\xi^\mu_1,\ldots, \xi^\mu_N\}$ are real $N$-dimensional vectors whose components are chosen independently at random 
from a Gaussian distribution with zero mean and unit variance.
The first term on the r.h.s. of Eq.~\eqref{def_h_perc} is a quantity of order one, since the normalization of $\underline x$ ensures that typical values of its entries are of order one
and therefore by central limit theorem $\underline \xi^\mu\cdot \underline x\sim \mathcal{O}(\sqrt{N})$ which justifies the factor $N^{-1/2}$ in Eq.~\eqref{def_h_perc}.
The constant $\sigma$ is a control parameter of order one.

It is instructive to compare the form of the constraint defined by Eq.~\eqref{def_h_perc} with the one in Eq.~\eqref{gap_spheres}.
The control parameter $\sigma$ plays a role similar to $R_i+R_j$ in Eq.~\eqref{gap_spheres} and it is easy to show that increasing $\sigma$ is equivalent to making the problem harder, as much as increasing the radii of individual spheres.
Indeed, the random variables $r_\mu = \underline \xi^\mu\cdot \underline x /\sqrt{N}$, evaluated on a random configuration $\underline x$ are Gaussian with zero mean and unit variance. 
Therefore the number of constraints in Eq.~\eqref{const_percp} that are violated on a random instance of $\underline x$ grows when increasing $\sigma$.

\begin{figure}
\centering
\includegraphics[scale=0.25]{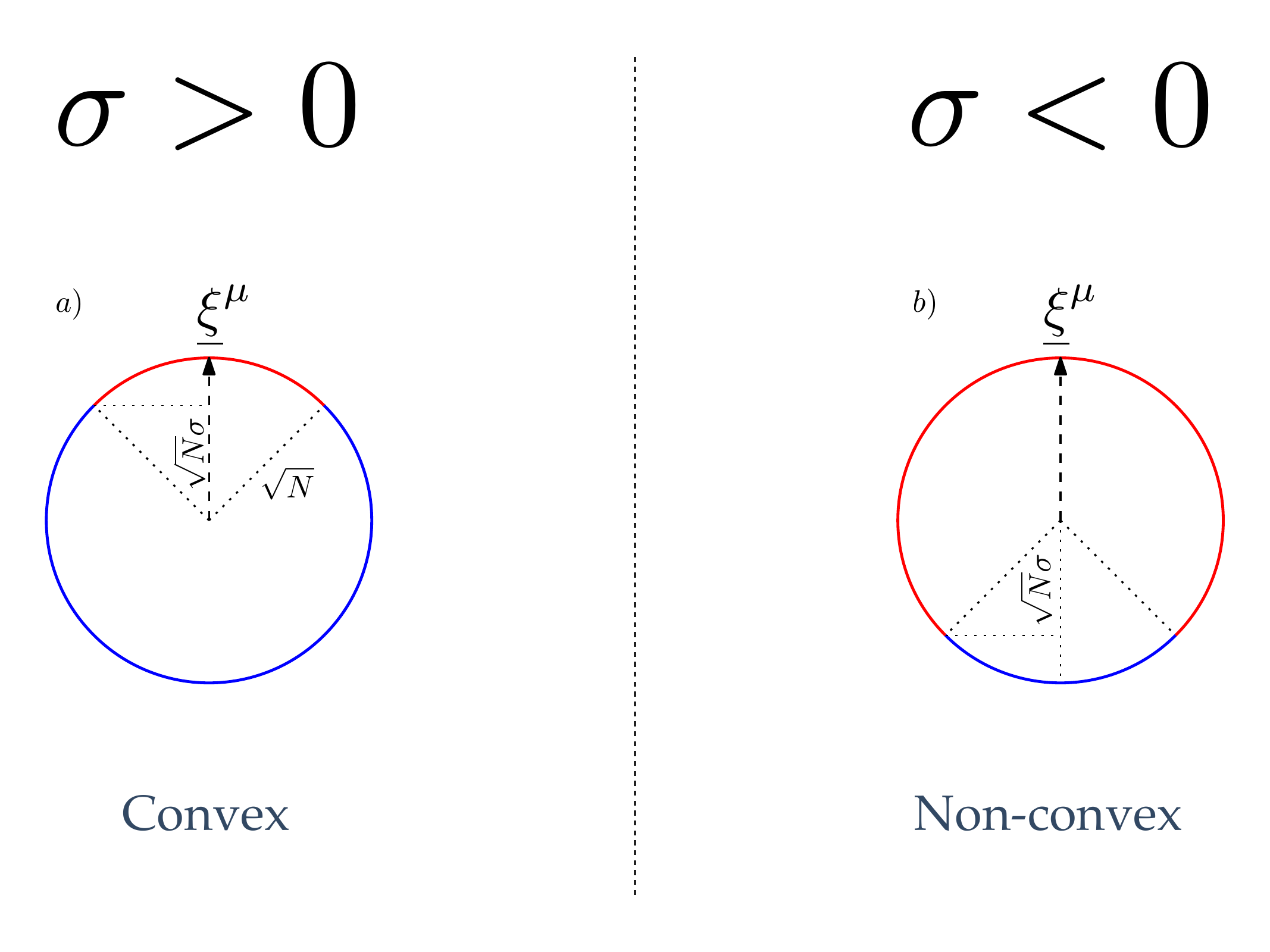}
\linespread{0.8}
\caption{\footnotesize{The geometry of the perceptron constraint as a function of $\sigma$. Panel $a)$: for $\sigma>0$ the region of phase space allowed by a single constraint (in red) is convex. Panel $b)$: the allowed region for a single constraint when $\sigma<0$ is non-convex.}}
\label{fig_space_pecpt}
\end{figure}

One of the most interesting aspects of the CCSP defined in Eqs.~\eqref{def_h_perc} and \eqref{const_percp} is that the nature of problem changes depending on the sign of $\sigma$.
This is summarized in Fig.\ref{fig_space_pecpt}.
If $\sigma>0$ each constraint $h_\mu\geq 0$ defines a convex portion of phase space and therefore the solution space of the CCSP is the intersection of convex domains which is convex.
In other words, in this case, for sufficiently small $\alpha$ one expect to have a volume of solutions defined by a convex region of phase space (a convex "lake").
Increasing $\alpha$, this convex space shrinks and at $\alpha = \alpha_J(\sigma)$ it reduces to a point beyond which no solutions can be found with high probability.

If we consider now the case $\sigma<0$, we loose the convexity of the allowed phase space identified by each constraint and therefore the whole CCSP is defined by the intersection of non-convex domains. 
In general, the intersection of non-convex domains is non-convex and can lead to a space of solutions which is disconnected (or poorely connected). {\it This fragmentation of the space of solutions is the essence of RSB}.
Given these considerations we expect that the CCSP defined by Eqs.~\eqref{def_h_perc} and \eqref{const_percp} and the packing problem of spheres are possibly related only when $\sigma<0$.

The CCSP defined in Eqs.~\eqref{def_h_perc} and \eqref{const_percp} is called the {\it spherical perceptron} problem.
It has been first studied via combinatorial methods for $\sigma=0$ by Cover in \cite{cover1965geometrical} who showed that the structure of the solution space undergoes to a phase transition at $\alpha=\alpha_J(0)=2$. This means  that the probability that a solution to the problem exists, jumps discontinuously from $1$ (satisfiable, SAT, phase) to 0 (unsatisfiable, UNSAT, phase) when increasing $\alpha$ at $\alpha=\alpha_J$ in the $N\to \infty $ limit.
The study of the problem for $\sigma>0$ was instead done by Gardner in \cite{gardner1988space} who derived the phase diagram for $\sigma>0$ by obtaining the exact equations determining the satisfiability threshold $\alpha_J(\s)$.
Soon after, Gardner and Derrida showed in \cite{gardner1988optimal} (see also Monasson's PhD thesis) that for $\sigma<0$, the satisfiability transition is in a RSB region of the phase diagram: in other words a RSB transition in the structure of the solution space happens before hitting the satisfiability transition point when increasing $\alpha$ at fixed $\sigma<0$. 
The link between the satisfiability transition of the spherical perceptron for $\sigma<0$ and the jamming point of spheres has been proposed by Franz and Parisi in \cite{franz2016simplest}. The full phase diagram of the model was derived in \cite{franz2017universality}  where the scaling theory for the satisfiability (jamming) point for $\sigma<0$ was studied showing that it coincides with the ones of spheres in infinite dimensions.

Together with the set of constraints of Eqs.~\eqref{def_h_perc} and \eqref{const_percp} one can define a cost function which penalizes configurations in phase space that violate them.
For example, if one chooses $v(h) = h^2/2$ together with Eq.~\eqref{generic_H} one obtains a cost function which closely follow the one of Harmonic Soft Spheres and therefore the spherical {\it harmonic} perceptron can be regarded as a mean-field abstract model of harmonic spheres.
This case was studied in \cite{franz2016simplest} and in \cite{franz2015universal} where an analysis of the Hessian of the ground state of $H$ was done. One of the main conclusions of \cite{franz2015universal} is that on approaching the jamming point from the UNSAT side, namely for $\alpha\to \alpha_J(\sigma)^+$, the density of states $D(\omega)$ is flat and positive for $\omega\to 0$ as much as in HSS.

If we choose $v(h)=|h|$ we obtain the {\it linear perceptron} problem which is therefore the corresponding abstract model for linear soft spheres.
The UNSAT phase of the model, $\alpha>\alpha_J(\sigma)$, was first studied in \cite{majer1993perceptrons} where it was understood that for $\sigma>0$ one can have a phase transition in the ground state structure of $H$ from a replica symmetric solution close to $\alpha_J(\sigma)$ to a RSB solution for larger values of $\alpha$. 
In \cite{franz2019critical} this problem was studied again and it was shown that the ground state of the linear perceptron in the RSB-UNSAT phase shares the same jamming criticality far from jamming observed in finite dimensional linear spheres. A scaling theory for the corresponding critical exponents was established generalizing the one for the jamming transition of hard spheres.

\subsubsection{Polydisperse perceptron and non-spherical particles}
To extend the spherical perceptron model to describe non-spherical particles, the following argument was suggested in \cite{brito2018universality, ikeda2019mean}.
Let us focus on ellipsoids for simplicity. The state of an ellipsoid can be determined by giving the position of its center and its orientation.
The orientation can be thought as a kind of {\it internal} degree of freedom.
Imagine now that we consider again spheres but we promote the radii of spheres to freely change with a certain chemical potential. 
The resulting model is a system of {\it breathing spheres} and the radii play the role of internal degrees of freedom. In \cite{brito2018universality} it was shown that the jamming point of this model has the same
properties of the corresponding one of ellipsoids. This is interesting because, following the analogy with the perceptron, it suggests a way to develop a new abstract model.

Indeed we consider the gap variables $h_\mu(\underline x)$
as defined by
\beq
h_\mu(\underline x) = \frac 1{\sqrt{N}}\underline \xi^\mu\cdot \underline x - \sigma -\Delta R_\mu
\label{h_ellissi}
\eeq
and we added a set of $M$ degrees freedom $R_\mu$. We will fix them so that
\beq
\sum_\mu R_\mu^2 =M\:.
\eeq
The control parameter $\Delta$ plays the role of asphericity of ellipsoids or the degree of "annealed" polydispersity in the breathing sphere model.
If we set $\Delta=0$ we are back to the simple spherical perceptron problem which describe a jamming transition in the same universality class as the one of spheres. 
Therefore the limit $\Delta\to 0$ is morally equivalent to the limit in which the asphericity of ellipsoids is sent to zero.

The polydisperse perceptron can be studied exactly via statistical mechanics methods and it has been shown in \cite{brito2018universality, ikeda2019mean} that at jamming
it reproduces all the key features of the jamming transition of non-spherical particles described in Sec.\ref{sec_ellipses}.

\subsubsection{Canyon Model and confluent tissues}
Confluent tissue models at zero temperature experience a rigidity transition which is akin a satisfiability transition for continuous degrees of freedom subjected to non-linear equality constraints.
Therefore if we want to model this behavior via the formalism described above we need to change the form of the function $h_\mu(\underline x)$ which has been considered in Eqs.~\eqref{def_h_perc} and \eqref{h_ellissi} since in both cases this is linear in the degrees of freedom. The simplest non-linear function which we can consider is
\beq
h_\mu(\underline x)=\frac 1N\sum_{i<j}J^\mu_{ij} x_ix_j - \sigma\:.
\label{gap_tissues}
\eeq
The matrices $J^\mu$ are independent identically distributed random matrices with entries extracted as independent Gaussian random variables with zero mean and unit variance.
The control parameter $\sigma$ now plays essentially the same role as the shape index in the Voronoi tissue model.

The form of the variables $h_\mu$ appearing in Eq.~\eqref{gap_tissues} can be further generalized.
In particular one can define
\beq
h_\mu(\underline x)=r_\mu(\underline x) - \sigma
\eeq
with $r_\mu(\underline x)$ a random Gaussian field with the following statistics
\beq
\overline r_\mu(\underline x) =0 \ \ \ \ \ \ \ \ \overline{r_\mu(\underline x)r_\nu(\underline y)} = \delta_{\mu\nu} G\left(\frac{\underline x\cdot \underline y}{N}\right)
\eeq
and the overline denotes the average over the realization of the random fields $r_\mu$s.
The covariance function $G(z)$ is rather arbitrary\footnote{but it should satisfy the properties of being a covariance function.}: in the particular case of Eq.~\eqref{gap_tissues} we have $G(z)=z^2/2$.

It is clear that a satisfiable phase can be only present if the model is {\it overparametrized} in the sense that $\alpha<1$. In this case the landscape of solutions looks like a {\it canyon} landscape and therefore we dub this model the Canyon Model.
This was studied and solved very recently in \cite{urbani2023continuous} considering a square loss function $l(h)=h^2/2$. A related model where $h_\mu(\underline x)$ is a linear function was instead studied in \cite{fyodorov2022optimization} and preliminary results on non-linear generalizations were also discussed in \cite{tublin2022few}. More recent studies in the mathematical literature have been presented in \cite{subag2023concentration, montanari2023solving}. 
Finally, recently in \cite{10.21468/SciPostPhys.15.5.219} the solution of the gradient descent dynamics of the model has been presented.

The phase diagram of the model can be understood at fixed $\alpha<1$ and changing $\sigma$. 
When $\sigma$ is small, one can argue that since the variables $r_\mu$ are Gaussian and centered in zero, the CCSP has many solutions.
Conversely, increasing $\sigma$ one can cross a satisfiability transition point at a critical value $\sigma_J>0$.

\subsection{A parallel presentation of the solution of the models}
\begin{figure}[h]
\centering
\includegraphics[width=\columnwidth]{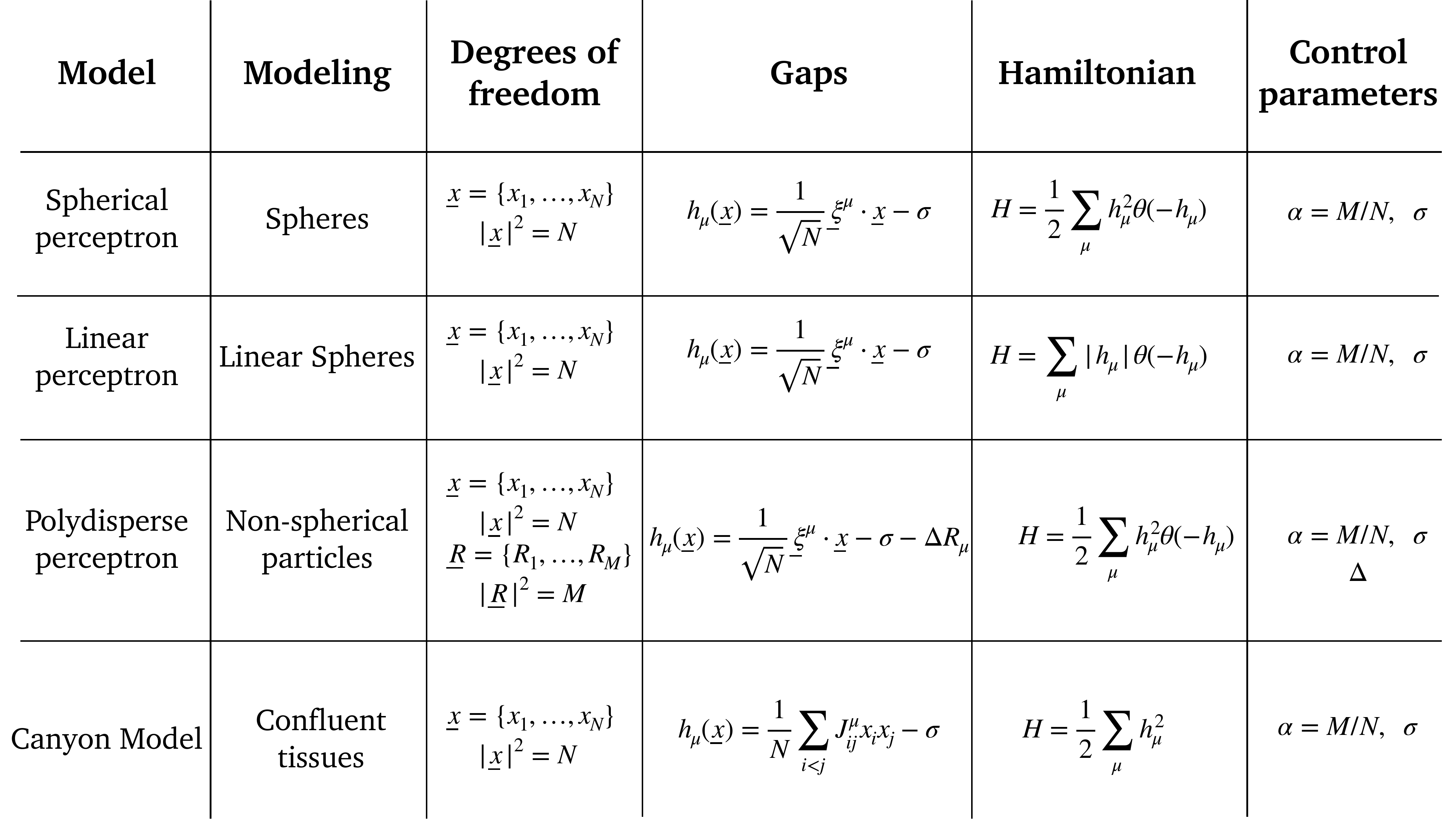}
\linespread{0.8}
\caption{\footnotesize{A table summarizing the main features of the models presented in Sec.\ref{sec_ab_models}.}}
\label{table_models}
\end{figure}

In this section we would like to provide a parallel solution of the models presented in the previous sections.
These models are summarized in Fig.\ref{table_models} where their main ingredients are reviewed.
We are interested in studying these models close to their satisfiability transition point.

It is important now to make a remark. In the previous section we have emphasized that the rigidity transitions
can be regarded as the point where simple greedy search algorithms stop finding solutions.
Most of the time we have been focusing on the Gradient Descent algorithm.
However it is clear that the location of the satisfiability transition is algorithm dependent. Therefore it is convenient to define
two kinds of satisfiability transitions:
\begin{itemize}
\item {\it Thermodynamic satisfiability (SAT/UNSAT) transition.} This is the point where the probability that a typical, random, large instance of a CCSP
has a solution, jumps discontinuously (for $N\to \infty$) from one (SAT phase) to zero (UNSAT phase).
In other words in the UNSAT phase there are no solutions to the problem with high probability in the large dimensional limit.
\item{\it Algorithmic SAT/UNSAT transition.} This is the point where the probability that a given algorithm (for example Gradient Descent) finds a solution of a typical, random, large instance of a CCSP,
 jumps discontinuously (for $N\to \infty$) from one (algorithmic SAT phase) to zero (algorithmic UNSAT phase).
 Therefore in the algorithmic UNSAT phase, the corresponding algorithm will not be able to find a solution to the problem with high probability in the thermodynamic limit.
\end{itemize}
It is clear that the thermodynamic transition is only a property of the constraints, their nature, their distribution, and the phase space geometry where the degrees of freedom can be tuned.
Therefore, this transition sets a bound to any other algorithmic transition: the thermodynamic UNSAT phase is by definition UNSAT for any algorithm.
On the other hand, the algorithmic transition depends also on the algorithm itself and therefore the corresponding UNSAT phase may be still satisfiable from the thermodynamic point of view.

The two types of satisfiability transitions may have different properties. 
However this is not what is found  since in many cases they share the same "universal" properties. 
This is something that can be checked in various ways empirically but for the moment there is no established theory for this fact\footnote{There is an exception to this statement: when the thermodynamic transition happens in a replica symmetric region of the phase diagram, see for example the spherical perceptron at $\sigma>0$, one can also make a theory for the algorithmic transition, for example with Gradient Descent. In this case the two transitions, thermodynamic and algorithmic coincide for the trivial reason that at the SAT/UNSAT point, only one solution exists and there is a convex landscape around it which therefore drives the gradient descent dynamics towards it.}. We will review this problem later in this chapter.

From the point of view of the theoretical analysis, studying algorithmic transitions is in general much harder than the thermodynamic ones
because in the former case one needs to have a control on the performances of search algorithms in the thermodynamic limit. This can be achieved
only in some particular cases. 
Therefore in this section we will focus on thermodynamic satisfiability transitions. 

A simple way to study this type of transitions is to develop a thermodynamic approach. 
Let us call $\underline X$ the set of degrees of freedom of the CCSP. Since we want be very general, we consider $\underline X=(\underline x, \underline R)$ with $|\underline x|^2=N$ and $|\underline R|^2=M$.
We define the following partition function at inverse temperature $\beta$:
\beq
Z = \int \de \underline X e^{-\beta H[\underline X]}
\label{partition_function}
\eeq
and the integral is over the phase space available to the degrees of freedom 
The Hamiltonian appearing in Eq~\eqref{partition_function} has the generic form described in Sec.\ref{sec_ab_models}. However its precise realization depends on the random variables which define the structure of the gaps. These are either the vectors $\xi^\mu$ of the random matrices $J^\mu$. These quantities are {\it quenched} in the sense that they are fixed in a given realization of the CCSP and therefore play the role of quenched disorder.
In the $\beta \to \infty $ limit, the partition function is dominated by different types of configurations depending on the control parameters of the problem.
If the corresponding CCSP is in a satisfiable phase, there exist configurations for which $H=0$ and therefore in this region, the partition function
measures the volume of solutions of the corresponding problem. Conversely, in the UNSAT phase, the integral in Eq.~\eqref{partition_function} is dominated, at zero temperature, by
the ground state of $H$ and therefore it depends on its properties.

The partition function $Z$ has a typical scaling which is exponential in $N$ namely $Z\sim e^{N \hat f}$. Indeed in the SAT phase we can expect that the volume of solutions of the CCSP is exponential in the dimension and in the UNSAT phase, saddle point method suggests that $Z$ is dominated by the ground state whose energy is extensive in the dimension of the problem.
Therefore it is useful to introduce the free energy defined as
\beq
\mathrm f = -\frac 1{\b N} \log Z =\mathrm e - T\mathrm s
\eeq
and we have denoted by $\mathrm e$ the intensive internal energy of the system and by $\mathrm s$ the corresponding entropy.
The zero temperature limit $T\to 0$ of the free energy changes when crossing the satisfiability transition. 
In particular we have
\beq
\mathrm f \simeq_{\b\to \infty} \begin{cases}
-T \mathrm s & \textrm{SAT phase}\\
\mathrm e_{GS} & \textrm{UNSAT phase}
\end{cases}
\eeq
and we have denoted by $\textrm e_{GS}$ the intensive ground state energy of the Hamiltonian.

The discussion we have been developing so far is at fixed realization of the disorder. However we are interested in the properties of a typical instance of the CCSP and therefore we would like to compute
 the average of the free energy over all sources of quenched disorder.
To do this we employ the replica method. Denoting with an overline the average over the random instances of the CCSP we can write
\beq
\overline f = -\frac 1{\beta N}\overline{\ln Z} = -\frac 1{\beta N}\lim_{n\to 0}\partial_n\overline{Z^n}\:.
\label{replica_method}
\eeq 
In all the expressions above we are assuming that we always take the thermodynamic limit $N\to \infty$ but we will keep omitting this detail in the formulas to simplify the notation.
Eq.~\eqref{replica_method} does not simplify the problem unless we look at the case where
the computation of $\overline{Z^n}$ is performed for integer values of $n$. The way the replica method works is to consider this case and then find the appropriate analytic continuation
down to $n\to 0$. 
The replica method is reviewed in several books, see \cite{MPV87, nishimori2001statistical, mezard2009information, parisi2020theory} and we are not going to discuss it at length, the interested reader can look at these references for further details.

Performing standard manipulations, see \cite{urbani2018statistical} for some detailed explanation on the general philosophy of the method for this particular class of models, one can write the average of the replicated partition function $\overline {Z^n}$ with the following integral
\beq
\overline{Z^n} \propto \int \de Q \de \tilde k \exp\left[ N \left(A_{\rm ent}(Q) + A_{\rm int}(Q, \tilde k) \right)\right]
\label{int_Z_rep}
\eeq
and we have neglected unimportant proportionality factors.
There are two integration variables. 
On one hand we have the overlap matrix $Q$ which is a symmetric $n\times n$ matrix whose diagonal elements are fixed to one, namely $Q_{aa}=1$.
The integration variable $\tilde k$ is instead relevant only for the polydisperse perceptron case when $\Delta>0$, and it is harmless in all the other cases.

The form of the two terms $A_{\rm ent}$ and $A_{\rm int}$ is rather simple.
The entropic term $A_{\rm ent}$ takes the very same form for all the models and it is given by
\beq
A_{\rm ent} = \frac 12 \log \det Q\:.
\eeq
This term is nothing but the Jacobian of the change of variables from $\underline x^{(a)}$, with $a=1,\ldots, n$ denoting the replica index, to $Q$.

The interaction term $A_{\rm int}$ instead as a form which depends on the particular model we are analyzing.
Its general form is given by
\beq
A_{\rm int} =-n\frac \a 2 \left(\log \tilde k- \tilde k\right)+ \left.\alpha \log \exp\left[\sum_{ab}G[Q_{ab}]\frac{\partial^2}{\partial r_a \partial r_b}\right] \gamma_{{\Delta^2}/{\tilde k}}\star  e^{-\beta l(r_a-\sigma)}\right|_{\{r_a=0\}_{a=1,\ldots, n}}\:.
\label{intera_term}
\eeq
The function $G(Q)$ is the correlation of the random variables $r_{\mu}$
which are defined as
\beq
r_\mu(\underline x) = \frac 1{\sqrt N} \xi^\mu\cdot \underline x
\eeq
in the case of the perceptron models, while they are general Gaussian variables in the case of the Canyon model used for confluent tissues.
Therefore we have that $G(z)=z$ in the case of the perceptron models, while it is $G(z)=z^2/2$ in the particular case of Eq.~\eqref{gap_tissues}. 
Moreover we have denoted by $\g_A$ a Gaussian function and the notation with $\star$ denotes a Gaussian convolution
\beq
\g_A \star f(x) = \int_{-\infty}^\infty\frac{\de z}{\sqrt{2\pi A}}e^{-z^2/(2A)}f(x-z)\:.
\eeq
Finally, the first term on the rhs of Eq.~\eqref{intera_term} is nothing but an entropic factor taking into account the volume of the $M$ dimensional sphere that describes the radii $R_\mu$.
If we consider the model where $\Delta\to 0$, this factor gives an additional constant entropic term to the free energy which could be removed by hand since it is irrelevant for all practical purposes.

The replica method proceeds now by evaluating the integral in Eq.~\eqref{int_Z_rep} via a saddle point over $Q$ and $\tilde k$.
While this procedure does not pose any problem for the case of $\tilde k$, the saddle point equations for $Q$ are rather complicated.
The major issue is that we have an $n\times n$ matrix and we would like to take the analytic continuation of the solution of the saddle point equations
when $n\to 0$. 
It is clear that one needs to resort to a parametrization of the matrix $Q$ which allows to take the analytic continuation for $n\to 0$.

In order to do this we will follow the method developed by Parisi \cite{parisi1980sequence, parisi1979infinite, parisi1980order, MPV87}
and consider a hierarchical parametrization of $Q$ encoded in a function $q(x)$ defined on the interval $x\in[0,1]$ with boundary condition $q_m=q(0)$ for $x\in[0,x_m]$
and $q(1)=q_M$ for $x\in [x_M,1]$. The function $q(x)$ is monotonously increasing and therefore it is convenient to define also its inverse $x(q)$.
The function $x(q)$ encodes the probability distribution of the overlap between two different replicas chosen at random from the GB measure:
\beq
P(q)= \overline{\frac 1{Z^2}\int \de X_1 \de X_2 e^{-\beta (H[X_1]+H[X_2])}\delta(q - \underline x_1\cdot \underline x_2/N)} =  \frac{\de x(q)}{\de q}\:.
\eeq

The saddle point equations for $Q$ become equations for the function $q(x)$ or equivalently $x(q)$.
We do not review the procedure to derive such equations, see \cite{urbani2018statistical} for a pedagogical presentation. However we report them in their full generality:
\beq
\begin{split}
&\frac {q_m}{\lambda^2(q_m)}+\int_{q_m}^q\frac{\de p}{\l^2(p)} = \alpha G'(q) \int_{-\infty}^\infty \de h P(q,h) [f'(q,h)]^2\\
&\lambda(q)=1-q_M+\int_{q}^{q_M}\de p x(p)\\
&\begin{cases}
\frac{\partial P(q,h)}{\partial q} &= -\frac {G'(q)}{2}\left[\frac{\partial^2 f}{\partial h^2}+x(q)\left(\frac{\partial f}{\partial h}\right)^2\right]\\
f(q_M,h)& = \log \gamma_{G(1)-G(q_M)+\Delta^2/\tilde k}\star e^{-\beta l(h)}
\end{cases}\\
&\begin{cases}
\frac{\partial f(q,h)}{\partial q} &= \frac {G'(q)}{2}\left[\frac{\partial^2 P}{\partial h^2}-2x(q)\frac{\partial}{\partial h}\left(P(q,h)\frac{\partial f}{\partial h}\right)\right]\\
P(q_m,h)& = \gamma_{G(q_m)}(h+\sigma)\:.
\end{cases}
\end{split}
\label{RSB_eqs_generic}
\eeq
In the case of the polydisperse perceptron problem we have an additional equation fixing the value of $\tilde k$ given by
\beq
1= \frac 1{\tilde k} +\frac{\Delta^2}{\tilde k^2}\int_{-\infty}^\infty \de h P(q_M,h) \left[\frac{\partial^2 f(q_M,h)}{\partial h^2}+\left(\frac{\partial f(q_M,h)}{\partial h}\right)^2\right]\:.
\eeq
These equations are valid for general $\beta$ which implies that they describe the finite temperature phase diagram. If we want to specialize them to study the satisfiability transition point
we need to take the additional limit $\beta \to \infty$. It is important to note that $\beta$ enters only in the boundary condition for the Parisi equation for $f(q,h)$.

Before discussing the outcome of the solution of these equations, it is useful to to mention the way in which they can be solved.
Despite their final structure is model dependent, the equations for $f$ and $P$ are rather universal. 
Indeed the Parisi equation for $f$ admits a natural interpretation in terms of a Bellmann equation, see \cite{MPV87, jagannath2016dynamic}
and it appears in all problems with fully connected topologies. 
Conversely, the equation for $P$ is a Fokker-Planck equation which describes the dynamics of the probability distribution of a {\it controlled} random walker.
Therefore if one knows a solution for $f$ one can in principle integrate
the corresponding equation for $P$. 
Once a solution of both equations is found, it can be used to update the current estimate of $x(q)$ in a carefully defined iterative scheme.
The bottleneck of this procedure is therefore the numerical integration of Eqs.~\eqref{RSB_eqs_generic} which is a nontrivial task.
This has been solved in several contexts, even beyond the models presented here, see \cite{pankov2006low, schmidt2008method, andreanov2012long, charbonneau2014exact, 10.21468/SciPostPhys.4.6.040, rainone2016following, maillard2023injectivity}. In general cases, the structure of the solution of these equations changes a lot at low temperature
or close to jamming due to scaling regimes which have to be taken into account explicitly to improve the accuracy of the numerical scheme \cite{pankov2006low, schmidt2008method, charbonneau2014exact}.
A case where such problems can instead be easily solved is the Canyon model of confluent tissues \cite{urbani2023continuous}.
In this particular case, the equation for $f$ can be solved analytically since its initial condition is a quadratic function when $l(h)=h^2/2$. 
Using the analytical solution for $f$ one can also show that the equation for $P$ is solved by a Gaussian function.
Therefore, in this particular case, the determination of the satisfiability point and the ground state energy can be done precisely.

Before discussing the results on the phase diagram and the critical behavior, I think that it is interesting to mention that many of the theoretical computations discussed in this section can be made rigorous.
In recent years a stream of works in statistics has focused on CCSP and many results and extensions have been developed.
This is for example the case of the spherical perceptron problem, see \cite{shcherbina2003rigorous, stojnic2013another, stojnic2013negative, montanari2021tractability, el2022algorithmic, stojnic2023generic, stojnic2023bilinearly, stojnic2023emph, stojnic2023capacity, stojnic2023fully, stojnic2023fully}.

\subsection{Results: phase diagrams and critical behaviors}

The solution of Eqs.~\eqref{RSB_eqs_generic} can take two, rather different, forms depending on whether they displays replica symmetry or its breaking.
If the solution is replica symmetric one has $q_m=q_M$ and everything simplifies given that the equations for $f$ and $P$ are solved by their initial conditions.
On the contrary, when replica symmetry breaking appears, one has that $q_m<q_M$ and $x(q)$ has a nontrivial profile.
In general, both types of solutions can be present in the SAT and UNSAT phase, and this depends on the specific features of the models.
In the following we discuss the properties of their phase diagrams as obtained by solving Eqs.~\eqref{RSB_eqs_generic}.

\vspace{0.3cm}
{\bf The SAT phase}\\
In the SAT phase, with probability one in the large $N$ limit, all models have a finite volume of solutions.
In the replica symmetric SAT phase (RS-SAT) the typical solutions, as sampled from the GB measure
in the zero temperature limit, are arranged in a simple way in phase space: their average overlap
probability distribution takes the simple form of $P(q)=\delta(q-q_{EA})$. 
In other words two typical solutions are found at overlap given by $q_{EA}\equiv q_M=q_m$.
The value of $q_{EA}$ is model and control parameter dependent.
This situation is typically present in all the models for small enough $\alpha$. Indeed,
regardless the convexity of the CCSP, if the number of constraints is very small, one expects a well connected
region of solutions in phase space which can give  rise naturally to a replica symmetric behavior\footnote{Note that having a small number of constraints, is not always a sufficient condition to induce replica symmetry in the structure of the space of solutions in generic constraint satisfaction problems. For example the perceptron problem with {\it Ising variables} is known to be glassy (dynamical RSB) at all positive and small $\alpha$ \cite{krauth1989storage}, see the recent literature \cite{perkins2021frozen, abbe2022binary, bolthausen2022gardner, baldassi2015subdominant, baldassi2016unreasonable, baldassi2016local}. However the mechanism leading to this behavior is related to the discreteness of the degrees of freedom and as far as I know, there is no example of the same behavior in a CCSP.}.
 
 Conversely, non-convex (or non-linear) models develop RSB when $\alpha$ is increased close to the thermodynamic satisfiability transition.
 The pattern of replica symmetry breaking can change depending on the control parameters.
 For example in the spherical perceptron problem when $\sigma<0$, RSB takes place in many forms depending on the value of $\sigma$, see \cite{franz2017universality}.

\vspace{0.3cm}
{\bf The jamming transition}\\
When $\alpha$ (or equivalently $\sigma$) crosses a critical value, all the CCSP discussed above undergo a satisfiability transition where the volume of solutions shrinks to zero. 
An important order parameter in this regard is $q_M$. This quantity measures the typical overlap between two configurations
in the same pure state of the zero temperature GB measure. When we reach the satisfiability point we expect to have $q_M\to 1$ continuously since a pure state is chrinking to a point in phase space.
Therefore it makes sense to study the behavior of $q_M$ as a function of the distance from jamming, let us call it $\epsilon$ \footnote{For example one can fix $\epsilon=|\alpha-\alpha_J|$, if $\alpha$ is used to drive the system close to the critical point.}. 
The behavior of $q_M$ is typically found to follow a power law behavior, $1-q_M\sim \epsilon^{\kappa}$, but the value of the exponent $\kappa$ changes depending on the nature of the jamming transition.
To discuss this point more carefully, it is useful to define a few quantities.
For the moment we will focus on models with inequalities constraints and come back at the end of this section to the Canyon model of confluent tissues which is instead based on equality constraints.

At the satisfiability point one can imagine that some constraints in Eq.~\eqref{const_percp} become marginally satisfied, namely $h_\mu=0$. In the language of sphere models, these are nothing but contacts and we denote by $C$ their total number.
Isostatic models are such that $C$ equals the number of degree of freedom while for hypostatic models $C$ is smaller than the number of degrees of freedom.
As for sphere models, contacts carry forces. These can be defined either as the Lagrange multipliers (slack variables) which enforce that the corresponding constraints are in fact equality constraints, or 
as the negative gaps which appear infinitesimally slightly in the UNSAT phase, normalized by their average value, as in the context of spheres, see the discussion in Sec.\ref{sec_spheres_general}.
We indicate by $P_f(f)$ the corresponding empirical distribution. In the large $N$ limit this becomes a continuous function and the comparison with sphere models suggests that it is important to look at the behavior of $P_f(f)$ for $f\to 0$. This is found to take the form $P_f(f)\sim f^{\theta}$ with $\theta$ a critical exponent.
Additionally, we can look at the constraints that are strictly satisfied, $h_\mu>0$, at the transition point and indicate with $g(h)$ their distribution. Again it is useful to study in detail the behavior of $g(h)$ close to zero and we assume that this takes the form $g(h)\sim h^{-\gamma}$ with $\gamma$ another critical exponent.

Given all these observables one can define two rather broad universality classes.
The first universality class includes 
\begin{itemize}
\item models that are hypostatic at jamming (both when the transition is in a RS or RSB phase of the phase diagram). This is the case of the spherical perceptron in the convex region, $\sigma=0$, where jamming appear in the replica symmetric phase, and the polydisperse perceptron which is always hypostatic;
\item models that are isostatic but for which jamming happens to be in a replica symmetric phase. This is the case of the spherical perceptron model at $\sigma=0$. 
\end{itemize}
For all these models one finds $\kappa=1$ and $P_f(f)$ and $g(h)$ are found to be regular at the transition ($\theta=\gamma=0$).

The second universality class is instead found for isostatic models displaying replica symmetry breaking at the transition. This is the case of the spherical perceptron problem at $\sigma<0$.
In this case $\kappa=1.41574\ldots $, $\theta=0.42311\ldots $ and $\gamma=0.41269\ldots $ and these three exponents are found to coincide with the ones describing jamming of spheres in infinite dimensions.

Finally, for what concerns the Canyon model of confluent tissues, this is by construction hypostatic at jamming since one finds a satisfiability transition only for $\alpha<1$. Coherently, one finds that $\kappa=1$ as for the other hypostatic models \cite{urbani2023continuous}. It is clear that in this model forces and positive gaps are not defined given that the model is defined out of equality constraints.
 
\vspace{0.3cm}
{\bf The UNSAT phase}\\
In the UNSAT phase, an arrangement of the degrees of freedom satisfying all constraints cannot be found with high probability in the large $N$ limit.
This implies that the GB measure at zero temperature is peaked on the ground state of the system which is found at positive energy density. 
Therefore, it this case, the solution of Eqs.~\eqref{RSB_eqs_generic} describe the properties of this ground state.
Also in the UNSAT phase we can distinguish two situations depending on whether the solution is replica symmetric or not.

In the RS-UNSAT phase the system has a unique ground state. This implies that $q_m=q_M=1-\chi T$ with $\chi>0$ a linear susceptibility
which depends on the control parameters of the models. However, $\chi$ must diverge if we approach the satisfiability point
from the UNSAT side of the transition. Indeed, crossing this point, we should have $q_M<1$ for $T\to 0$.

This simple picture changes when the UNSAT phase is RSB.
In this case one can show that the RSB equations develop a scaling regime where one can define  a scaling function $y(q)=\beta x(q)$
which describes the organization of pure states on the ultrametric tree close to the ground state.
The properties of this scaling function depend on the nature of the ground state.
For example, an important difference is found between the scaling solution of the harmonic perceptron problem in the RSB-UNSAT phase \cite{franz2017universality}
and the corresponding one of the linear perceptron problem \cite{franz2019critical}.
The hallmark of this different behavior is the behavior of $q_M$ as a function of the temperature when $T\to 0$. In the case of the harmonic perceptron
one finds that $q_M\simeq 1-\chi T$, \cite{franz2017universality} while in the linear perceptron case one finds $q_M\simeq 1-\chi T^{\kappa}$ with an anomalous exponent $\kappa$ which turns out to be the same as the one controlling the 
behavior of $q_M$ close to jamming coming from the SAT side of the transition. The reason behind this behavior is that local minima and the ground state of the linear perceptron in the UNSAT-RSB phase are isostatic. 

Therefore whenever isostaticity appears with a non-convex energy landscape, it leads to a well defined universality class controlled by the critical exponent of the jamming transition.

\subsection{Extensions}
In this section  we want to briefly mention  some extension of the models already presented.
\begin{itemize}
\item {\it Multilayer perceptron.} In \cite{franz2019jamming} a different class of CCSP has been considered. This class corresponds to modeling the gap variables via more complex non-linear functions which are inherited from multilayer perceptron architectures \cite{engel2001statistical}. 
The corresponding CCSP is again non-convex and for the class of models considered in \cite{franz2019jamming} it has been found that at the satisfiability transition point they display the non-convex isostatic universality class
which give rise to the same jamming critical exponents of hard spheres.
\item{\it Piecewise linear perceptron.} A generalization of the linear perceptron has been considered in  \cite{sclocchi2021proliferation}.
This is defined by a different cost function 
\beq
v(h)=\begin{cases}
-2h - H_0 & h<-H_0\\
-h & -H_0<h<0\:.
\end{cases}
\eeq
Local minima of the Hamiltonian defined out of $v(h)$ have the properties that some gaps are found to be either zero ($h_\mu=0$) or are exactly equal to $-H_0$ ($h_\mu=-H_0$).
We call $C_0$ the total number of vanishing gaps and $C_-$ the total number of gaps equal to $-H_0$. It turns out that these quantities fluctuate across different local minima.
However in the RSB-UNSAT phase the sum $C_0+C_-$ equals to $N$ and this is a new version of isostaticity\footnote{More precisely, it is found that $C_0+C_-=N-1$ given that the total number of degrees of freedom is $N-1$ because the vector $\underline x$ is constrained to be on the sphere defined by $|\underline x|^2=N$.} Correspondingly one has a more complex pattern in the force and gap probability distributions but they are still controlled by the same critical exponents found at the jamming point of hard spheres.
\item{\it Optimization over non-convex excluded volume constraints}. Consider the Hamiltonian
\beq
H=\frac 12 \sum_{i=1}^N (x_i-1)^2\:.
\eeq
We would like to study its ground state when $\underline x$ is constrained to be a solution of the spherical perceptron constraint satisfaction problem. Therefore we assume that we are in a region in the $(\alpha,\sigma)$ phase diagram where such solutions exist with probability one (SAT phase) and we would like to find the one that minimizes $H$. 
This problem has been considered in \cite{sclocchi2022high} where it has been shown that the ground state of $H$ lies always at the boundary of the space of solutions of the perceptron CCSP. Moreover the ground states undergoes a RSB transition in a region where the phase space of the CCSP is described by a replica symmetric solution. At the RSB transition point of the ground state of $H$ it is found that it lies on an isostatic boundary of the solution space of the perceptron CCSP, namely the number of constraints of the CCSP that are marginally satisfied is equal to the number of degrees of freedom. A careful study of the RSB phase has not been performed but it has been conjectured that there the ground state is described by the same universality class of the jamming transition of hard spheres.
\end{itemize}
All in all, these extensions shows that the isostatic and hypostatic universality class are broad and one can organize different models according to their behavior at the jamming point.

\section{Scaling theory of the isostatic criticality} \label{Sec_isostaticity}\label{Sec_iso_criticality}
In this section we would like to discuss briefly the way in which the critical exponents describing the jamming transition of hard spheres emerge in the context of the satisfiability transition point of the spherical perceptron. 
The discussion will be coincise but further details can be found in \cite{franz2017universality, urbani2018statistical}.

Since we focus on the spherical perceptron problem we can consider the Eqs.~\eqref{RSB_eqs_generic}. 
We study the approach to jamming from the SAT side of the transition. Therefore we can safely take the $T\to 0$ limit.
In this limit the initial condition for the equation for $f(q,h)$ becomes
\beq
\begin{split}
f(q_M,h) &= \log \gamma_{1-q_M}\star \theta(h) = \log \Theta\left[\frac{h}{\sqrt{2(1-q_M)}}\right]\\
\Theta(x) &= \frac 12 \left(1+{\rm Erf}(x)\right)
\end{split}
\eeq
The $\Theta$ function is close to a sigmoid fucntion. When $q_M\to 1$ it is clear that $\Theta(x)$ approaches $\theta(x)$.
This implies that $f(q_M,h)$ develops a non-analytic behavior for $h\simeq 0$. This non-analyticity is smoothed out for all values of $q_M<1$.
Therefore in order to study the jamming limit in which $q_M\to 1$ one needs to understand how the non-analyticity is smoothed out for small but finite values of $1-q_M$.
In order to do that we assume that $1-q_M\sim \epsilon^\kappa$ being $\epsilon$ the linear distance in control parameter space from the satisfiability transition point.
We need to understand the scaling of $x(q)$ and $P$ and $f$ with $\epsilon$.
Therefore we define the following scaling functions
\beq
\begin{split}
y(q) &= \epsilon^{-1}x(q)\\
\hat \lambda(q) &= \epsilon^{-1}\lambda(q)
\end{split}
\eeq
Given that $1-q_M\sim \epsilon^\kappa$, we assume that
\beq
y(q) \sim (1-q)^{-1/\kappa} \ \ \ \ \ q\to 1
\eeq
We now need to study the behavior of $f(q,h)$ and $P(q,h)$.
First of all it is useful to introduce
\beq
m(q,h) = \lambda(q)f(q,h)
\eeq
which obeys the following PDE
\beq
\begin{cases}
\dot m(q,h) &=-\frac 12 m''(q,h)-\frac{y(q)}{\hat \lambda(q)}m(q,h)\left[1+m'(q,h)\right]\\
m(q_M,h)&=-h\theta(-h)
\end{cases}
\eeq
and we have denoted with primes the derivatives with respect to $h$ and with dots the ones with respect to $q$.
We would like to construct a scaling ansatz for $m$ when $q\to 1$. If we look at the regime where $1-q\gg \epsilon^\kappa$
we first have that
\beq
\frac{y(q)}{\hat \l(q)} \simeq \frac{k-1}{k}\frac{1}{1-q}\:.
\eeq
It is then easy to conjecture the following scaling form
\beq
m(q,h) = -\sqrt{1-q}\ \MM\left(\frac{h}{\sqrt{1-q}}\right)\ \ \ \ \ \ \MM(t\to \infty)=0\ \ \ \ \ \ \ \MM(t\to -\infty)\simeq t 
\eeq
with a scaling function $\MM$ which satisfies the following differential equation
\beq
\MM(t)-t\MM'(t) = \MM''(t)+2 \frac{\k-1}{\k}\MM(t) \left(1-\MM'(t)\right)\:.
\eeq
The solution of this equation depends on the exponent $\kappa$.
In order to determine it we need to consider the equation for $P$. 
It can be shown that a good scaling ansatz for this function in the $q\to 1$ limit, comparable with its initial conditions, is
\beq
P(q,h) =\left\{
\begin{array}{lcl}
(1-q)^{(1-\k)/\k}P_-(h(1-q)^{(1-\k)/\k}) & h\sim -(1-q)^{-(1-\k)/\k}\\
(1-q)^{-a/\k}P_0(h(1-q)^{-1/2}) & |h|\sim \sqrt{1-q}\\
P_+(h) & h\gg \sqrt{1-q}
\end{array}
\right.
\label{scaling_P}
\eeq
where we had to introduce three scaling regimes and an additional critical exponent $a$.
The matching conditions between the three scaling regimes are controlled by the matching asymptotics
\beq
\begin{cases}
P_-(t) \sim |t|^{\theta} & t\to 0^-\\
P_0(t) \sim |t|^{\theta} & t\to -\infty
\end{cases}
\ \ \ \ \ \ \ \ 
\begin{cases}
P_0(t) \sim t^{-\g} & t\to \infty\\
P_+(t) \sim t^{-\g} & t\to 0^+
\end{cases}
\eeq
with the following scaling relations
\beq
\th=\left(2(1-\k+a)\right)/(\k-2)\ \ \ \ \ \ \ \ \g=\frac{2a}{\k}\:.
\eeq
While the scaling ansatz for $P$ is made out of three regimes, one can show, see \cite{charbonneau2014exact}, that plugging it inside the equation for $P$ 
can give access to only a scaling equation for the intermediate scaling regime $P_0$: 
\beq
\begin{split}
&\frac a\k P_0(t) +\frac 12 t P_0'(t) = \frac 12 P''_0(t) +\frac{k-1}{k} \left(P_0(t)\MM(t)\right)'\\
&P_0(t\to -\infty) \sim |t|^{\left(2(1-\k+a)\right)/(\k-2)}\ \ \ \ \ \ \ \ \ P_0(t\to \infty) \sim t^{-2a/\k}
\end{split}
\label{eqP0}
\eeq
The structure of this equation is quite interesting. 
The additional exponent $a$ appears as an eigenvalue for a linear equation. 
It is found that given $\k$ and a solution of the equation for $\MM$ only one solution for $a$ exists for which $P_0$ satisfy
the boundary conditions. The eigenvalue $a$ therefore is a function of $\k$ which is still undertermined.
To fix it we need to consider an equation that can be derived directly from the Eqs.~\eqref{RSB_eqs_generic}
\beq
x(q) = \frac{\l(q)}{2}\frac{\int_{-\infty}^\infty \de h P(q,h) \left(f'''(q,h)\right)^2}{\int_{-\infty}^\infty \de h P(q,h) \left(f''(q,h)\right)^2\left(1+\l(q) f'(q,h)\right)}\:.
\eeq
In the scaling limit $q\to 1$, this equation becomes
\beq
\frac{\k-1}{\k} = \frac 12 \frac{\int_{-\infty}^\infty \de t P_0(t) \left(\MM''(t)\right)^2}{\int_{-\infty}^\infty \de t P_0(t) \left(\MM'(t)\right)^2\left(1-\MM'(t)\right)}\:.
\eeq
This equation closes the scaling theory and gives a solution for $\kappa$ from which we can obtain $\theta$ and $\gamma$.
A detailed computation \cite{charbonneau2014exact} shows that indeed $\theta$ and $\gamma$ are the exponents controlling forces and gaps.
The scaling theory derived is exactly the same the one for infinite dimensional hard spheres at jamming \cite{charbonneau2014exact}.
The RSB-UNSAT phase of the linear perceptron gives rise to a richer scaling theory which can be reduced to the one presented above \cite{franz2019critical}.

Finally it is also useful to discuss how isostaticity appears. 
A direct consequence of the RSB equations is the following relation
\beq
\frac 1{\l^2(q)} = \a \int_{-\infty}^\infty\de h P(q,h) (f'(q,h))^2
\label{replicon_eq}
\eeq
This relation encodes the {\it marginal stability} of the RSB solution. Indeed one can show \cite{charbonneau2014exact} that the RSB saddle point solution, as view from the larger space of all $Q_{ab}$ matrix elements, is a stationary point of the replica action whose stability is controlled precisely by a quantity, the replicon eigenvalue, which can be written as the lhs minus the rhs of Eq.~\eqref{replicon_eq}. Therefore Eq.~\eqref{replicon_eq} implies that the saddle point is marginally stable. Furthermore, \eqref{replicon_eq} depends on $q$ which shows that there is a continuous of marginally stable stability eigenvalues.
Plugging the scaling ansatz of Eq.~\eqref{scaling_P}  in Eq.~\eqref{replicon_eq} one can easily show that
\beq
1=\alpha \int_{-\infty}^0\de t P_-(t)
\eeq
and since $P$ for $q\to 1$ is essentially the distribution of gaps (this is clear from the behavior of $P_+$), one gets that the total number of gaps that are (vanishingly) negative and that enter in the distribution of contact forces is equal to the number of degrees of freedom.
Therefore the marginal stability condition of the RSB solution is essentially equivalent to isostaticity. 

Finally the analysis presented in this section has been generalized to the linear perceptron case in \cite{franz2019critical} where a more complex scaling solution involving five different scaling regimes in the behavior of $P(q,h)$ has been presented. However due to emerging symmetries in the problem, the scaling theory of the linear perceptron case reduces to the one presented in this section leading therefore to the same critical exponents and isostaticity.

\subsection{Self-intermediate asymptotics of the second kind} \label{math_jamming}
We would like to report an analogy that to the best of my knowledge was not presented elsewhere.
The form of the scaling in Eq.~\eqref{scaling_P} involves a scaling regime for $|h|\sim \sqrt{1-q}$.
This scaling regime depends on an anomalous exponent $a$ which must be determined by solving a {\it non-linear eigenvalue problem}. Indeed the scaling Eq.~\eqref{eqP0} contains $a$ as a sort of eigenvalue: however it must be determined by a solution which satisfies the right boundary conditions which depend on $a$ themselves. Therefore $a$ is said to be a non-linear eigenvalue.
Without introducing $a$ in Eq.~\eqref{scaling_P} there is no scaling solution of the equation for $P$ in the scaling regime $h\sim \sqrt{1-q}$. Furthermore, the scaling form $P_-$ is non universal and cannot be fixed unless the full solution of the PDEs is found \cite{charbonneau2014exact}.

This situation has a deep parallel with asymptotic solutions of PDEs. Consider the classical example of the Barenblatt equation \cite{barenblatt1996scaling, goldenfeld2018lectures}. In its simplest version this is a diffusion-type equation for a function $u(t,x)$:
\beq
\begin{cases}
&\partial_t u(t,x) = - Dk\partial_x^2u(t,x)\\
&u(0,t) = \frac{1}{l} \phi(x/l)\:.
\end{cases}
\eeq
However the diffusion term $D$ is not constant as in standard diffusion equations but it depends on $u$ itself:
\beq
D =\begin{cases}
1+\epsilon & \partial_t u(t,x)<0\\
1 & \partial_t u(t,x)>0
\end{cases}
\eeq
If $\epsilon=0$ one recovers a standard diffusion equation whose asymptotic solution is a Gaussian profile with a scaling variable fixed by dimensional analysis $z=x/\sqrt t$. Furthermore, the scaling solution is {\it memoryless} in the sense that the initial condition of the equation enters only through the total mass $m=\int \de x u(0,x)$ which is conserved. In this case it is said that the asymptotic solution is of the first kind and essentially is fixed by dimensional analysis.

However as soon as $\epsilon\neq 0$, things change completely. A simple solution with the same scaling variable $z$ is not found in this case and one needs to assume that the asymptotic form of $u$ is
\beq
u(t,x)\sim t^{-\a -1/2} f\left(\frac{x}{\sqrt t}\right)\:.
\label{bare_II}
\eeq
The anomalous exponent $\a$ is fixed by a non-linear eigenvalue problem of the very same kind as the one fixing $a$.
The exponent $\alpha$ can be computed either numerically by solving the corresponding non-linear problem, together with the scaling function $f$. Alternatively it can be shown to admit an expansion in power series of $\epsilon$. The perturbative series can be computed via a renormalization group scheme and it can be shown that a resummation of the expantion leads to the same numerical value of $\a$ computed numerically. We do not discuss this point here, but the interested reader can look at \cite{goldenfeld1989intermediate, goldenfeld2018lectures}.

The scaling solution of the form in Eq.~\eqref{bare_II} is called a self-intermediate asymptotic solution of the second kind. In fact, many PDEs show this kind of behavior and probably the most celebrated one is the {Fisher-Kolmogorov-Petrovsky–Piskunov} equation with the famous velocity selection problem \cite{barenblatt1972self}.
In a nutshell, the scaling solution at jamming corresponds to a self-intermediate solution of the second kind of the RSB PDEs.

While this analysis does not add much to the previous one, a question is left. Indeed for the Barenblatt equation there is a clear parameter $\epsilon$ that can be used to control how much the scaling solution of the second kind is close to one of the first kind. In other words, $\epsilon$ coincides with the distance from the upper critical dimension in a renormalization group setting where the computations are done via dimensional regularization. 
It is unclear in the case of the jamming transition what is the corresponding parameter which can be used to transform the full non-perturbative solution of the scaling equations into a perturbative scheme. 

The renormalization group analogy discussed in this section is very evocative.
It suggests that scaling at jamming corresponds to having a situation in which the landscape of pure states becomes self-similar close to the jamming point and in jamming critical situations (linear perceptron and variants).

In other words, to the best of my knowledge, this is the first example where renormalization is performed in the space of pure states rather than in time and space as it is usually done in standard critical phenomena settings. 
This connects directly this research line with the one of the previous chapter. The problem of understanding RSB in finite dimensional systems is intimately linked to the quest for having a statistical mechanics framework where one can mix local spatio-temporal fluctuations with fluctuations and self-similarity of the organization of a complex pure state landscape.

\subsection{Avoiding isostatic criticality}
It is interesting to understand the source of the critical behavior discussed in the previous section by comparing it with the situations where isostaticity is not found.
In this regard we can consider the polydisperse perceptron problem which is RSB but not isostatic at jamming.
It is useful to look at the initial condition for the equation for $f(q_M,h)$. In this case, for $\beta\to \infty$ we have
\beq
f(q_M,h) = \log \gamma_{1-q_M+\Delta^2/\tilde k}\star \theta(h)\:.
\eeq
As far as $\Delta>0$, the singularity at $h=0$ present for $\Delta=0$ is smoothed out. It can be shown that this smoothing destroys the isostatic scaling solution and therefore $\Delta$ is a sort of relevant perturbation in the renormalization group analogy discussed above.
The crossover regime from $\Delta>0$ to $\Delta\to 0$ can be fully worked out and it has been done in \cite{ikeda2019mean}.
The same phenomenon is even more evident if one considers the case of the Canyon model for confluent tissues. 
In this case the initial condition for $f$ is a quadratic function because for $l(h)=h^2/2$ the convolution in Eq.~\eqref{RSB_eqs_generic} gives a Gaussian function whose logarithm is a parabola.
Therefore the singularity is smoothed out and one obtains a simple diffusive scaling. 
 
\newpage
\section{On the off-equilibrium dynamics problem}\label{off_equilibrium_dynamics}
The scaling solution presented in the previous section is heavily based on the replica theory of the jamming transition.
This means that it is supposed to work to describe the properties of the GB measure close to the 
thermodynamic satisfiability point.
Therefore there is no reason to believe that it should work beyond the thermodynamic formalism from which it descends
and in particular it may be totally irrelevant from the point of view of the algorithmic satisfiability transitions.
However it turns out that both isostaticity and the non-trivial critical exponents $\k, \ \theta$ and $\gamma$ are found, within the precision of numerical simulations,
to describe also the algorithmic version of the jamming transition.

The same scenario is also found away from jamming when considering linear spheres and the linear perceptron.
In this case the scaling theory is supposed to describe the ground state of the system\footnote{Or the endpoint of algorithms that are described by the same equations with the same form as \eqref{RSB_eqs_generic}. This is the case of the AMP algorithm when generalized to RSB, see \cite{montanari2021optimization, alaoui2020algorithmic}.}.
However numerical simulations are limited to local minima of the cost function which are anyway found, within the precision of numerical simulations,
to display the same jamming criticality described by the same exponents derived from the scaling theory.

This clearly poses the question on the universality of the scaling theory developed above.
There are two options: either these critical exponents can be derived in a different way with a totally different formalism unrelated to RSB, 
or the scaling theory derived above works also for the algorithmic satisfiability transition.
To clarify this important point there are three possibilities.

\vspace{0.3cm}
{\it Direct solution of the algorithmic jamming transition or minimization algorithms - }
A theory of Gradient Descent for harmonic soft spheres can be developed in the limit of infinite dimensions, see \cite{agoritsas2019out}.
The same can be done for the perceptron problem, see \cite{agoritsas2018out}.
In both cases, one can solve the dynamics through dynamical mean-field theory which allow in principle to get the properties of the algorithmic jamming point.
However this leads to a set of self consistent stochastic processes that are hard to integrate numerically, see the recent work \cite{manacorda2020numerical, manacorda2022gradient}.
This is a big limitation because a long time integration may be very useful to guess the scaling limit of the DMFT equations \cite{cugliandolo1993analytical}.
However while I believe that such a dynamical derivation of the exponents may be useful, it is also true that gradient descent is just one algorithm
and the question on the universality of the critical exponents is just shifted to other algorithmic strategies to get to jamming.
Moreover numerical simulations never rely entirely on purely gradient descent dynamics: a large number of techniques
is used to study packing at jamming via heuristic greedy optimization procedures, see for example \cite{charbonneau2015jamming, franz2021surfing}. 
Therefore while DMFT would provide an alternative to RSB to study the transition, I believe that the superuniversality (with respect to the algorithms)
of the critical exponents would not be anyway fully explained.

\vspace{0.3cm}
{\it Study of the energy landscape - }
If we consider the linear perceptron case, it is clear that the greedy minimization strategies lead to local minima of the energy landscape
that are unrelated in general to the ground state. One possibility to show that such local mimima share the same critical properties of the ground
state would be to analyze the landscape of the Hamiltonian away from the ground state.
There are several ways to do this, see \cite{ros2023high} for a recent review. 
However, crucially all methods known for the moment share two fundamental limitations: first they are limited to study typical (most numerous) minima at a given energy
level, and it is unclear that these are the local minima accessed by the algorithms. Most importantly, they have severe limitations when local minima are marginally stable
and this is precisely the situation of the local minima of the linear perceptron\footnote{However see \cite{kent2023count} for a recent proposal.}. 

\vspace{0.3cm}
{\it Stochastic stability - } Both solving the dynamics of minimization algorithms and analyzing the full landscape of the Hamiltonian is a complicated task.
One may be even more ambitious: indeed isostatic criticality is the same across various models and therefore it would be desirable
to have a theory which is totally independent of the model. This suggests to use renormalization group ideas to claim universality with respect to the underlying microscopic models but, unfortunately,
as far as I know, there is no field theory for jamming criticality and therefore it is unclear how to implement any renormalization group argument.

A possible way out would be change the perspective by postulating a physical principle which should hold for all local
minima accessible by local algorithms, and use it to construct a theory for the critical exponents. A conjecture along these lines has been proposed in \cite{franz2021surfing}.
One may first suppose that all the minima of the Hamiltonian that can attract local optimization algorithms should be marginally stable.
For jamming critical systems, marginal stability coincides with isostaticity but it is not sufficient alone to provide a recipe to
compute the critical exponents. 
To go beyond one may postulate that local minima are stochastically stable: small perturbations of the Hamiltonian change local minima without
affecting their critical properties. Marginal stability and stochastic stability are intrinsically encrypted in the RSB formalism but may hold also away from the ground state.
A crucial observation is that we do not expect that these physical principles should lead to the Eqs.~\eqref{RSB_eqs_generic}.
Rather, they may just lead to the scaling theory derived in the previous section: 
this means that out-of-equilibrium, while the Parisi ultrametric picture of pure states
loose relevance globally, still local accessible minima are organized according to a $y(q)$, for $q\to 1$ which satisfies the scaling theory derived above.
In other words one may interpret the scaling solution as a sort of theory of the {\it local} organization of nearby accessible local minima.
I believe that not only these ideas can be put in a solid mathematical construction, but there should be a framework where universality in complex systems can be studied more deeply in a model-independent way. I come back to this point at the very end of this manuscript.

\section{Perspectives}
In this chapter I have presented an overview of my research activity aiming at understanding the physics of the jamming and rigidity transition of several systems.
I have discussed the main salient features of these critical points in particle based models and I also showed how to study them by an abstract mapping to CCSP. 
These problems are mean-field in nature and allow the development of a detailed theory for the thermodynamic rigidity transition points.

The critical behavior at the jamming transition can be found in two broad universality classes. 
In the simplest one, jamming is hypostatic while the more interesting case corresponds to isostatic non-convex models which display anomalous dimensions controlling the microstructure of the configurations at the critical point.

The isostatic universality class is also found to be super-universal in two rather different ways.
First of all, numerically it is found that off-equilibrium (algorithmic) jamming points display the same critical behavior as the thermodynamic one. 
We have argued that this should be related to the stochastic stability of the organization of local minima of the energy landscape which leads to self-similarity.
However understanding better this point may be important.

The second form in which isostatic criticality is super-universal is that it is robust to finite dimensional fluctuations. Indeed, accurate numerical simulations on off-equilibrium finite dimensional frictionless hard spheres at jamming and linear spheres in the jammed phase \cite{franz2020critical, charbonneau2021finite} clearly show that the anomalous exponents found within the mean-field scaling theory describe accurately also finite dimensional systems. 
It is rather difficult to understand why this is so and little progress has been made in the last years.
The main bottleneck here is that the current description of the jamming transition is not done in terms of standard field theory arguments: the only exception to this, as far as I know, is the scaling theory of bulk critical exponents as presented in \cite{goodrich2016scaling, liu2023universal} but this approach has not been extended to capture the universal microstructure of jammed packings and it is unclear whether it works when these properties affects the scaling of bulk quantities close to jamming as it happens for the linear sphere case, see \cite{franz2021surfing}. 

The lack of a standard field theory approach for jamming connects again this research line with the one of the previous chapter.
From an intuitive point of view, my understanding of the robustness of the isostatic criticality to finite dimensional fluctuations can be traced back to isostaticity. 
Indeed it is found, see \cite{hexner2018two, hexner2019can}, that fluctuations away from isostaticity are suppressed at jamming.
This implies that relevant perturbations that destabilize packings are single contact breaking \cite{muller2015marginal}. 
It is clear that in this situation one has a sort of {\it ultraviolet-infrared coupling}: the (ultraviolet) microscopic structure determines the global (infrared) response. 
However this is a coincidence and one may expect that generic disordered systems, may be susceptible (marginally stable) only on much stronger perturbations: in other words, destabilizing the system could require the perturbation of a fraction of degrees of freedom controlled by some fractal dimension. For the case of spheres, this fractal dimension is zero because of isostaticity and this would imply that mean-field theory is correct.

Before concluding I would like to mention two other interesting research directions I have explored more recently.
As I previously mentioned in this chapter, a theory of simple search algorithms may be not sufficient to understand the jamming point. However it is also true that gradient based algorithms are at the core of open purpose optimization problems and having a better understanding of their properties is desirable. This will become evident in the next chapter where we will discuss the dynamics of learning. 
One important point which I believe is very interesting is what are the properties of the configurations reached by gradient based algorithms (think about gradient descent) in CCSP in the SAT phase. 
For example one may want to understand how typical configurations at zero energy are connected in phase space and what are their geometrical properties. 
This is especially relevant from the deep learning point of view, something that we will review again in the next chapters. 
Understanding gradient descent dynamics is typically hard and while a dynamical mean-field theory treatment is generally possible, it is also true that this leads to a set of equations which are rather hard to integrate.

However very recently it was shown that the Canyon model represents an interesting exception to this situation. In this particular case one can still solve the gradient descent dynamics via dynamical mean-field theory but the corresponding equations take a much simpler form which allows an efficient numerical integration \cite{10.21468/SciPostPhys.15.5.219}. I believe that this direction can be explored further and I am currently working on this to address the properties of the SAT landscape of the Canyon model as seen from gradient based algorithms. An interesting conjecture that came out in \cite{10.21468/SciPostPhys.15.5.219} is that for the particular form of the non-linearity of the constraints as described in Eq.~\eqref{gap_tissues}, the numerical integration of the DMFT equations suggests that gradient descent seems to have a satisfiability threshold which coincides with the thermodynamic one. If true, this would be a non-trivial statement because the model is non-linear. Further investigations are needed to understand better this point.

Finally I would like to mention that while from the point of view of physics of colloidal glasses, quantum fluctuations are essentially irrelevant (hard spheres are driven by entropic forces being thermal and quantum fluctuations very small), however quantum hard spheres are rather interesting from the conceptual point of view since they represent many body quantum billiards and represent ideal systems where one can study problems related to quantum chaos, ergodicity and its breaking. 
While quantum hard spheres can be studied in the infinite dimensional limit, see \cite{winer2023glass} for a very recent work on this subject, the line of reasoning presented in this chapter suggests to use simplified models. 

The quantum spherical perceptron was introduced in \cite{franz2019impact} and further studied in \cite{artiaco2021quantum}, see also \cite{baldassi2018efficiency} for a related study, to understand the effect of quantum fluctuations of the jamming transition. 
However these pioneering studies are limited by the fact that to include quantum fluctuations in the model one is led to an {\it impurity problem} in a particular dynamical mean-field theory formulation whose level of difficulty mirrors the one found in the problem of integrating numerically the DMFT equations for the gradient descent dynamics. 
 
In \cite{urbani2023quantum} I instead proposed to shift the problem from the perceptron to the Canyon model. Again in this case one can study much more in detail the quantum partition function of the model and can get a lot of information which is much harder to get in the case of the quantum spherical perceptron.
I believe that this research direction is very interesting and it would lead to simplified models for quantum billiards in infinite dimensions.
Moreover one of the reasons I find the quantum Canyon Model appealing is that it is a proxy for the quantum version of the finite dimensional model it comes from. Indeed the original model of Voronoi tessellations of confluent tissues is, in the end, a model of random dynamical geometry \cite{ambjorn1997quantum} where one can tune the local tension through the shape index. This can be used as a way to control fluctuations of a carefully defined metric \cite{grossman2022instabilities}. 
Therefore a quantization of the mean-field model would be a proxy to understand how a quantum model of random metrics behaves and therefore may connect naturally with high-energy physics.

Finally, in a rather far perspective I would like to mention that non-linear equations in high dimension arise in, possibly, uncountable situations. However one case that connects again with high-energy physics is the case of conformal field theories. These theories can be defined in terms of a set of degrees of freedom which are the conformal dimensions of operators and the structure constants defining the operator product expansion (OPE) \cite{poland2019conformal}. These quantities are constrained by crossing symmetry which essentially consists in writing four point correlators in two different but equivalent ways using the OPE. It turns out that these constraints are generally non-linear equalities in these degrees of freedom and it makes sense to understand the properties of the space of solutions  of the corresponding bootstrap equations. While this has not been achieved in general, it must be noted that such type of CCSP is high-dimensional given that number of degrees of freedom is formally infinite.
In infinite dimensions it may be possible that the real structure of solutions does not differ much (from a statistical point of view) from random models (unless if one focuses on {\it crystalline} solutions, a situation that happens with minimal models in 2D CFTs) and it may be interesting to study variations of the Canyon model closer to bootstrap constraints, something which is ongoing.
It is clear that given the non-linear nature of the constraints the space of solutions may be rather complicated. The analogy with the present chapter and its body of works suggests that this space may experience replica symmetry breaking.
It is also interesting to mention that this perspective resonates with a recent study where a tentative sampling algorithm has been developed \cite{laio2022monte} in a practical situation.
Finally if RSB is relevant one may ask if one can define some form of dynamical exploration on this space of solutions and/or if one can introduce a proper metric. These are very interesting questions and toy problems as the one considered in this chapter may turn to be useful.

\fontfamily{palatino}\fontseries{ppl}\fontsize{11}{11}\selectfont

\chapter{Statistical physics of learning algorithms}

In the last six years I developed a research line on problems related to high-dimensional inference and machine learning.
These problems have a long tradition in statistical physics of disordered systems and have a large overlap with a research activity pursued in mathematics, statistics and computer science.

I believe that it is fair to say that most of the research activity on these subjects, at least for what concerns the community
of people working at the crossroads between these problems and the statistical mechanics of disordered systems, has been focused on trying to understand 
the phase diagrams of high-dimensional inference and machine learning prototypical problems. 
From the algorithmic point of view most of the works have been devoted to elucidate the performance of message passing algorithms.
These algorithms have the peculiarity to be relatively simple to analyze and are directly related to the cavity method in statistical physics of disordered systems. 
A lot of research activity has been pushed to make rigorous statements on the performances of these algorithms in prototypical hard high-dimensional inference problems.

My main activity in this field has been to analyze a different class of algorithms, based on gradient descent dynamics and its variations.
These algorithms are used daily in machine learning and inference problems given that they are open purpose and can be easily adapted to many settings.
One of these algorithm is Stochastic Gradient Descent which is the workhorse algorithm to train artificial neural networks.

My main interest has been focused on developing theoretical frameworks where one can analyze in detail such class of algorithms
to try to benchmark their performances.
The content of this chapter summarizes this research activity.
As for the other chapters of this manuscript I limit myself to review the models and the frameworks I worked on leaving all the details
to the original papers while reporting only the  main conclusions of this research.

The chapter is organized as follows.
In Sec. \ref{sec_infe} I will review the physics of prototypical high-dimensional inference problems
and discuss the performances of gradient based algorithms as compared to message passing ones.
The main message of this section will be that the latter are typically better in terms of performances with respect to the former.

In Sec.\ref{sec_GD_percp} we will instead discuss the performances of gradient based algorithms in the context of supervised learning problems.
The main idea will be to focus on the stochastic gradient descent algorithm and to characterize its noise and its properties in terms of the exploration of the loss landscape.
We will be very much interested in benchmarking its performances as compared to gradient descent in prototypical hard high-dimensional
optimization problems.

\section{High-dimensional inference}\label{sec_infe}
A prototypical inference problem consists in reconstructing a vector from a set of noisy observations.
When it comes to low dimensions, this is a simple problem which can be attacked in various ways.
Essentially most of basic science is based on this problem.

In the last century, the possibility to sense high-dimensional vectors has become more and more widespread
and so has been the corresponding reconstruction problem.
These high-dimensional inference problems take the form of a particular class of disordered systems \cite{zdeborova2016statistical}
in prototypical settings.

To fix the ideas it is useful to describe a simple yet paradigmatic model.
Imagine to consider a vector $\underline x^*=\{x_1^*,\ldots, x_N^*\}$ which we call the signal.
The components of this vector are independent random variables that can take three values, $x_i\in\{-1,0,1\}$.
We assume that the density of non-zero components is parametrized by a control parameter which we call $\rho$
and that the signal itself is extracted from a probability distribution given by
\beq
\begin{split}
P_X(\underline x^*)&= \prod_{i=1}^N p_x(x_i^*)\\
p_x(x_i^*)&=(1-\rho)\delta(x_i^*)+\frac \rho 2 \left[\delta(x_i^*+1)+\delta(x_i^*-1)\right]\:.
\end{split}
\eeq

Given an instance of the signal, one can imagine a situation in which this is unknown but we have access to it only through noisy observations of the following kind
\beq
M_{ij} = \frac{x_i^*x_j^*}{\sqrt N} + \sqrt{\Delta}\xi_{ij}\ \ \ \ \ \ \ i< j
\label{spiked_matrix}
\eeq
where the variables $\xi_{ij}$ are random, Gaussian distributed, with zero mean and unit variance.
Therefore a natural inference problem consists in reconstructing the signal given the matrix $M$. 
We also assume that the statistical structure of the measurement process as in Eq.~\eqref{spiked_matrix} is known.

The inference problem just defined is called the {\it spiked matrix model}.
Indeed the matrix $M$ is essentially a Gaussian random matrix with a small perturbation provided by a projector on the signal.
It is interesting to observe that the part of the matrix $M$ which contains the information on the signal has a scaling proportional to $N^{-1/2}$.
Therefore, naively,  one may expect that if we take $N\to \infty$ the problem cannot be solved.
However in this limit one has also an infinite number of measurements.
These scalings conspire together and therefore one can expect that for $N\to \infty$ one has a well defined signal to noise ratio, which to a first approximation can be assume to be $1/\Delta$,
above which the reconstruction of the signal can be done up to some accuracy.
The goal of the next sections is to (i) understand and characterize the critical value of the signal to noise ratio above which the inference problem can be solved and with which accuracy and (ii) what are the performances of practical reconstruction algorithms.

\newpage
\subsection{The glassy phase of high-dimensional inference problems}\label{sec_glassy}
A way to study the spiked matrix model defined in the previous section is to assume that one has all the information on the measurement process except the signal itself.
This is called the Bayes optimal situation and, as the name suggests to do, a way to solve the reconstruction problem is to apply the Bayes formula.
The posterior probability over the signal $\underline x$ given the observation $M$ is given by 
\beq
P(\underline x|M) \propto P_X(\underline x) P(M|\underline x)
\label{Bayes}
\eeq
and we have neglected irrelevant normalization factors.
In Eq.~\eqref{Bayes} we have denoted by $P(M|\underline x)$ the probability of observing the matrix $M$ given the signal $\underline x$.
This probability is easily estimated given that the variables $\xi_{ij}$ are Gaussian.
We have
\beq
 P(M|\underline x) \propto \exp\left[-\frac{1}{2\Delta} \sum_{i< j} \left(M_{ij}-\frac{x_ix_j}{\sqrt N}\right)^2\right]\:.
\eeq
Using Eq.~\eqref{spiked_matrix} and Eq.~\eqref{Bayes} we arrive at the following measure over the signal
\beq
\begin{split}
P(\underline x|M)&\propto P_X(\underline x) \exp\left[-\frac 1{2\D} \sum_{i< j} \left(\frac{x_i^*x_j^*}{\sqrt N} + \xi_{ij} -\frac{x_ix_j}{\sqrt N} \right)^2\right]\:.
\end{split}
\eeq
This is a random measure since it depends on the realization of the noise matrix $\xi$. In other words, this matrix plays the role of quenched disorder 
in the model.
It can be shown that the best estimate of the signal can be given by considering
\beq
\hat x_i =\langle x_i \rangle
\label{thermo_recon}
\eeq
where the average is performed over the posterior measure in Eq.~\eqref{Bayes} \cite{zdeborova2016statistical}. 
In order to quantify the performances of the estimator in Eq.~\eqref{thermo_recon} 
it is useful to define the mean squared error (MSE) given by
\beq
MSE = \frac 1N \sum_{i=1}^N |x_i^*-\hat x_i|^2\:.
\eeq
This quantity can be also written as
\beq
MSE = \rho -2m +q_0
\eeq
where we have defined the magnetization and the overlap order parameters as
\beq
\begin{split}
m&= \frac 1N \sum_{i=1}^N \hat x_ix_i^*\\
q_0&=\frac 1N \sum_{i=1}^N \hat x_i^2\:.
\label{order_parameters}
\end{split}
\eeq
While the reconstructed signal $\hat {\underline x}$ can be defined theoretically, from an algorithmic point of view computing it requires
to sample the posterior probability distribution which is something that may be difficult to do depending on the properties of the measure itself.
In order to understand better this point it is important to characterize the posterior measure.

A thermodynamic study of this probability distribution can be done by defining the free energy as
\beq
\begin{split}
f(\Delta) &= -\frac 1N \overline{\ln Z}\\
Z=&\int \de \underline x P_X(\underline x) \exp\left[-\frac 1{2\D} \sum_{i< j} \left(\frac{x_i^*x_j^*}{\sqrt N} + \xi_{ij} -\frac{x_ix_j}{\sqrt N} \right)^2\right]
\end{split}
\eeq
and the overline here stands for the average over a typical realization of the observation matrix $M$ (therefore we are averaging over the random instances of the noise matrix $\xi$).
The free energy $f$ is a function of two control parameters that are $\rho$, namely the density of non-zero elements in the signal, and $\Delta$ which measures the strength of the noise in the observations.
At fixed $\rho$ we expect that the reconstruction problem is easy when $\Delta$ is small and hard when $\Delta$ is large.
We are interested in understanding more precisely this behavior and how it affects the free energy as a function of $\Delta$.
This problem has been analyzed in \cite{lesieur2017constrained, antenucci2019glassy} and here we summarize the main findings.

The free energy can be computed using the replica method to perform the average over the realization of the matrix $\xi$.
Using standard manipulations one can show that $f(\Delta)$ can be obtained through a variational principle
\beq
f(\Delta) = \min_{m,q_0} \phi_{RS}(\Delta, m,q_0)
\label{variational_fe}
\eeq
where $\phi_{RS}(m,q)$ is a replica symmetric estimation of $f$. This quantity depends 
on two order parameters whose physical interpretation coincides with the expressions in Eq.~\eqref{order_parameters}.
In particular the minimizers of Eq.~\eqref{variational_fe} can be shown to give the thermodynamic value of the magnetization and the overlap as defined in Eq.~\eqref{order_parameters}.

If we are in a situation in which we have full knowledge of the statistical properties of the noise $\xi$, we can show that the solution of the minimization problem
in Eq.~\eqref{variational_fe} is obtained for $m=q_0$ and this condition reflects a symmetry, which is typically called Nishimori symmetry (or equivalently Nishimori condition)
between the generation of the measurement matrix and the statistical properties of the estimation process itself. 
In general one may not have the full information of the generation process (for example the value of $\Delta$ may be unknown) and this clearly generates an asymmetry
between the posterior measure and the generation of the signal. It can be shown that when the Nishimori condition is met, there is no RSB in the free energy, \cite{nishimori2001statistical, zdeborova2016statistical}.

We now discuss the phase diagram as it comes out from the solution of the variational problem in Eq.~\eqref{variational_fe}.
The pattern of phase transitions as a function of $\Delta$ depends in general on $\rho$.
We will consider the richest situation in which the inference problem has an {\it hard phase} whose meaning will be clarified in the following.
Such hard phase appears for small values of $\rho$, see \cite{lesieur2017constrained} for more details.

For $\Delta>\Delta_{\rm d}$, the replica symmetric free energy $\phi_{RS}(m,q_0=m)$ has a unique minimizer at $m=0$.
This paramagnetic phase corresponds to a regime in which the noise is too large for reconstruction to be possible.
For $\Delta\in (\Delta_{\rm IT}, \Delta_{\rm d}]$, a secondary minimum in $\phi_{RS}$ appears at a high value of the magnetization $m>0$
and it describes configurations of phase space that are correlated with the signal $\underline x^*$.
However this minimum is metastable with respect to the paramagnetic one $m=0$ and therefore if we estimate the signal via Eq.~\eqref{thermo_recon}
this is totally blind to this metastable minimum in $\phi_{RS}$.
Therefore the phase $\Delta>\Delta_{IT}$ corresponds to the region where the reconstruction of the signal is information theoretically impossible.

The situation changes at $\Delta=\Delta_{IT}$. At this point there is a first order phase transition where the free energy $\phi_{RS}$ gets a global minimum for $m>0$.
This implies that the reconstruction of the signal is information theoretically possible and therefore for $\Delta<\Delta_{\rm IT}$
the posterior measure is dominated by phase space configurations that are highly correlated with the signal itself.
We denote by $m_{\rm sig}$ the magnetization the global minimum of $\phi$ for $\Delta<\Delta_{\rm IT}$.
It is important to note that for $\Delta<\Delta_{\rm IT}$ and in particular for $\Delta \in (\Delta^*,\Delta_{\rm IT})$, $\phi_{RS}$ has still a spurious metastable
minimum at $m=0$. This minimum undergoes a second order transition at $\Delta^*$ where the corresponding position shifts to $m>0$ albeit $m<m_{\rm sig}$.
Finally for $\Delta<\Delta_{\rm AMP}$, this metastable minimum disappears and $\phi_{RS}$ has a unique minimum highly correlated with the signal.

As far as the posterior measure is concerned, this picture is sufficient to establish when it is possible to reconstruct theoretically the signal which happens at $\Delta_{\rm IT}$.
However in order to perform the reconstruction task one needs to sample the posterior measure to compute the averages in Eq.~\eqref{thermo_recon}.
The analysis above says nothing about the existence of polynomial time algorithm capable to perform such estimation.

It turns out in fact that it is an open problem to establish whether there exist polynomial time algorithms capable to sample configurations
highly correlated with the signal up to $\Delta_{\rm IT}$.
The best known algorithm in this regard is the Approximate Message Passing (AMP) algorithm \cite{donoho2009message} which is an iterative scheme to compute $\hat x_i$ avoiding the direct sampling of the posterior measure.
It can be shown that the performances of this algorithm are exactly tracked by the evolution of the structure of $\phi_{RS}$.
In particular this algorithm is able to output a configuration highly correlated with the signal only for $\Delta<\Delta_{\rm AMP}$. This range in $\Delta$ is therefore said to be the AMP-easy phase.
For $\Delta\in (\Delta_{\rm AMP}, \Delta^*)$ the AMP algorithm provides a configuration which is weakly correlated with the signal and 
for $\Delta<\Delta^*$ it cannot provide any correlation with the signal itself.
In other words the AMP algorithm is always attracted by the less informative spurious minimum of $\phi_{RS}$ whenever this exists.
The phase where $\Delta\in(\Delta_{\rm AMP}, \Delta_{\rm IT})$ is said to be AMP-hard because while the signal could be reconstructed from an information theory viewpoint,
this algorithm fails at doing so. 

It turns out that while AMP is an extremely efficient algorithm from the computational viewpoint and it can be analyzed in detail from the mathematical perspective, 
it is of limited applicability.
Indeed it is guaranteed to converge only when the inference problem has the precise structure as the one designed above.
In practical settings, the noise matrix $\xi$ may have unknown statistical properties and the Gaussianity assumption may be not relevant.
Therefore it makes sense to consider more open purpose algorithms to estimate the signal via the posterior measure. A natural one would be a Monte Carlo Markov Chain to sample the posterior
and to compute the averages in Eq.~\eqref{thermo_recon}. While analyzing such algorithm is in general hard to do, glass physics teaches us that
a proxy to understand its performances is the complexity of the structure of the free energy landscape it is supposed to sample.

This perspective was considered in \cite{antenucci2019glassy} where an analysis of the free energy landscape of the posterior measure was performed.
First of all we note that that $P(\underline x|M)$ is a high-dimensional random probability measure which depends on $\xi$. Given that, it may happen that this  measure develops metastable states poorly correlated with the signal which can trap sampling algorithms.
In order to establish whether this can happen one needs to asses the existence of these spurious metastable states.

The landscape analysis of the posterior measure can be done via a method developed by Monasson \cite{monasson1995structural} and a detailed computation applied to the 
the inference problem described by Eq.~\eqref{spiked_matrix} has been presented in \cite{antenucci2019glassy}.
The results of this study shows that the hard phase of the AMP algorithm is actually glassy in the sense that there are exponentially many in $N$ spurious
metastable states in the posterior measure that are uncorrelated with the signal.
Crucially, glassiness extends in the AMP-easy region, $\Delta<\Delta_{\rm AMP}$ and therefore this suggests a conjecture 
according to which sampling algorithms are suboptimal with respect to AMP in the sense that the former are able to reconstruct the signal at noise level $\Delta$
which is strictly smaller than $\Delta_{\rm AMP}$.

In order to investigate this conjecture one needs to construct a framework in which sampling algorithms can be analyzed explicitly
and this will be reviewed in the next section.

Before concluding, it is important to underline that the structure of the phase diagram outlined for the spiked matrix model considered in this section is not universal.
In particular, the first order behavior of the information theoretic transition can become second order (this happens, for example, for the problem analyzed here at sufficiently large values of $\rho$, see \cite{lesieur2017constrained} and it leads to the absence of an hard phase). The pattern of possible phase transitions can be analyzed extensively by looking at the properties of $\phi_{RS}$ and a careful classification has been presented in \cite{ricci2019typology} using Landau-type arguments.

\subsection{Suboptimality of gradient based algorithm with respect to Approximate Message Passing}\label{Mixed_MT_model}
In the previous section we have pointed out that the posterior measure of hard high-dimensional inference problems can be glassy.
This suggests that sampling algorithms may slow down and fail to reconstruct the signal.
In order to test this possibility we need to construct a prototype high-dimensional hard inference problem where
the dynamics of such algorithms can be tracked explicitly to benchmark their performances.
This has been done in \cite{mannelli2020marvels} where  a different inference problem was introduced.
This is the so called mixed spiked matrix-tensor problem.

Consider an $N$-dimensional signal $\underline x^*$ extracted uniformly from the $N-1$ dimensional hypersphere fixed by the condition that $|\underline x^*|^2=N$.
We measure this signal via the following two quantities
\beq
\begin{split}
M_{ij} &=\frac{x_i^* x_j^*}{\sqrt N} +\sqrt{\D_2}\ \xi_{ij}^{(2)} \ \ \ \ \ \ i< j\\
T_{i_1\ldots i_p} &=\frac{\sqrt{(p-1)!}}{N^{(p-1)/2}}x_i^*\ldots i_p^* +\sqrt{\D_p}\ \xi_{i_1\ldots i_p}^{(p)} \ \ \ \ \ \ i_1< i_2<\ldots< i_p\:.
\end{split}
\eeq
The matrix $M$ and the $p$-tensor $T$ are noisy observations of the signal given that the matrix $\xi^{(2)} $ and tensor $\xi^{(p)}$ 
have Gaussian entries with zero mean and unit variance.
The two control parameters $\Delta_2$ and $\Delta_p$ set the level of the noise in the observations.
We will assume that $p\geq 3$.

Given the matrix $M$ and the tensor $T$ the goal of the inference problem is to reconstruct the signal.
This problem is clearly a variant of the one considered in the previous section.
However the present setting has a simplification related to the fact that the components of the signal are continuous variables
and, as we will see, this will allow to define simple sampling algorithms.
As in the previous section, we can set up a Bayesian framework where we can write a posterior measure over the signal as
\beq
P(\underline x|M, T) \propto P_X(\underline x) P(M, T|\underline x)
\eeq
and now the prior $P_X$ is nothing but a flat measure over the hypersphere defined by $|\underline x|^2=N$.

The posterior measure can be analyzed in the same way as it has been done in the previous section and one can write a phase diagram in terms of two
control parameters: $\Delta_2$ and $\Delta_p$.
We will focus on a regime in which the model has an hard phase in the same way as defined previously.
This happens if $\Delta_p<1$ and therefore we will fix $\Delta_p<1$ and change $\Delta_2$.
The form of the phase diagram of the model can be studied by writing a replica symmetric free energy $\phi_{RS}$ 
and following its local and global minima as a function of $\Delta_2$.
For $\Delta_2>\Delta_{\rm d}$ the posterior measure is paramagnetic and the free energy has a global minimum at $m=0$, totally uncorrelated with the signal.
Here, as in the previous section, the magnetization is defined as in Eq.~\eqref{order_parameters}.
For $\Delta_2<\Delta_{\rm d}$ a local minimum at $m>0$ appears and becomes the global one at $\Delta_2=\Delta_{\rm IT}$.
This point represents the information theoretic transition.
However the paramagnetic minimum survives up to a critical point\footnote{Differently from the problem defined in Sec.\ref{sec_glassy} the paramagnetic minimum is always found at $m=0$.} $\Delta_{2}=\Delta_{\rm AMP} =1$.
As in the previous section, the paramagnetic minimum, when it exists, is the dominant attractor of the AMP algorithm
and therefore, also for this problem, this algorithm has an hard phase.
The structure of the free energy landscape of this problem is still glassy and therefore one can conjecture that glassiness may hamper 
the performance of sampling algorithms.

To study this point better we first write more explicitly the posterior measure.
This takes the form of a GB measure
\beq
P(\underline x|M, T)\propto e^{-\HH_{M,T}[\underline x]}
\eeq
where the Hamiltonian is given by
\beq
\HH_{M,T}[\underline x] = -\frac 1{\Delta_2\sqrt N}\sum_{i<j}M_{ij}x_ix_j-\frac{\sqrt{(p-1)!}}{\Delta_p N^{(p-1)/2}}\sum_{i_1<\ldots<i_p}T_{i_1\ldots i_p}x_{i_1}\ldots x_{i_p}\:.
\label{Hpspin}
\eeq
This suggests to consider the following Langevin dynamics as a sampling algorithm
\beq
\dot x_i(t) = -\mu(t)x_i(t) -\frac{\partial \HH_{M,T}}{x_i} + \eta_i(t) \ \ \ \ \ i=1,\ldots, N
\label{Langevin_algo}
\eeq
where the thermal noise $\underline \eta$ is Gaussian with
\beq
\begin{split}
&\langle \eta_i(t)\rangle =0 \ \ \ \ \ \forall i\\
&\langle \eta_i (t)\eta_j(t')\rangle = 2\sigma\delta_{ij}\delta(t-t')  \:.
\end{split}
\label{noise_lange}
\eeq
The Lagrange multiplier $\mu(t)$ appearing in Eq.~\eqref{Langevin_algo} is used to project the Langevin dynamics so that at each time step one has $|\underline x(t)|^2=N$.
We will assume that the initial condition of this algorithm is sampled uniformly with the prior measure.
It can be shown that the stationary measure of the Langevin algorithm for $\sigma=1$ is precisely the posterior measure over the signal
and therefore it can be used to sample configurations from this measure so that one can compute the averages in Eq.~\eqref{thermo_recon}. In the following we will keep $\sigma$ in the equations since it will be useful for the rest of the chapter.

While the Langevin algorithm can sample the posterior, the relaxation time to this stationary measure may depend on the control parameters of the problem
Therefore the performances of this algorithm are controlled by its relaxation time to the posterior measure.
It may happen that infinitely lived glassy states could trap the dynamics on finite timescales\footnote{One can show that if we wait an exponential timescale in the system size $N$ we are sure that this algorithm is able to converge to its stationary equilibrium measure. However this statement is of no practical interest when $N$ is large.} hampering therefore the performances of the Langevin dynamics. 
To understand whether this happens one should track exactly the corresponding dynamics.
This can be done explicitly on this problem through dynamical mean field theory (DMFT).

\begin{figure}
\centering
\includegraphics[width=0.475\columnwidth]{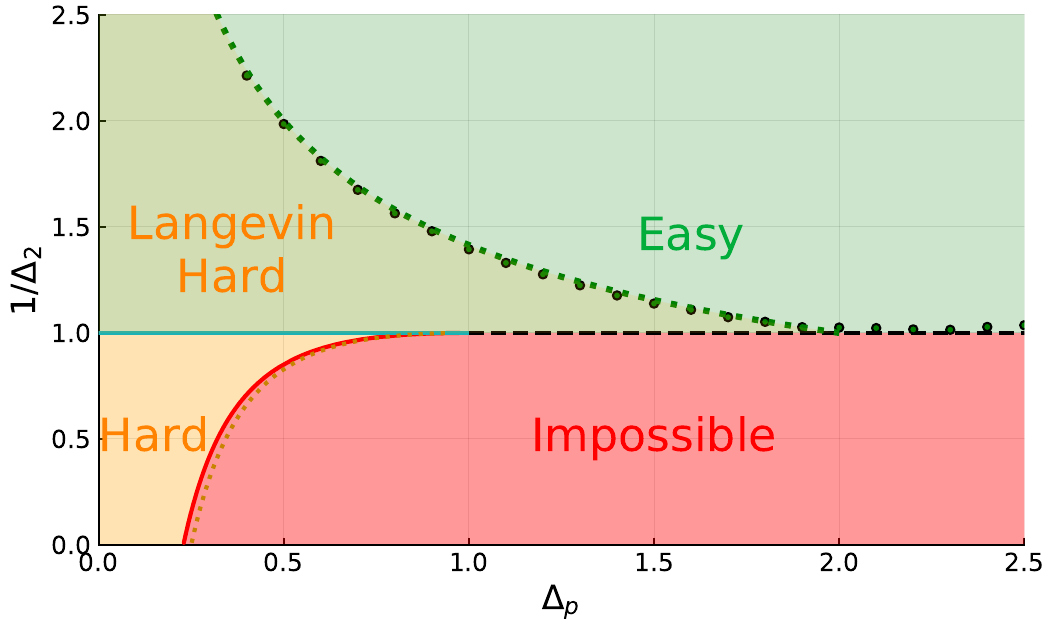}
\includegraphics[width=0.475\columnwidth]{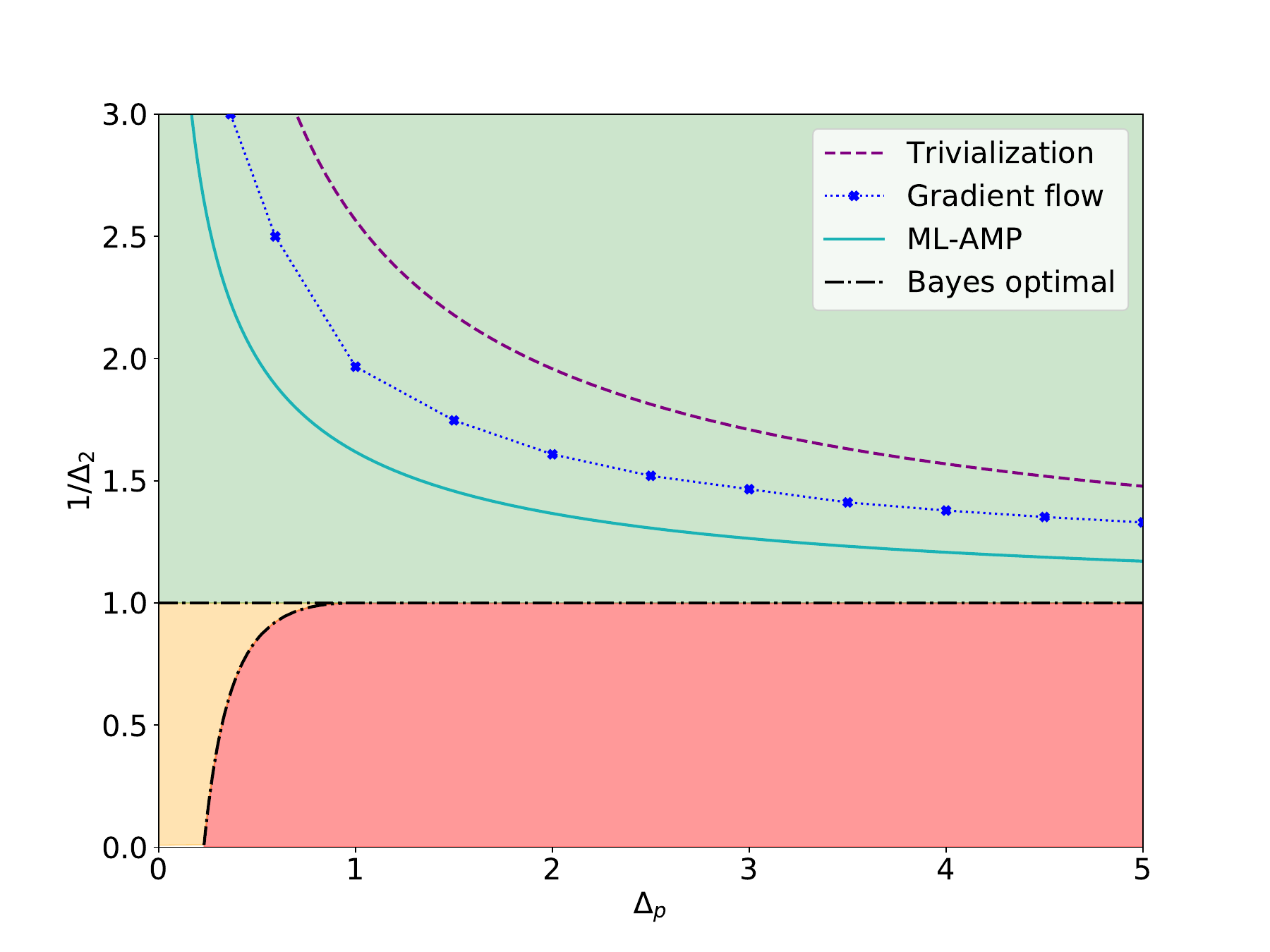}
\linespread{0.8}
\caption{\footnotesize{The phase diagram of the spiked matrix-tensor model. Left Panel (taken from \cite{mannelli2020marvels}): the phase diagram for the Langevin algorithm for $p=3$. The hard phase corresponds to the region where the AMP algorithm fails to reconstruct the signal. The olive region is instead the Langevin hard phase. The continuous red line corresponds to the information theoretic transition while the red dashed line is where an informative metastable minimum in $\phi_{RS}$ appears. Right Panel (taken from \cite{mannelli2019passed}): the phase diagram for the Gradient Descent algorithm for $p=3$. The green line represents the point above which the AMP algorithm starts finding the ground state of $\HH_{M,T}$. The dashed purple line represents instead the line above which the landscape of $\HH_{M,T}$ becomes topologically trivial in the sense that there is just a single minimum highly correlated with the signal. The blue points instead denote the line above which gradient descent is able to find a configuration highly correlated with the signal.   }}
\label{fig_AMP_langevin}
\end{figure}

The main idea is that the Langevin algorithm is tracked exactly in the $N\to \infty$ limit by a set of partial differential equations for a set of dynamical order parameters.
These quantities are dynamical correlations functions defined as
\beq
\begin{split}
m(t) &= \frac{1}{N}\sum_{i=1}^N  x_i^* \langle x_i(t)\rangle\\
C(t,t') &= \frac{1}{N}\sum_{i=1}^N  \langle x_i(t) x_i(t')\rangle
\end{split}
\label{mag_corr}
\eeq
and we have denoted with the brackets the average over the noise $\underline \eta$.
Note that the spherical constraint on the vector $\underline x(t)$ implies that $C(t,t)=1$ at all times.
In addition to the quantities defined in \eqref{mag_corr} we can introduce the response function $R(t,t')$ defined as
\beq
R(t,t') =\frac 1N \sum_{i=1}^N \left\langle \frac{\partial x_i(t)}{\partial \eta_i(t')}\right \rangle\:.
\label{resp}
\eeq
It can be easily shown that $R(t,t')$ has a causal structure in the sense that $R(t,t'\geq t)=0$.
The DMFT technique allows to construct a set of partial differential equations for the quantities defined in \eqref{mag_corr} and \eqref{resp}. 
Defining $Q(z) = z^2/(2\Delta_2)+z^p/(p\Delta_p)$, they are given by
\beq
\begin{split}
\partial_t m(t) &= -\mu(t) m(t) + Q'(m(t))+\int_0^t \de s R(t,s)Q''(C(t,s))m(s)\\
\partial_t C(t,t')&=-\mu(t) C(t,t')+Q'(m(t))m(t')+\int_0^t \de s R(t,s)Q''(C(t,s))C(s,t')\\
&+\int_0^{t'}\de s R(t',s)Q'(C(t,s))\\
\partial_t R(t,t')&= -\mu(t)R(t,t')+\int_{t'}^t\de s R(t,s)Q''(C(t,s))R(s,t')\\
\mu(t) &= \sigma+Q'(m(t))m(t)+\int_0^t \de s R(t,s)Q''(C(t,s))C(s,t)\\
&+\int_0^{t}\de s R(t,s)Q'(C(t,s))
\end{split}
\label{DMFT_pspin}
\eeq
and fixing $\sigma=1$ one obtains the DMFT equations for the Langevin algorithm.
These equations can be integrated numerically quite efficiently since they have a causal structure.
A crucial observation is that the equation for $m(t)$ admits always a solution $m(t)=0$ and therefore if we choose to initialize the dynamics
with $m(0)=0$ this will not change as time passes.
However if we initialize the dynamics with a random initial condition, we have that the initial magnetization will be of the order  $N^{-1/2}$.
Therefore the right way to integrate numerically Eqs.~\eqref{DMFT_pspin} is to assume a small initial magnetization and study numerically the limit in which this is sent to zero.
This can be done systematically, see \cite{mannelli2020marvels}.

Here we summarize the main findings of this analysis.
For $\Delta_2\ll 1$ one can expect that the Langevin algorithm will be able to sample the posterior measure on relatively short times.
This practically means that the magnetization $m(t)$ reaches a plateau value for $t\to \infty$ which coincides with the global minimum of the RS free energy discussed before.
However, increasing the noise strength, one can observe that the time it takes for the magnetization to reach the signal increases. 
We call $\tau(\Delta_2)$ this relaxation time. 
It is found that $\tau(\Delta_2)$ diverges when $\Delta_2$ reaches a critical value $\Delta_2^*$ above which the magnetization cannot increase on timescales
that are not exponential in the system size.
Most importantly, one finds that for $\Delta_p<1$, $\Delta_2^*<\Delta_{\rm AMP}$.
This confirms that the glassy phase of the posterior measure hampers the performance of the Langevin algorithm and this confirms the conjecture outlined in \cite{antenucci2019glassy}.
The phase diagram of the model is summarized in the left panel of Fig.~\ref{fig_AMP_langevin}.

\subsection{Gradient descent recovery threshold and topological complexity of the energy landscape}
An interesting modification of the Langevin algorithm is the one that can be obtained by setting $\sigma=0$ in Eqs.~\eqref{Langevin_algo},\eqref{noise_lange}.
In this case one essentially obtains an algorithm, Gradient Descent, trying to maximize the log-likelihood of the signal. Indeed the log-likelihood is nothing but $-\HH_{M,T}$ in Eq.~\eqref{Hpspin}.
The performances of this algorithm can be analyzed as well and the corresponding DMFT equations follow directly from Eqs.\eqref{DMFT_pspin} by setting $\sigma=0$.
These equations can be integrated numerically and a precise analysis has been presented in \cite{mannelli2019passed}.
In a nutshell one can still compute the relaxation time of gradient descent dynamics on the signal and extract the value of $\Delta_2$, call it $\Delta_2^{\rm GD}$ where this diverges.
A precise phase diagram can be found in the right panel of Fig.\ref{fig_AMP_langevin}.

The main results of this analysis are two and can be summarized as follows.
First, one observes the same behavior that was observed for the Langevin algorithm in the sense
that the AMP algorithm, that can be developed to minimize the Hamiltonian in Eq.~\eqref{Hpspin}, is able to give a correlated configuration with the signal
at noise levels $\Delta_2$ that are larger than Gradient Descent.
The second important result is that the Gradient Descent algorithm works also in the regime in which a (annealed) landscape analysis based on the Kac-Rice method, see \cite{ros2023high} for a review,
would predict that $\HH_{M,T}$ contains exponentially many spurious minima uncorrelated with the signal. Indeed the topology trivialization transition of the landscape of $\HH_{M,T}$ happens
at a value of $\Delta_2$ that is strictly smaller than $\Delta_2^*$.
Therefore this shows that having exponentially many spurious minima in the {\it loss} landscape is not sufficient to trap the dynamics of gradient descent.
A generalization of these results to the case in which Gradient Descent dynamics is augmented with inertial terms, a method called {\it momentum acceleration}, has been presented in \cite{sarao2021analytical}.

A mechanism to understand the location of the recovery transition for gradient descent in the mixed spiked matrix-tensor model has been proposed in \cite{sarao2019afraid} (see \cite{mannelli2020marvels} for the corresponding one in the case of the Langevin dynamics).
Below the recovery threshold, the attractors of the gradient descent dynamics which are uncorrelated with the signal are a particular class of local minima of the Hamiltonian. The recovery transition corresponds to the point where the Hessian of these minima develops an unstable direction towards the signal. 
While this mechanism is interesting, I believe that its powerfulness is limited to the present model for the following reasons. First, it assumes that the attractors of the dynamics become unstable towards the signal all at the same signal-to-noise ratio. Moreover it assumes that no other attractor of the dynamics exists other than the signal itself when these minima become unstable. These conditions are in general not verified and the present model is the (pathological) exception rather than the rule. A more realistic situation where these assumptions are violated, can be found in another inference problem, phase retrieval, which will be reviewed in the next section of this chapter, see also \cite{sarao2020complex}.

I think that a general theory for the recovery threshold of gradient descent is hard to construct because there are several indications that the asymptotic properties of gradient descent dynamics on high-dimensional rough landscapes cannot be predicted being completely dependent on arbitrary short time dynamics \cite{folena2020rethinking}. 
In other words the asymptotic behavior of GD is not stable upon perturbations of the initial transient dynamics.
However Gradient Descent (or Langevin) are never employed in artificial neural network applications where the main optimization algorithm
is stochastic gradient descent (SGD). 
As we will see, SGD introduces an out-of-equilibrium driven dynamics and this leads to a non-equilibrium stationary state which is rather different from the relaxational dynamics of gradient descent.

\section{Neural networks}\label{sec_GD_percp}
In the previous section we described how gradient based optimization/sampling algorithms
perform to solve high-dimensional inference problems and how they compare to other algorithms, in particular AMP.
The purpose of this section is to investigate the same kind of questions in the context of supervised learning problems.
The workhorse optimization algorithm in this case is stochastic gradient descent and a substantial part of my research activity
has been to study this algorithm in detail using simple model systems.

Consider a training dataset $\DD$ of $P$ points, each consisting of an $N$-dimensional real vector $\underline \xi^\mu$ and its associated label $y_\mu$. 
The index $\mu$ runs over the datapoints in the dataset, $\mu=1,\ldots,P$.
We assume that the vectors and the labels are correlated in some way through an unknown, possibly non-linear, rule which we would like to detect.
The goal of the supervised learning task is to estimate a predictor of the label $y$ given a vector $\underline \xi$ using the points given in the training dataset \cite{engel2001statistical}.

A way to do that is by considering a class of functions
\beq
y = f(\underline w, \underline \xi)
\label{neural_network}
\eeq
which take as an input an $N$-dimensional vector $\underline \xi$ and outputs the corresponding label $y$. 
These functions are parametrized by a set of weights $w_i$ collected in the weight vector $\underline w\in {\mathbb R}^D$. In general there is no connection between the dimension $N$, $P$ and $D$.
The goal of the training phase is to estimate the best weight vector $\underline w$ which would minimize the errors in predicting the label on unseen datapoints.
The structure of the functions $f$ in Eq.~\eqref{neural_network} is rather general and is typically non-linear both as a function of $\underline w$ and $\underline \xi^\mu$.
To fix the ideas one can imagine to have a multilayer neural network in which the input $\underline \xi$ goes through a set of linear and non-linear transformations operated by successive hidden units organized in layers \cite{goodfellow2016deep}.

To find the weights $\underline w$ it is useful to define a loss function, also called empirical risk, $H$ as 
\beq
H= \sum_{\mu\in \DD} \ell(y_\mu, f(\underline w,\underline \xi^\mu))\:.
\eeq
The sum on the lhs runs over the datapoints in the training dataset and the local cost function $\ell$ is designed to penalize configurations of the weights which provide a wrong label prediction on a datapoint.
A simple function that illustrates this idea is the square loss:
\beq
\ell \left(y_\mu, f(\underline w,\underline \xi^\mu)\right) = \frac 12 \left(y_\mu-f(\underline w,\underline \xi^\mu)\right)^2\:.
\label{loss_square_nn}
\eeq
It is important to stress the similarity between Eq.~\eqref{loss_square_nn} and the loss function considered in the case of the Canyon model of confluent tissues reviewed in the previous chapter.
Indeed, as in the Canyon model, the training problem can be seen as a CCSP with equality constraints. These are nothing but the input-output associations of the labels with the corresponding datapoints. The role of degrees of freedom of the CCSP is played by the weight vector $\underline w$ while each datapoint is providing a constraint through the function $f$ and the datapoint $\{\underline \x^\mu,y_\mu\}$. The main difference between this setting and the Canyon one is that in the latter, the function $f$ depends {\it linearly} on the disorder (the random matrices $J^\mu$ for example); instead in the current case $f$ is in general a {\it non-linear} function of the datapoint vectors $\underline \xi^\mu$. In other words, the gap variables in the Canyon model are Gaussian random functions while in this context, even if we assume that the vectors $\underline \xi^\mu$ are Gaussian, the resulting gap variable $h_\mu=y_\mu-f(\underline w,\underline \xi^\mu)$ will not be Gaussian due to the non-linearity of $f$ as a function of $\underline \xi^\mu$.

The minimization of $H$ can be done via simple algorithms such as Gradient Descent (GD). This can be written as
\beq
\underline{\dot w}  =- \frac{\partial H}{\partial \underline w} = -\sum_{\mu=1}^P\frac{\partial \ell(y_\mu, f(\underline w,\underline \xi^\mu))}{\partial \underline w}\:.
\label{GD_nn}
\eeq
In general this is a continuous time algorithm but it is typically run using an Euler discretization with a fixed time step $\de t$, also called the learning rate.
There are in general two bottlenecks to run GD. 
First, each single term in the sum on the right hand side of Eq.~\eqref{GD_nn} must be computed and this is typically a hard computational task given that the dimension of $\underline w$ is huge.
It turns out that in standard artificial neural networks which have a multilayer structure, the computation of the derivatives in Eq.~\eqref{GD_nn} can be done efficiently  
through the {\it backpropagation} algorithm. We will not review this in detail but the interested reader can find more information in \cite{goodfellow2016deep}.
The second bottleneck is that the derivative in Eq.~\eqref{GD_nn} has to be taken for each datapoint in the training set.
This complicates the algorithm given that the dimension of the latter, $P$, is huge too.
An empirical solution to this problem has been suggested in \cite{bottou2009curiously, bottou2012stochastic, lecun2002efficient,bengio2012practical} and leads to the Stochastic Gradient 
Descent (SGD) algorithm.
This amounts to replace Eq.~\eqref{GD_nn} with 
\beq
\underline w(t+\de t) = \underline w(t) - {\de t}\sum_{\mu \in  \BB_t} \frac{\partial \ell(y_\mu, f(\underline w,\underline \xi^\mu))}{\partial \underline w}\:.
\label{SGD_nn}
\eeq
The main difference between Eq.~\eqref{GD_nn} and Eq.~\eqref{SGD_nn} is that in the latter the computation of the gradient of the loss is approximate.
One considers at each time step a portion, also called a {\it minibatch}, of the training dataset and approximates the full gradient of the loss with a sum over the datapoints contained in the minibatch.
The minibatches change at each time step and a full exploration of the training dataset is called an {\it epoch} of training dynamics.
The datapoints in the minibatches are selected randomly and so is the way in which minibatches are presented along the dynamics. 
Therefore the SGD algorithm contains a sort of built-in noise which is rather uncommon in physics settings. 
This makes the stationary measure of the whole dynamics hardly understandable from a theoretical viewpoint.
However it has been shown in numerical simulations (and daily use) that in overparametrized neural networks where $P\ll D$, 
the algorithm converges to configurations of $\underline w$ at zero loss.
Therefore even if the gradient is computed in an approximate way, this does not hamper in general
the efficacy of the dynamics to reach the bottom of the loss landscape.
Understanding the success of SGD in optimizing artificial neural network is part of the big question on how to understand the deep learning technology and it has become central in theoretical machine learning research.

In the last few years several approaches have been trying to give better theoretical foundations to the SGD algorithm and to benchmark its performances.
Here I just summarize some research directions undertaken in this regard:
\begin{itemize}
\item {\it Online SGD.} An extreme case of the SGD algorithm can be considered if minibatches are composed by single datapoints. If one assumes that the dataset is huge and that datapoints
are seen only once along the dynamics, one obtains what is called the online SGD algorithm. It turns out that this algorithm can be studied in detail and a lot of theoretical works have been devoted to understand its performances \cite{saad1995line, saad1995exact, saad1997globally, saad2009line, coolen2000dynamics, coolen2000line, mei2018mean, veiga2022phase, ben2022high, misiakiewicz2023six}. However it is fair to say that this setting is rather far from practical situations: indeed there is no notion of training epoch and even the loss function becomes irrelevant since one can show that the algorithm is essentially performing a gradient descent on the {\it population loss} which corresponds to the case in which the sum over the samples in the dataset in Eq.~\eqref{loss_square_nn} is replaced by a true average over the statistical distribution of the dataset itself. 
\item {\it Sharpness and Levy flights.} An empirical research line has instead tried to characterize the properties of SGD in real settings by looking either at (i) the statistical properties of the configurations reached at long times or (ii) at the behavior of the noise and the minibatch gradient along the dynamics itself. In some cases it has been found that SGD drives the dynamics towards regions at zero loss which are {\it wide} in the sense that if one looks at the spectrum of the Hessian  of the loss, this is typically populated by many small eigenvalues \cite{feng2021inverse} and that wideness positively correlates with the performances of the network, see also \cite{baldassi2016unreasonable}. However this sharpness conjecture has been recently questioned by new experiments, see \cite{andriushchenko2023modern}. Furthermore it has been argued that the nature of the SGD noise may share some empirical similarities with Levy flights \cite{gurbuzbalaban2021heavy, simsekli2019tail}.
\item {\it Implicit regularization.} A popular research line has been trying to relate the properties of the noise and algorithm with special types of configurations of the loss landscape suggesting that training dynamics, gradient descent and SGD (mostly online) drive the weights towards regions which fulfill specific criteria that are not built-in in the algorithms. One particular idea is that the dynamics leads to configurations of the weights that have zero loss and at the same time minimize the length of the weight vector itself. This {\it implicit regularization} phenomenon is very appealing from the theoretical point of view, it correlates with good generalization properties and some evidences have been put forward in simple settings \cite{neyshabur2014search, neyshabur2017implicit, gunasekar2018implicit, soudry2018implicit, arora2019implicit}.
\item {\it Linear networks.} Finally, the full-batch gradient descent algorithm has been fully analyzed in deep linear networks \cite{saxe2013exact, advani2020high}. This has the disadvantage that linear networks have limited expressivity
but on the other hand one can understand in detail the effect of the depth of the architecture. 
This approach has been extremely instructive and it has been shown that many features of the dynamics in deep linear networks can capture the training dynamics in practical situations.
However the extension of this analysis to SGD has not been accomplished yet due to the fact that it relies on random matrix type  computations which do not go through easily when the  noise of SGD is present.
\end{itemize}
Despite all these approaches it is fair to say that a theory to understand how SGD explores non-convex complex loss landscapes is still to be developed.
My contribution to this field has been to analyze this problem through the lens of statistical physics in simple yet paradigmatic settings.
In the next sections I will describe how dynamical mean field theory (DMFT) can be employed to track the SGD dynamics exactly in simple models.
The main advantage of this approach to this problem is that it allows to get results that are exact in the infinite dimensional limit.
This permits to establish, for example, the beneficial effects of the SGD noise for optimization and it provides a very good starting point to
understand the properties of the configurations reached by this algorithm when optimizing non-convex loss functions.
However this approach has also some drawbacks the main one being that the DMFT can be developed only in simple enough settings.
In particular its main limitation is that it is hard to include in the analysis the depth of neural network architectures apart from, in principle, just a single hidden layer neural networks.

In the next two sections I will review the DMFT approach to SGD dynamics in the context of two supervised learning problems. 
In Sec.~\ref{classification_Section} I will discuss the case of a simple supervised classification task while in Sec.~\ref{SGD_inference} I will consider a high-dimensional optimization problem as a setting
to benchmark the performances of SGD against GD.

\subsection{Stochastic Gradient Descent and persistent-SGD}\label{classification_Section}
The simplest supervised learning problem is the one of classification.
Consider a training set made of $P$ points.
Each of them is composed by an $N$-dimensional vector $\underline \xi^\mu$ and its corresponding label, $y_\mu=\pm 1$.
We assume that these vectors form two Gaussian clouds of the form
\beq
\underline \xi^\mu=y_\mu\frac{\underline w^*}{N} +\sqrt \Delta \frac{\underline \eta^\mu}{\sqrt N}\:.
\eeq
The vectors $\pm \underline w^*$ denote the centers of the Gaussian clouds while the control parameter $\Delta$ sets
the width of the clouds. The vectors $\underline \eta^\mu$ are random with independent identically distributed entries extracted from a Gaussian distribution
with zero mean and unit variance.
The size of the training dataset is controlled by $P$ which is assumed to be proportional to $N$, $P=\alpha N$ with $\alpha$ another control parameter called the {\it sample complexity}.
We can assume that if $\Delta$ and $\alpha$ are small enough there is a regime where the two Gaussian clouds are sufficiently far apart and can be separated by an hyperplane.

The simplest predictor allowing to perform the classification task can be defined by a single layer neural network, namely a perceptron:
\beq
y = {\rm sign} \left(\underline w \cdot \underline \xi\right)\:.
\label{perce_nn}
\eeq
The purpose of the training task is to find the best estimate of the $N$-dimensional weight vector $\underline w$ which classifies correctly the dataset.
To do that one can define a loss function in the following way.
Exploiting the properties of the $\rm sign$ function, Eq.~\eqref{perce_nn} can be rewritten as
\beq
y \underline w \cdot \underline \xi>0\:.
\eeq
Using the training dataset, the training problem corresponds to find a solution for $\underline w$ which satisfies the following set of inequalities
\beq
y_\mu \underline w \cdot \underline \xi^\mu>0\ \ \ \ \ \ \forall \mu=1,\ldots, P\:.
\eeq
This is precisely the perceptron CCSP considered in the previous chapter, for $\sigma=0$.
The main difference between the present setting and the previous one is that here the datapoints are structured and there is a correlation between the vector $\underline \xi^\mu$
and its label $y_\mu$. This differs with what was done in the previous chapter where we considered unstructured Gaussian vectors (and therefore there was no label to be taken into account).
To solve the training problem one can define the loss function as
\beq
\begin{split}
H&=\sum_{\mu=1}^P v(h_\mu) \theta(-h_\mu)\\
h_\mu&=y_\mu \underline w\cdot \underline \xi^\mu\:.
\end{split}
\label{loss_nn_perc}
\eeq
One typically considers the squared hinge loss where $v(z)=z^2/2$ and this is analogous to the harmonic cost function considered in the previous chapter to model harmonic soft spheres.
Otherwise one can take the hinge loss $v(z)=|z|$ which corresponds to the linear cost function considered in the linear perceptron case to describe linear soft spheres.
It is clear that the predictor in Eq.~\eqref{perce_nn} is invariant under rescaling of the length of $\underline w$. Therefore we can either
consider the case in which we explicitly regularize the norm of this vector or we let it free to evolve along the training dynamics.
The explicit regularization can be either imposed by adding a penalty cost to the Hamiltonian proportional to the length of $\underline w$, the simplest form being a Ridge regularization term which amounts to add to $H$ a term of the form
\beq
\delta H= \frac \l 2 |\underline w|^2\:,
\eeq
otherwise one can consider the case in which the norm of $\underline w$ is fixed at all times, for example $|\underline w|^2=N$.

To minimize the loss in Eq.~\eqref{loss_nn_perc} one can first consider GD
\beq
\underline w(t+\de t)=\underline w(t) +\de t \left[-\lambda(t) \underline w(t) -\sum_{\mu=1}^P \frac{\partial v(h_\mu)}{\partial \underline w} \theta(-h_\mu)\right]\:.
\eeq
The regularization term $\lambda(t)$ can be either time dependent and self-consistently fixed to guarantee $|\underline w(t)|^2=N$ or it can be kept constant when the Ridge regularization trick is used.
We assume that the GD dynamics is initialized from a random initial condition in which the coordinates $\underline w(0)$ are sampled {\it i.i.d.} from a Gaussian distribution with zero mean and a fixed variance.
If the spherical constraint is employed then this variance has to be one; otherwise it can be chosen at will in the context of the Ridge regularization where the norm of $\underline w$ is free to change with time.

While understanding the performance of the GD algorithm is already a very interesting problem, we would like to gain insights on the SGD algorithm.
To study this case we need to define a set of random minibatches. In \cite{mignacco2020dynamical} it was proposed to consider a set of binary selection variables $s_\mu(t)=\{0,1\}$ and to define the SGD dynamics as
\beq
\underline w(t+\de t)=\underline w(t) +\de t \left[-\lambda(t) \underline w(t) -\frac 1b \sum_{\mu=1}^P s_\mu(t) \frac{\partial v(h_\mu)}{\partial \underline w} \theta(-h_\mu)\right]\:.
\label{SGD_percp}
\eeq
The idea behind this change is that the selection variables account for the datapoints that are taken into account in the computation of the approximate gradient.
The simplest choice is to extract them independently and randomly at each timestep as
\beq
s_\mu(t) =\begin{cases}
1 & \textrm{ with probability } b\\
0& \textrm{ with probability } 1-b
\end{cases}
\eeq
so that on average we have
\beq
\langle \sum_{\mu=1}^P s_\mu(t) \rangle = bP
\eeq
being $b$ a control parameter for setting the size of the minibatches. The limit $b\to 1$ corresponds to Gradient Descent.
Note that $b$ enters explicitly in Eq.~\eqref{SGD_percp} as a renormalization factor introduced to have a gradient which has a comparable strength for different values of $b$.
The SGD algorithm described in Eq.~\eqref{SGD_percp} defines a discrete time dynamics which does not admit a continuous time limit.
Indeed the idiosyncratic dynamics of the selection variables $s_\mu(t)$ does not allow for an easy definition of incremental changes for $\de t\to 0$.
To overcome this difficulty in \cite{mignacco2020dynamical} it was proposed to change the dynamics of the selection variables according to the following Markov jump process
\beq
\begin{split}
\textrm{Prob}(s_\mu(t+\de t)=1|s_\mu(t)=0) &= 1-\textrm{Prob}(s_\mu(t+\de t)=0|s_\mu(t)=0)=\frac{ \de t}{\tau}\\
\textrm{Prob}(s_\mu(t+\de t)=0|s_\mu(t)=1)&= 1- \textrm{Prob}(s_\mu(t+\de t)=1|s_\mu(t)=1) = \frac{1-b}{b\tau}\de t\:.
\end{split}
\eeq
This new dynamics for the selection variables has an additional parameter $\tau$ which tunes the persistence time spent by a pattern out of the minibatch.
Therefore the corresponding algorithm is called {\it persistent-SGD}.

It turns out that for both definition of SGD dynamics in Eq.~\eqref{SGD_percp} and the persistent version, one can develop a DMFT analysis\footnote{This is in general true unless the spherical constraint is imposed on $\underline w$, where the DMFT can be performed only for the persistent-SGD in the continuous time limit \cite{mignacco2022effective}. }.
This has been started in \cite{agoritsas2018out} in the context of the perceptron CCSP where the DMFT was developed to study the behavior of gradient based dynamics, and it was adapted in \cite{mignacco2020dynamical} to take into account the effect of SGD noise 
see also \cite{mignacco2022effective, mignacco2021stochasticity} for additional works.
The form of the DMFT equations differs a lot from the one describing the Langevin or GD dynamics in the mixed spiked matrix-tensor model analyzed in the first part of this chapter. 
In the present case the equations for the correlation and response functions are rather involved because their integration
requires the solution of a self-consistent stochastic process.
We will not detail neither the structure of the equations nor the way in which they are solved numerically but the interested reader can look at \cite{mignacco2020dynamical} for more details.
We will instead describe the results of this analysis.

In \cite{mignacco2020dynamical} it has been shown that SGD, in its plain and persistent versions, allows to reach zero loss when the two Gaussian clouds are sufficiently well separated.
The performances of these two algorithms differ from GD and the small batch size acts in general as an effective regularizer when the norm of the weight vector $\underline w$
is not constrained ($\lambda=0$).

In \cite{mignacco2022effective} the DMFT analysis has been used to characterize the effective noise of SGD. It has been shown that in the phase where the sample complexity $\alpha$ is sufficiently large and the two Gaussian clouds cannot be separated by an hyperplane (UNSAT phase), the loss function is bounded away from zero and the SGD dynamics lands on a non-equilibrium stationary state which can be characterized by a sort of effective temperature which depends on the learning rate, the batch size and the persistence time. 
This effective temperature can be measured in the steady state from the behavior of the correlation and response functions through the violation of the fluctuation-dissipation theorem, in a way which is reminiscent to what is done in glasses \cite{cugliandolo1997energy, cugliandolo2011effective}. However it is important to note that in the current situation, differently from glasses, there is no actual separation of timescales, the dynamics is not slow and the relaxation to a non-equilibrium steady state is fast. Therefore the effective temperature defined in this way is not a true temperature unless one is close to the gradient descent limit where it becomes trivially zero. The situation is closer to active matter systems \cite{loi2008effective} or driven disordered systems \cite{berthier2000two} rather than to glasses.
This picture changes when the sample complexity is sufficiently small and the two Gaussian clouds can be well separated.
In this case one can show that the SGD dynamics stops when it reaches zero loss so that it performs a sort of self-annealing. 
Trivially in this case the effective temperature, as measured from the correlation and response functions, is zero both for GD and SGD (standard or persistent). 
Therefore in order to characterize the noise of SGD other observables need to be constructed. 

A possibility studied in \cite{mignacco2022effective} is to consider the average distance between two instances of the SGD dynamics starting from the same point but undergoing two different realizations of the minibatch noise. Interestingly it is found that such measure of the strength of the noise is non-monotonic in the batch size. However it is currently unclear why this is so and this finding remains still unexplained.

It is important to conclude this section by mentioning that the training setting discussed so far represents a convex optimization problem. However the DMFT analysis can be extended to non-convex settings and a simple model belonging to this class has been explored in \cite{mignacco2020dynamical}.
Finally a rigorous derivation of the DMFT equations has been presented in \cite{celentano2021high, gerbelot2022rigorous}.

\subsection{Effectiveness of SGD}\label{SGD_inference}
In the previous section we discussed how the SGD algorithm can be analyzed using DMFT to track its behavior in simple classification problems.
The purpose of this section is to discuss how to benchmark the performances of this algorithm and to compare it to GD.
It is clear that to do so one should consider hard high-dimensional non-convex optimization problems.
The natural candidates are therefore high-dimensional inference problems possessing an hard phase, as the mixed spiked
matrix-tensor model discussed in Sec.\ref{Mixed_MT_model}. 
However it is hard to define the SGD algorithm for the mixed spiked matrix-tensor given that the model lacks the bibartite structure 
where one can write the Hamiltonian as a sum of contributions each corresponding to a datapoint.

In \cite{mignacco2021stochasticity}  a different problem called phase retrieval was investigated to this purpose.
Consider an $N$-dimensional vector $\underline w^*$ with norm fixed to $|\underline w^*|^2=N$.
Additionally a set of $P=\alpha N$ random $N$-dimensional vectors $\underline \xi^\mu$ is introduced, with $\mu=1,\ldots, P$.
In the following the control parameter $\alpha$ will play the role of the sample complexity.
The coordinates of these vectors are i.i.d. Gaussian variables with zero mean and unit variance.

Given the signal vector $\underline w^*$ and the vectors $\underline \xi^\mu$, we can construct a dataset via the following rule
\beq
y_\mu=\left(\frac 1{\sqrt N} \underline w^* \cdot \underline \xi^\mu\right)^2\:.
\label{phase_retrieval_dataset}
\eeq
This allows to define a set of datapoints in which the vector $\underline \xi^\mu$ is correlated with its corresponding label $y_\mu$ through the signal vector $\underline w^*$.

Given this dataset we can assume that it comes from a rule with the same form as in Eq.~\eqref{phase_retrieval_dataset} but where the signal $\underline w^*$ is unknown.
Therefore defining a predictor as
\beq
y=\left(\frac 1{\sqrt N}\underline w \cdot \underline \xi\right)^2\:,
\label{phase_retrieval_inference}
\eeq
the supervised learning task is to find $\underline w$ that best fits the training dataset. It is clear that the best performances will be reached when $\underline w=\pm \underline w^*$.
We note that given that the vectors $\underline \xi^\mu$ are Gaussian, statistical rotational invariance implies that we can consider $\underline w^*$ to be $\underline w^*=\{1,\ldots, 1\}$ without losing generality.
A natural cost function to perform this task is given by the square loss as
\beq
H[\underline w]  =\frac 12 \sum_{\mu=1}^P \left(y_\mu-\left(\frac 1{\sqrt{N}} \underline w\cdot \underline \xi^\mu\right)^2\right)^2 = \frac 12 \sum_{\mu=1}^P \left(\left(\frac 1{\sqrt{N}} \underline w^*\cdot \underline \xi^\mu\right)^2-\left(\frac 1{\sqrt{N}} \underline w\cdot \underline \xi^\mu\right)^2\right)^2\:.
\label{loss_PR}
\eeq

We first observe that this cost function defines an high-dimensional non-convex loss surface. This is because the rule in Eq.~\eqref{phase_retrieval_inference}
is non-linear, both in the data vector and on the fitting parameters.
The name of the problem, phase retrieval, comes from the fact that it is a special case of a problem where all vectors are complex variables, and the non-linearity 
in Eq.~\eqref{phase_retrieval_dataset} is preceded by an absolute value so that the measured labels $y_\mu$ are insensitive to the phase of the vectors \cite{dong2023phase}.
The present special case is also referred in the literature as the {\it sign} retrieval problem.

The minimization of the loss function in Eq.~\eqref{loss_PR} can be done either via GD or via SGD both in its plain version or in the persistent one.
The performances of these algorithms, also compared to a Langevin version of simulated annealing, have been investigated in \cite{mignacco2021stochasticity}.
When the corresponding algorithms can take a continuous time limit, it has been shown that the DMFT can track exactly their performances.

Ideally, we would like to achieve a precise characterization of the recovery threshold of these algorithms.
Imagine that $\alpha$ is sufficiently large (even scaling with $N$, see \cite{sarao2020complex}). We can naively expect that all these algorithms 
will be able to recover the signal $\underline w^*$ since they will have enough information from the training dataset. 
However it is reasonable to think that the convergence of these algorithms to the ground state of $H$ will depend on $\alpha$ itself
and in particular it should exist a recovery threshold $\alpha^*$ such that for $\alpha<\alpha^*$ these algorithms are not able to find back the signal vector.
The recovery threshold  $\alpha^*$ can be expected to be algorithm-dependent and therefore a natural question is what is the best algorithm which allows to reconstruct
the signal with the smallest possible sample complexity, namely with the smallest amount of information.

This question has been investigated in \cite{sarao2020complex, mignacco2021stochasticity}. In particular in \cite{mignacco2021stochasticity} 
it has been shown that extensive numerical simulations seem to hint that SGD and especially its persistent version are significantly better than GD
in the sense that their recovery threshold seem to be smaller than the pure GD algorithm. This suggests that SGD provides an out-of-equilibrium noise
which may be helpful for the optimization dynamics.
However the results of \cite{mignacco2021stochasticity} are essentially limited: establishing the performances of these algorithms in the $N\to \infty$ limit is hard to do given that 
\begin{itemize}
\item numerical simulations are plagued by huge finite size effects which prevent a safe finite size scaling analysis of their relaxation time;
\item the DMFT equations, which would allow an analysis of the performances of the algorithms in the thermodynamic limit, cannot be integrated efficiently over long timescales in this particular case.
\end{itemize}
Therefore while empirical evidence suggested that SGD may be significantly better than GD, it was not possible to establish this result
from a clear theoretical viewpoint in the phase retrieval problem.

An interesting progress came out by considering a different (and simpler) model whose study was initiated by Fyodorov in \cite{fyodorov2019spin}.
The main idea is to relax the phase retrieval rule so that the labels $y_\mu$ are Gaussian variables when seen from the perspective of the dataset.
Practically one changes the dataset according to the following construction. 
Consider a set of $N\times N$ symmetric random matrices $J^\mu$ whose entries are independent Gaussian random variables with zero mean and unit variance.
Introducing again a signal vector $\underline w^*$ we can define a dataset as
\beq
y_\mu=\frac 1N \sum_{i<j}J^\mu_{ij}w_i^*w_j^*
\label{fyo_model}
\eeq
Therefore the dataset  consists now in the pairs $\{J^\mu,y_\mu\}$ and $\mu=1,\ldots,P=\alpha N$ being $\alpha$ the sample complexity.

At this point one can play the same game as in the phase retrieval problem trying to infer the signal vector from the dataset and knowing the structure of the non-linear measurement process.
To do that a square loss function can be employed:
\beq
H[\underline w]  =\frac 12 \sum_{\mu=1}^P \left(y_\mu-\frac 1N \sum_{i<j}J^\mu_{ij}w_iw_j\right)^2 = \frac 12 \sum_{\mu=1}^P \left(\frac 1N \sum_{i<j}J^\mu_{ij}w_i^*w_j^*-\frac 1N \sum_{i<j}J^\mu_{ij}w_iw_j\right)^2
\label{loss_fyo}
\eeq

It is clear that this problem is similar to the Canyon model of confluent tissues. The main difference is that what previously was a control parameter, $\sigma$, here becomes a random variable $y_\mu$ which is correlated with the random matrix $J^\mu$ and is constraint dependent.
In \cite{fyodorov2019spin} the ground state of the loss function has been analyzed suggesting that this model has a replica symmetric ground state for $\alpha>1$ which is given by the signal (or $-\underline w^*$ given the ${\mathbb Z}_2$ symmetry of the problem).
However much less is know from the algorithmic point of view. 

The performances of GD and SGD to minimize the loss function in Eq.~\eqref{loss_fyo} were analyzed in \cite{kamali2023stochastic} where a DMFT approach to the problem was developed.
Interestingly, it turns out that for this particular problem, the DMFT analysis can be performed much more efficiently with respect to phase retrieval and the corresponding equations can be integrated for sufficiently long times to allow a reliable extrapolation of the relaxation times of the algorithms.
This analysis has shown that for this particular problem the recovery threshold of the SGD algorithm is at smaller sample complexity of the corresponding one of GD and this unambiguously shows
in a high-dimensional non-convex hard optimization problem that the out-of-equilibrium noise of SGD can be helpful to improve the performances of GD.

\section{Perspectives}
Simple high-dimensional inference problems provide a natural framework where to benchmark optimization algorithms.
In this chapter I have reviewed my research activity in this context.
I showed that dynamical mean field theory can be used to track exactly the dynamics of gradient based algorithms in sufficiently simple, yet paradigmatic, high-dimensional non-convex inference problems.
This analysis has allowed to show that
\begin{itemize}
\item Algorithms that try to sample the posterior measure such as the Langevin algorithm are typically worse that approximate message passing and this is due to the underlining glassiness of the posterior measure.
\item The out-of--equilibrium noise of the SGD algorithm allows to improve the performances of pure GD in terms of the recovery threshold or signal to noise ratio.
\end{itemize}
It is important to note that these results, including the ones concerning the performances of SGD, have been mainly obtained in the context of inference problems.
Indeed if we focus for a moment on the model defined in Eq.~\eqref{fyo_model}, it can be shown, see \cite{kamali2023stochastic} that the recovery threshold $\alpha^*$ is larger than one.
In other words the corresponding model is in an underparametrized phase given that the size of the dataset is larger than the dimension of the weight vector $\underline w$.

Standard artificial neural network models are instead used in the overparametrized phase where the size of the dataset is smaller than the number of degrees of freedom.
Therefore a natural perspective of the current research line is to develop a DMFT analysis to study simple models of artificial neural networks where we can introduce the effect of overparamtrization.
One possibility is to generalize the model in Eq.~\eqref{fyo_model} to include a simple one hidden layer structure as in \cite{sarao2020optimization}.

However I believe that the most interesting perspective is to develop simple variants of the same model to study the phase space structure of the loss landscape and how the generalization properties
of SGD are correlated with the landscape itself. 
This class of models offers a true exactly soluble way to pinpoint all the research lines outlined at the beginning of Sec.\ref{sec_GD_percp}. Questions around the flatness of the landscape, its correlation to the generalization properties, and the way in which GD-based algorithms explores the loss surface can be fully understood in simple yet fully high-dimensional and non-convex settings and work is in progress to address these points.

\fontfamily{palatino}\fontseries{ppl}\fontsize{11}{11}\selectfont

\chapter{Conclusions}
In this manuscript I tried to cover the essential points of my research activity around three rather different research fields
in the statistical physics of disordered systems and their interdisciplinary applications.
In a nutshell this activity turns around three main fundamental questions:
\begin{itemize}
\item {\it What is the nature of amorphous solids at low temperature? Can we understand their properties in a unifying perspective?}
\item {\it What are the universal properties of satisfiability points of continuous constraint satisfaction problems? Their universality classes? The effect of finite dimensionality?}
\item {\it How to benchmark optimization and inference algorithms in paradigmatic high-dimensional non-convex optimization problems? Can we understand the properties of Stochastic Gradient Descent?}
\end{itemize}
While I believe that the present body of work is far from closing these questions, I also think that it has allowed to uncover a set of new ideas and useful insights to try to clarify them.
The main take home messages of this manuscript can be summarized as follows.
\begin{itemize}
\item A theory of structural glasses at low temperature cannot be fully developed if the problem of the spin glass transition in a field is not clarified.
These two problems are intimately related through the Gardner transition and its presence or lack thereof in three dimensional systems has become a central problem in the field.
The main proposal that came out through the work presented in the first chapter is that there are two mechanisms for the emergence of a spin glass transition in a field at zero temperature. One of them, associated to the emergence of non-linear excitations in the form of two level systems, has more 
plausibility to be realized in finite dimensions. {\it It does not require a divergent spin glass susceptibility} and therefore may also explain numerical and experimental findings
where the signs of such divergence are either not found or not strong.
\item Simple non-convex, high-dimensional continuous constraint satisfaction and optimization problems can serve as good mean field paradigms for a number of finite dimensional systems.
An important outcome of this research line is that many models which differ for their microscopic interactions lead to the same universal physics when they are pushed close
to their satisfiability/rigidity transitions, or when it comes to investigating accessible local minima. 
This concept of universality is rather different from what is typically found in phase transitions of clean systems. It rather concerns some universal self-organization of accessible minima of high-dimensional non-convex functions. This aspects of disordered system physics was previously referred to as {\it marginal stability} but, as the case of jamming critical system clearly shows, marginal stability concepts are not enough to capture quantitatively the behavior of these systems. The scaling theory reviewed in the second chapter instead suggests that {\it self-similarity/stochastic stability in the space of accessible configurations} (rather than in real space) is the key ingredient to understand these systems in more detail and to build a theory which goes beyond the replica approach. 
Moreover I believe that one of the most important messages of this body of works is that continuous constraint satisfaction problems offer a simple yet paradigmatic set of models to investigate the energy landscape of several systems, from glassy systems to neural networks. The employment of continuous variables brings new physical behaviors related to the organization of such landscapes, their responses under perturbations and the way in which local algorithms explore them. 
\item Benchmarking the performances of optimization algorithms in high-dimensional optimization problems is possible in well controlled simple, yet paradigmatic, settings. 
Approximate message passing algorithms have been shown to be better in toy models where gradient based exploration/sampling algorithms are hampered by the underlying glassiness of the energy landscape. A theory of stochastic gradient descent can be developed and has lead to show in a well controlled setting that its out-of-equilibrium noise can be helpful for optimization.
I believe that the use of dynamical mean field theory can serve as a tool to answer several well posed questions about how practical algorithms explore phase space and how their performances are related to it.
This research line is still in the infancy but I think that the main outcome of the works reviewed in this manuscript is that we are now in the position where to address several questions and answer to them in simple settings. 
\end{itemize}
One of the main message of this manuscript is that many of the ideas developed here emerged from the introduction of new models. 
In particular in the first chapter the KHGPS model was discussed and I believe that this belongs to a new class of models in which local heterogeneities mix with disordered interactions. These models turn out to be interesting from several points of view (see for example \cite{mergny2021stability} for an application to ecological models).
In chapter 2 instead many interesting results have been derived by mapping several finite dimensional models of glasses to perceptron problems. 
There are two main classes of models differing by the nature of the constraints imposed to the degree of freedom, being them inequalities or equalities.
While perceptrons with inequality constraints were previously studied since the seminal work of Gardner, I think that the Canyon model to describe high-dimensional non-linear equality constraints provides a new interesting model to study energy landscapes with a nature different from the ones considered up to now. Given the nature of the constraints, this model is a natural candidate to address questions in the realm of high-dimensional non-linear regression which is essentially at the basis of all artificial neural network computations.
It also allows to define non-linear and rough manifold models which can either be at the basis of {\it overparametrized landscape models} or {\it data manifold models} with tunable regularity structure.

Finally I would like to make a comment on the way in which many of the key ideas described in this manuscript have been developed. 
The KHGPS model was introduced to solve an apparent discrepancy between the density of vibrational states in structural glasses and the predictions of mean field theory. It turned out
to be a very interesting model giving a new perspective on the problem of the spin glass transition in  a field in finite dimensional systems.
The Canyon model was instead introduced as a model to describe the rigidity transition in confluent tissues but I think that its relevance goes well beyond these particular systems 
due to the paradigmatic properties of the energy landscape it defines and to the benefits in terms of theoretical solubility it offers.
These are two among several examples of ideas developed in one direction and turned out to be useful in others.
This conceptual {\it self-attention} on my research activity across the topics reviewed here is, on a personal note, what I have been enjoying the most
during these years.

\section{Perspectives}
I conclude this manuscript by mentioning what are the perspective that I find more interesting about the results described here.
The main of them can be summarized as follows.
\begin{itemize}
\item Concerning the problem of the spin glass transition in a field, there are two main questions that come next. First, I think that the question about how to characterize the size of non-linear excitations is still open from a theoretical viewpoint. We know how to solve models where these excitations are localized in single degrees of freedom (single spin flip in mean field spin glasses or opening single contacts in sphere systems). However it is clear that this is not general and I think that one needs to understand how to incorporate the size of non-linear excitations in the replica theory developed so far. Next to this question is the one about how to extract more information from the zero temperature field theory described in Chapter 1. It is likely that these two questions are related, and that understanding the former may help with the latter.
\item Concerning continuous constraint satisfaction problems, I see two main research directions that are clearly important and whose solution may be accessible within the next years. First, the development of a theory of self-similarity and stochastic stability of accessible configurations beyond the replica theory should be possible. A consistent theory should lead to a classification of possible scaling solutions and should be constructive. In a nutshell we would like to have the analog of what the conformal bootstrap represents for second order phase transitions. Moreover a more detailed analysis of the energy landscape of these problems as probed by out-of-equilibrium algorithms is certainly within reach and I believe that the most interesting model where to develop such studies is the Canyon model.
\item Concerning optimization algorithms, I think that the works reviewed in the third chapter of this manuscript laid down the basis for a more careful study of SGD and its properties. Most of the questions that have been asked in the field about the properties of the loss landscape, its correlation to generalization and the way in which the SGD noise interplays with them can be attacked within the model systems I have discussed. It is also clear that DMFT will be among the most important tools to get precise theoretical results.
\end{itemize}

To conclude well this thesis I think that it is important to mention the other research directions that will attract my attention in the next years and that are rather far from the topics reviewed here.

The activity around problems related to inference and learning has driven me towards the question of understanding how in general non-conservative dynamical systems can self-organize to learn to perform tasks.
It is clear that most biological systems can be ascribed to this setting. In this context, the problem of processing information via non-conservative stochastic dynamical processes is key to adaptation and survival.
Non-conservative dynamics is rather different from stochastic descent on energy (loss) landscapes. When non-conservative forces are at play the performances of these systems cannot be understood as the fixed points of Lyapunov functions. Nevertheless nature shows that learning is possible also in this framework. 
One of the prominent examples that are believed to be in this class of systems are biological neural networks where asymmetric interactions between neurons can drive the collective dynamics towards high-dimensional chaotic attractors. It is widely believed that chaos can be controlled in this setting and can be helpful for computation and information processing, see for example \cite{sussillo2009generating}. 
Understanding the underlining biophysical mechanisms for learning in such high-dimensional non-linear dynamical systems is very important and I have started a research program in this direction, see \cite{fournier2023statistical}.
It is clear that these settings require different paradigms, models and ideas to make progresses and are interesting challenges for which the toolbox and experience summarized in this manuscript may be of some help.

\linespread{0.9}
\AtNextBibliography{\footnotesize}
\setlength\bibitemsep{0.5\itemsep}
\printbibliography

\end{document}